\pdfoutput=1

\documentclass[12pt,a4paper]{article}

\usepackage{ifthen} 
\newboolean{pdflatex}
\setboolean{pdflatex}{true} 

\newboolean{articletitles}
\setboolean{articletitles}{true} 

\newboolean{uprightparticles}
\setboolean{uprightparticles}{false} 

\newboolean{inbibliography}
\setboolean{inbibliography}{false} 

\newcommand{\mpr}        {\mbox{$m^\prime$}}
\newcommand{\thpr}       {\mbox{$\theta^\prime$}}

\usepackage{microtype}
\usepackage{lineno}  
\usepackage{xspace} 
\usepackage{caption} 

\usepackage{graphicx}  
\usepackage{rotating}
\usepackage{color}
\usepackage{colortbl}
\graphicspath{{./figs/}} 

\usepackage{amsmath} 
\usepackage{amssymb}
\usepackage{amsfonts}
\usepackage{upgreek} 

\newcommand*\patchAmsMathEnvironmentForLineno[1]{%
\expandafter\let\csname old#1\expandafter\endcsname\csname #1\endcsname
\expandafter\let\csname oldend#1\expandafter\endcsname\csname
end#1\endcsname
 \renewenvironment{#1}%
   {\linenomath\csname old#1\endcsname}%
   {\csname oldend#1\endcsname\endlinenomath}%
}
\newcommand*\patchBothAmsMathEnvironmentsForLineno[1]{%
  \patchAmsMathEnvironmentForLineno{#1}%
  \patchAmsMathEnvironmentForLineno{#1*}%
}
\AtBeginDocument{%
\patchBothAmsMathEnvironmentsForLineno{equation}%
\patchBothAmsMathEnvironmentsForLineno{align}%
\patchBothAmsMathEnvironmentsForLineno{flalign}%
\patchBothAmsMathEnvironmentsForLineno{alignat}%
\patchBothAmsMathEnvironmentsForLineno{gather}%
\patchBothAmsMathEnvironmentsForLineno{multline}%
\patchBothAmsMathEnvironmentsForLineno{eqnarray}%
}

\usepackage{hyperref}    
\usepackage[all]{hypcap} 

\def\dkpi {\ensuremath{\Dp\Km\pim}\xspace}
\def\dpipi {\ensuremath{\Dp\pim\pim}\xspace}

\def\Btodkpi {\ensuremath{\Bm\to\dkpi}\xspace}
\def\Btodpipi {\ensuremath{\Bm\to\dpipi}\xspace}
\def\Btodstarpipi {\ensuremath{\Bm\to\Dstarp\pim\pim}\xspace}
\def\Btodstarkpi {\ensuremath{\Bm\to\Dstarp\Km\pim}\xspace}
\def\Btodanddstarkpi {\ensuremath{\Bm\to D^{(*)+}\Km\pim}\xspace}

\def\kpipi {\ensuremath{\Km\pip\pip}\xspace}
\def\Dtokpipi {\ensuremath{\Dp\to\kpipi}\xspace}

\def\swave {S-wave\xspace}
\def\Dpi {\ensuremath{\Dp\pim}\xspace}
\def\Dpiswave {\Dpi\swave}
\def\dstarv       {\ensuremath{\D^{*}_{v}(2007)^{0}}\xspace} 
\def\olddstarzero {\ensuremath{\D^{*}_{0}(2400)^{0}}\xspace} 
\def\olddstartwo {\ensuremath{\D^{*}_{2}(2460)^{0}}\xspace} 
\def\newdstarone {\ensuremath{\D^{*}_{1}(2680)^{0}}\xspace} 
\def\newdstartwo {\ensuremath{\D^{*}_{2}(3000)^{0}}\xspace} 
\def\newdstarthree {\ensuremath{\D^{*}_{3}(2760)^{0}}\xspace} 

\def\msqdpimin {\ensuremath{m^2(\Dpi)_{\rm min}}\xspace}
\def\msqdpimax {\ensuremath{m^2(\Dpi)_{\rm max}}\xspace}

\def\lhcb {\mbox{LHCb}\xspace}

\def\babar  {\mbox{BaBar}\xspace}

\def\MagUp {\mbox{\em Mag\kern -0.05em Up}\xspace}

\ifthenelse{\boolean{uprightparticles}}%
{

 \def\Ppi         {\ensuremath{\uppi}\xspace}

 \def\PDelta      {\ensuremath{\Delta}\xspace}                 
 \def\PXi      {\ensuremath{\Xi}\xspace}                 
 \def\PLambda      {\ensuremath{\Lambda}\xspace}                 
 \def\PSigma      {\ensuremath{\Sigma}\xspace}                 
 \def\POmega      {\ensuremath{\Omega}\xspace}                 
 \def\PUpsilon      {\ensuremath{\Upsilon}\xspace}

 \def\PB      {\ensuremath{\mathrm{B}}\xspace}                 
                  
 \def\PD      {\ensuremath{\mathrm{D}}\xspace}

 \def\PK      {\ensuremath{\mathrm{K}}\xspace}

 \def\Pb      {\ensuremath{\mathrm{b}}\xspace}                 
 \def\Pc      {\ensuremath{\mathrm{c}}\xspace}

 \def\Pi      {\ensuremath{\mathrm{i}}\xspace}

 \def\Pn      {\ensuremath{\mathrm{n}}\xspace}                 
                  
 \def\Pp      {\ensuremath{\mathrm{p}}\xspace}

 \def\Ps      {\ensuremath{\mathrm{s}}\xspace}

}
{

 \def\Ppi         {\ensuremath{\pi}\xspace}

 \mathchardef\PDelta="7101
 \mathchardef\PXi="7104
 \mathchardef\PLambda="7103
 \mathchardef\PSigma="7106
 \mathchardef\POmega="710A
 \mathchardef\PUpsilon="7107
                  
 \def\PB      {\ensuremath{B}\xspace}                 
                  
 \def\PD      {\ensuremath{D}\xspace}

 \def\PK      {\ensuremath{K}\xspace}

 \def\Pb      {\ensuremath{b}\xspace}                 
 \def\Pc      {\ensuremath{c}\xspace}

 \def\Pi      {\ensuremath{i}\xspace}

 \def\Pn      {\ensuremath{n}\xspace}                 
                  
 \def\Pp      {\ensuremath{p}\xspace}

 \def\Ps      {\ensuremath{s}\xspace}

}

\makeatletter
\ifcase \@ptsize \relax
  \newcommand{\miniscule}{\@setfontsize\miniscule{4}{5}}
\or
  \newcommand{\miniscule}{\@setfontsize\miniscule{5}{6}}
\or
  \newcommand{\miniscule}{\@setfontsize\miniscule{5}{6}}
\fi
\makeatother

\DeclareRobustCommand{\optbar}[1]{\shortstack{{\miniscule (\rule[.5ex]{1.25em}{.18mm})}
  \\ [-.7ex] $#1$}}

\def\squark    {{\ensuremath{\Ps}}\xspace}

\def\cquark    {{\ensuremath{\Pc}}\xspace}

\def\bquark    {{\ensuremath{\Pb}}\xspace}

\def\pion   {{\ensuremath{\Ppi}}\xspace}
\def\piz    {{\ensuremath{\pion^0}}\xspace}

\def\pip    {{\ensuremath{\pion^+}}\xspace}
\def\pim    {{\ensuremath{\pion^-}}\xspace}

\def\kaon    {{\ensuremath{\PK}}\xspace}
  \def\Kbar    {{\kern 0.2em\overline{\kern -0.2em \PK}{}}\xspace}

\def\KorKbar    {\kern 0.18em\optbar{\kern -0.18em K}{}\xspace}

\def\Kp      {{\ensuremath{\kaon^+}}\xspace}
\def\Km      {{\ensuremath{\kaon^-}}\xspace}

  \def\Dbar    {{\kern 0.2em\overline{\kern -0.2em \PD}{}}\xspace}
\def\D       {{\ensuremath{\PD}}\xspace}

\def\DorDbar    {\kern 0.18em\optbar{\kern -0.18em D}{}\xspace}
\def\Dz      {{\ensuremath{\D^0}}\xspace}
\def\Dzb     {{\ensuremath{\Dbar{}^0}}\xspace}
\def\Dp      {{\ensuremath{\D^+}}\xspace}

\def\Dstar   {{\ensuremath{\D^*}}\xspace}

\def\Dstarp  {{\ensuremath{\D^{*+}}}\xspace}

\def\Dsp     {{\ensuremath{\D^+_\squark}}\xspace}

\def\B       {{\ensuremath{\PB}}\xspace}
\def\Bbar    {{\ensuremath{\kern 0.18em\overline{\kern -0.18em \PB}{}}}\xspace}

\def\BorBbar    {\kern 0.18em\optbar{\kern -0.18em B}{}\xspace}
\def\Bz      {{\ensuremath{\B^0}}\xspace}
\def\Bzb     {{\ensuremath{\Bbar{}^0}}\xspace}
\def\Bu      {{\ensuremath{\B^+}}\xspace}
\def\Bub     {{\ensuremath{\B^-}}\xspace}
\def\Bp      {{\ensuremath{\Bu}}\xspace}
\def\Bm      {{\ensuremath{\Bub}}\xspace}

  \def\Y#1S{\ensuremath{\PUpsilon{(#1S)}}\xspace}

\def\FourS {{\Y4S}}

\def\proton      {{\ensuremath{\Pp}}\xspace}

\def\neutron     {{\ensuremath{\Pn}}\xspace}

\def\Lz          {{\ensuremath{\PLambda}}\xspace}
\def\Lbar        {{\ensuremath{\kern 0.1em\overline{\kern -0.1em\PLambda}}}\xspace}
\def\LorLbar    {\kern 0.18em\optbar{\kern -0.18em \PLambda}{}\xspace}

\def\Lc      {{\ensuremath{\Lz^+_\cquark}}\xspace}

\def\to                 {\ensuremath{\rightarrow}\xspace}

\def\AT#1     {\ensuremath{A_{\mathrm{T}}^{#1}}\xspace}           

\def\C#1      {\ensuremath{\mathcal{C}_{#1}}\xspace}                       
\def\Cp#1     {\ensuremath{\mathcal{C}_{#1}^{'}}\xspace}                    
\def\Ceff#1   {\ensuremath{\mathcal{C}_{#1}^{\mathrm{(eff)}}}\xspace}        
\def\Cpeff#1  {\ensuremath{\mathcal{C}_{#1}^{'\mathrm{(eff)}}}\xspace}       
\def\Ope#1    {\ensuremath{\mathcal{O}_{#1}}\xspace}                       
\def\Opep#1   {\ensuremath{\mathcal{O}_{#1}^{'}}\xspace}                    

\newcommand{\tev}{\ifthenelse{\boolean{inbibliography}}{\ensuremath{~T\kern -0.05em eV}\xspace}{\ensuremath{\mathrm{\,Te\kern -0.1em V}}}\xspace}
\newcommand{\gev}{\ensuremath{\mathrm{\,Ge\kern -0.1em V}}\xspace}
\newcommand{\mev}{\ensuremath{\mathrm{\,Me\kern -0.1em V}}\xspace}
\newcommand{\kev}{\ensuremath{\mathrm{\,ke\kern -0.1em V}}\xspace}
\newcommand{\gevnsp}{\ensuremath{\mathrm{Ge\kern -0.1em V}}\xspace}
\newcommand{\mevnsp}{\ensuremath{\mathrm{Me\kern -0.1em V}}\xspace}
\newcommand{\kevnsp}{\ensuremath{\mathrm{ke\kern -0.1em V}}\xspace}
\newcommand{\ev}{\ensuremath{\mathrm{\,e\kern -0.1em V}}\xspace}
\newcommand{\gevc}{\ensuremath{{\mathrm{\,Ge\kern -0.1em V\!/}c}}\xspace}
\newcommand{\mevc}{\ensuremath{{\mathrm{\,Me\kern -0.1em V\!/}c}}\xspace}
\newcommand{\gevcc}{\ensuremath{{\mathrm{\,Ge\kern -0.1em V\!/}c^2}}\xspace}
\newcommand{\gevgevcccc}{\ensuremath{{\mathrm{\,Ge\kern -0.1em V^2\!/}c^4}}\xspace}
\newcommand{\mevcc}{\ensuremath{{\mathrm{\,Me\kern -0.1em V\!/}c^2}}\xspace}

\def\mm   {\ensuremath{\rm \,mm}\xspace}

\def\mum  {\ensuremath{{\,\upmu\rm m}}\xspace}

\def\fm   {\ensuremath{\rm \,fm}\xspace}

\def\invfb   {\ensuremath{\mbox{\,fb}^{-1}}\xspace}

\def\gsim{{~\raise.15em\hbox{$>$}\kern-.85em
          \lower.35em\hbox{$\sim$}~}\xspace}
\def\lsim{{~\raise.15em\hbox{$<$}\kern-.85em
          \lower.35em\hbox{$\sim$}~}\xspace}

\newcommand{\Real}{\ensuremath{\mathcal{R}e}\xspace}

\def\ptot       {\mbox{$p$}\xspace}
\def\pt         {\mbox{$p_{\rm T}$}\xspace}

\def\evtgen     {\mbox{\textsc{EvtGen}}\xspace}

\def\geant      {\mbox{\textsc{Geant4}}\xspace}

\def\photos     {\mbox{\textsc{Photos}}\xspace}

\def\pythia     {\mbox{\textsc{Pythia}}\xspace}

\def\tell1  {TELL1\xspace}
\def\ukl1   {UKL1\xspace}

\newcommand{\ie}{\mbox{\itshape i.e.}\xspace}

\usepackage{cite} 
\usepackage{mciteplus}
\usepackage{multirow}
\usepackage{afterpage}
\allowdisplaybreaks

\begin{document}

\renewcommand{\thefootnote}{\fnsymbol{footnote}}
\setcounter{footnote}{1}

\begin{titlepage}
\pagenumbering{roman}

\vspace*{-1.5cm}
\centerline{\large EUROPEAN ORGANIZATION FOR NUCLEAR RESEARCH (CERN)}
\vspace*{1.5cm}
\hspace*{-0.5cm}
\begin{tabular*}{\linewidth}{lc@{\extracolsep{\fill}}r}
\\
 & & CERN-EP-2016-184 \\  
 & & LHCb-PAPER-2016-026 \\  
 & & August 3, 2016 \\ 
 & & \\
\end{tabular*}

\vspace*{2.5cm}

{\bf\boldmath\huge
\begin{center}
  Amplitude analysis of \Btodpipi decays
\end{center}
}

\vspace*{1.5cm}

\begin{center}
The LHCb collaboration\footnote{Authors are listed at the end of this paper.}
\end{center}

\vspace{\fill}

\begin{abstract}
  \noindent
  The Dalitz plot analysis technique is used to study the resonant substructures of \Btodpipi decays in a data sample corresponding to $3.0\invfb$ of $pp$ collision data recorded by the LHCb experiment during 2011 and 2012.
  A model-independent analysis of the angular moments demonstrates the presence of resonances with spins 1, 2 and 3 at high $\Dp\pim$ mass.
  The data are fitted with an amplitude model composed of a quasi-model-independent function to describe the \Dpiswave\ together with virtual contributions from the $\Dstar(2007)^{0}$ and $\B^{*0}$ states, and components corresponding to the \olddstartwo, \newdstarone, \newdstarthree\ and \newdstartwo\ resonances.
  The masses and widths of these resonances are determined together with the branching fractions for their production in \Btodpipi decays.
  The \Dpiswave\ has phase motion consistent with that expected due to the presence of the \olddstarzero\ state.
  These results constitute the first observations of the \newdstarthree\ and \newdstartwo\ resonances, with significances of $10\,\sigma$ and $6.6\,\sigma$, respectively.
\end{abstract}

\vspace*{1.5cm}

\begin{center}
  Submitted to Phys.~Rev.~D
\end{center}

\vspace{\fill}

{\footnotesize 
\centerline{\copyright~CERN on behalf of the \lhcb collaboration, licence \href{http://creativecommons.org/licenses/by/4.0/}{CC-BY-4.0}.}}
\vspace*{2mm}

\end{titlepage}


\newpage
\setcounter{page}{2}
\mbox{~}

\cleardoublepage


\renewcommand{\thefootnote}{\arabic{footnote}}
\setcounter{footnote}{0}



\pagestyle{plain} 
\setcounter{page}{1}
\pagenumbering{arabic}


\section{Introduction}
\label{sec:introduction}

There is strong theoretical and experimental interest in charm meson spectroscopy because it provides opportunities to study QCD predictions within the context of different \mbox{models~\cite{Godfrey:1985xj,Isgur:1991wq,Colangelo:2012xi,Mohler:2012na,Moir:2016srx}}.
Experimental knowledge of the masses, widths and spins of the charged and neutral orbitally-excited (1P) charm meson states has been gained through analyses of both prompt production~\cite{delAmoSanchez:2010vq,LHCb-PAPER-2013-026} and three-body decays of $B$ mesons~\cite{Abe:2003zm,Aubert:2009wg,Kuzmin:2006mw,LHCb-PAPER-2014-070,LHCb-PAPER-2015-007,LHCb-PAPER-2015-017}.
Progress has been equally strong for excited charm-strange ($c\bar{s}$) mesons~\cite{Brodzicka:2007aa,LHCb-PAPER-2014-035,LHCb-PAPER-2014-036,Lees:2014abp,LHCb-PAPER-2015-052}.
These studies have in addition revealed several new states at higher masses, most of which have not yet been confirmed by analyses of independent data samples.
Moreover, quantum numbers are only known for states studied in amplitude analyses of multibody \B meson decays, since analyses of promptly produced excited charm states only determine whether the spin-parity is natural (\ie $J^P = 0^+, 1^-, 2^+, ...$) or unnatural (\ie $J^P = 0^-, 1^+, 2^-, ...$), not the resonance spin.
The experimental status of the neutral excited charm states is summarised in Table~\ref{tab:PDG} (here and throughout the paper, natural units with $\hbar = c = 1$ are used).
The \olddstarzero, $D_1(2420)^0$, $D_1^\prime(2430)^0$ and \olddstartwo\ mesons are generally understood to be the four 1P states.
The spectroscopic identification for heavier states is not clear.

\begin{table}[!b]
\centering
\caption{\small
  Measured properties of neutral excited charm states.
  World averages are given for the 1P resonances (top part), while all measurements are listed for the heavier states (bottom part).
  Where two uncertainties are given, the first is statistical and second systematic; where a third is given, it is due to model uncertainty.
  The uncertainties on the averages for the \olddstarzero\ mass and the $D_1(2420)^0$ and \olddstartwo\ masses and widths are inflated by scale factors to account for inconsistencies between measurements.
  The quoted \olddstartwo\ averages do not include the recent result from Ref.~\cite{LHCb-PAPER-2015-007}.
} 
\begin{tabular}{ccccc}
  \hline \\ [-2.5ex]
  Resonance & Mass $(\mevnsp)$ & Width $(\mevnsp)$ & $J^P$ & Ref. \\
  \hline \\ [-2.5ex]
  \olddstarzero & $\phantom{.0}2318 \pm 29\phantom{.}$ & $267 \pm 40$ & $0^+$ & \cite{PDG2014} \\
  $D_1(2420)^0$ & $2421.4 \pm 0.6$ & $27.4 \pm 2.5$ & $1^+$ & \cite{PDG2014} \\
  $D_1^\prime(2430)^0$ & $\phantom{.0}2427 \pm 40\phantom{.}$ & $384\,^{+130}_{-110}$ & $1^+$ & \cite{PDG2014} \\
  \olddstartwo & $2462.6 \pm 0.6$ & $49.0 \pm 1.3$ & $2^+$ & \cite{PDG2014} \\
  \hline \\ [-2.5ex]
  $D^*(2600)$ & $2608.7 \pm 2.4 \pm 2.5$ & $\phantom{1.}93 \pm \phantom{1.}6 \pm \phantom{1.}13$ & natural & \cite{delAmoSanchez:2010vq} \\
  $D^*(2650)$ & $2649.2 \pm 3.5 \pm 3.5$ & $\phantom{.}140 \pm \phantom{.}17 \pm \phantom{1.}19$ & natural & \cite{LHCb-PAPER-2013-026} \\
  $D^*(2760)$ & $2763.3 \pm 2.3 \pm 2.3$ & $60.9 \pm 5.1 \pm \phantom{1}3.6$ & natural & \cite{delAmoSanchez:2010vq} \\
  $D^*(2760)$ & $2760.1 \pm 1.1 \pm 3.7$ & $74.4 \pm 3.4 \pm 19.1$ & natural & \cite{LHCb-PAPER-2013-026} \\
  $D_1^*(2760)^0$ & $2781 \pm 18 \pm 11 \pm 6$ & $177 \pm 32 \pm 20 \pm 7$ & $1^-$ & \cite{LHCb-PAPER-2015-007} \\
  \hline
\end{tabular}
\label{tab:PDG}
\end{table}

The $\Bm\to\Dp\pim\pim$ decay mode has been previously studied in Refs.~\cite{Abe:2003zm,Aubert:2009wg}. 
The inclusion of charge-conjugate processes is implied throughout the paper.
The Dalitz plot (DP) models that were used contained components for two excited charm states, the \olddstarzero\ and \olddstartwo\ resonances, together with nonresonant amplitudes.
More recently, a DP analysis of $\Bm\to\Dp\Km\pim$ decays~\cite{LHCb-PAPER-2015-007} included, in addition, a contribution from the $D_1^*(2760)^0$ state.
The properties of this state indicate that it belongs to the 1D family~\cite{Chen:2015lpa,Godfrey:2015dva}.
The $D_1^*(2760)^0$ width is found to be larger than in previous measurements based on prompt production, which may be due to a contribution from an additional resonance, as would be expected if both 2S and 1D states with spin-parity $J^P = 1^-$ are present in this region.
There should also be a 1D state with $J^P = 3^-$ at similar mass, as seen in the charm-strange system~\cite{LHCb-PAPER-2014-035,LHCb-PAPER-2014-036}.
As yet there is no evidence for such a neutral charm state, but a DP analysis of $\Bzb\to\Dz\pip\pim$ decays~\cite{LHCb-PAPER-2014-070} led to the first observation of the $D_3^*(2760)^+$ state.

One challenge for DP analyses with large data samples is the modelling of broad resonances that interfere with nonresonant amplitudes in the same partial wave.
Inclusion of both contributions in an amplitude fit can violate unitarity in the decay matrix element, and also gives results that are difficult to interpret due to large interference effects.
In the case of \Btodpipi decays this is particularly relevant for the \Dpiswave, where both the \olddstarzero\ resonance and a nonresonant contribution are expected.
In the $\pip\pim$ and $\Kp\pim$ systems such effects can be handled with a K-matrix approach or specific models such as the LASS function~\cite{lass} inspired by low-energy scattering data, respectively.
In the absence of any \Dpi\ scattering data, a viable alternative approach is to use a quasi-model-independent description, in which the partial wave is fitted using splines to describe the magnitude and phase as a function of $m(\Dpi)$. 
Determination of the phase depends on interference of the \swave\ with another partial wave, so that some model dependence remains due to the description of the other amplitudes in the decay.
This approach was first applied to the $K\pi$ \swave using $\Dtokpipi$ decays~\cite{Aitala:2005yh}. 
Subsequent uses include further studies of the $K\pi$ \swave~\cite{Bonvicini:2008jw,Link:2009ng,delAmoSanchez:2010fd,Lees:2015zzr} as well as the $\Kp\Km$~\cite{Aubert:2007dc} and $\pip\pim$~\cite{Aubert:2008ao} \swave{s}, in various processes.
Similar methods have been used to determine the phase motion of exotic hadron candidates~\cite{LHCb-PAPER-2015-038,LHCb-PAPER-2015-029}.
Quasi-model-independent information on the \Dpiswave\ could be used to develop better models of the dynamics in the \Dpi\ system~\cite{Kolomeitsev:2003ac,Vijande:2006hj,Guo:2006fu,Gamermann:2006nm}.

In this paper, the DP analysis technique is employed to study the contributing amplitudes in \Btodpipi decays, where the charm meson is reconstructed through $\Dtokpipi$ decays.
The analysis is based on a data sample corresponding to an integrated luminosity of $3.0\invfb$ of data collected with the LHCb detector during 2011 when the $pp$ collision centre-of-mass energy was $\sqrt{s} = 7 \tev$, and 2012 with $\sqrt{s} = 8 \tev$.

The paper is organised as follows.
Section~\ref{sec:detector} provides a brief description of the LHCb detector and the event reconstruction and simulation software.
The selection of signal candidates is described in Sec.~\ref{sec:selection} and the determination of signal and background yields is presented in Sec.~\ref{sec:mass-fit}.
The angular moments of \Btodpipi decays are studied in Sec.~\ref{sec:moments} and are used to guide the amplitude analysis.
The DP analysis formalism is reviewed briefly in Sec.~\ref{sec:dalitz-generalities}, and implementation of the amplitude fit is given in Sec.~\ref{sec:dalitz}.
Experimental and model-dependent systematic uncertainties are evaluated in Sec.~\ref{sec:systematics},
and the results and a summary are presented in Sec.~\ref{sec:results}.

\section{LHCb detector}
\label{sec:detector}

The \lhcb detector~\cite{Alves:2008zz,LHCb-DP-2014-002} is a single-arm forward
spectrometer covering the \mbox{pseudorapidity} range $2<\eta <5$,
designed for the study of particles containing \bquark or \cquark
quarks. The detector includes a high-precision tracking system
consisting of a silicon-strip vertex detector 
surrounding the $pp$ interaction region, a large-area silicon-strip detector
located upstream of a dipole magnet with a bending power of about
$4{\rm\,Tm}$, and three stations of silicon-strip detectors and straw
drift tubes placed downstream of the magnet.
The polarity of the dipole magnet is reversed periodically throughout data-taking.
The tracking system provides a measurement of momentum, \ptot, of charged particles with relative uncertainty that varies from $0.5\,\%$ at low momentum to $1.0\,\%$ at $200\gev$.
The minimum distance of a track to a primary vertex, the impact parameter (IP), is measured with a resolution of $(15+29/\pt)\mum$,
where \pt is the component of the momentum transverse to the beam, in~\gev.
Different types of charged hadrons are distinguished using information
from two ring-imaging Cherenkov detectors. 
Photon, electron and hadron candidates are identified by a calorimeter system consisting of scintillating-pad and preshower detectors, an electromagnetic calorimeter and a hadronic calorimeter. 
Muons are identified by a system composed of alternating layers of iron and multiwire proportional chambers. 

The trigger consists of a hardware stage, based on information from the calorimeter and muon systems, followed by a software stage, in which all tracks with $\pt>500~(300)\mev$ are reconstructed for data collected in 2011 (2012).
The software trigger line used in the analysis reported in this paper requires a two-, three- or four-track secondary vertex with significant displacement from the primary $pp$ interaction vertices~(PVs). 
At least one charged particle must have $\pt > 1.7\gev$ and be inconsistent with originating from the PV.
A multivariate algorithm~\cite{BBDT} is used for the identification of secondary vertices consistent with the decay of a \bquark hadron.

In the offline selection, the objects that fired the trigger are associated with reconstructed particles.  
Selection requirements can therefore be made not only on the trigger line that fired, but on whether the decision was due to the signal candidate, other particles produced in the $pp$ collision, or a combination of both.
Signal candidates are accepted offline if one of the final state particles created a cluster in the hadronic calorimeter with sufficient transverse energy to fire the hardware trigger. 

Simulated events are used to characterise the detector response to signal and
certain types of background events.
In the simulation, $pp$ collisions are generated using
\pythia~\cite{Sjostrand:2006za,*Sjostrand:2007gs} with a specific \lhcb
configuration~\cite{LHCb-PROC-2010-056}.  Decays of hadronic particles
are described by \evtgen~\cite{Lange:2001uf}, in which final state
radiation is generated using \photos~\cite{Golonka:2005pn}. The
interaction of the generated particles with the detector and its
response are implemented using the \geant
toolkit~\cite{Allison:2006ve, *Agostinelli:2002hh} as described in
Ref.~\cite{LHCb-PROC-2011-006}.

\section{Selection requirements}
\label{sec:selection}

The selection criteria are the same as those used in Ref.~\cite{LHCb-PAPER-2015-007}, where a detailed description is given, with the exception that only candidates that are triggered by at least one of the signal tracks are retained in order to minimise the uncertainty on the efficiency.
First, loose requirements are applied in order to obtain a visible peak in the \B candidate invariant mass distribution.
These criteria are found to be $91\,\%$ efficient on simulated signal decays.
The remaining data are then used to train two artificial neural networks~\cite{Feindt2006190} that separate signal from different categories of background.
The first is designed to distinguish candidates that contain real $\Dtokpipi$ decays from those that do not; the second separates signal \Btodpipi decays from background combinations.
The {\it sPlot} technique~\cite{Pivk:2004ty} is used to statistically separate signal decays from background combinations using the \D (\B) candidate mass as the discriminating variable for the first (second) network. 
The first network takes as input properties of the \D candidate and its decay product tracks, including information about kinematics, track and vertex quality.
The second uses a total of 27 input variables, including the output of the first network, as described in Ref.~\cite{LHCb-PAPER-2015-007}.
The neural network input quantities depend only weakly on the position in the DP, so that training the networks with the same data sample used for the analysis does not bias the results.
A requirement that reduces the combinatorial background by an order of magnitude, while retaining about $75\,\%$ of the signal, is imposed on the second neural network output.

Particle identification (PID) requirements are applied to all five final state tracks to select pions or kaons as necessary.
Background from $\Dsp \to \Km \Kp \pip$ decays, where the \Kp is misidentified as a \pip meson, are suppressed using a tight PID criterion on the higher momentum \pip from the \Dp decay. 
The combined efficiency of the PID requirements on the five final state tracks is determined using $D^{*+} \to D^{0}\pip$, $D^{0} \to \Km \pip$ calibration data~\cite{LHCb-DP-2012-003} and found to be around $70\,\%$.

Potential background from $\Lc \to \proton \Km\pip$ decays, misreconstructed as $\Dp$ candidates, is removed if the invariant mass lies in the range $2280$--$2300\mev$ when the proton mass hypothesis is applied to the low momentum pion track.
Decays of \Bm mesons to the $\Km\pip\pip\pim\pim$ final state that do not proceed via an intermediate charm state are removed by requiring that the
\D and \B candidate decay vertices are separated by at least $1 \mm$. 
The signal efficiency of this requirement is approximately $85\,\%$.

To improve mass resolution, the momenta of the final state tracks are rescaled~\cite{LHCb-PAPER-2012-048,LHCb-PAPER-2013-011} 
using weights obtained from a sample of $J/\psi \to \mu^{+} \mu^{-}$ decays where the measured mass peak is matched to the known value~\cite{PDG2014}.
Additionally, a kinematic fit~\cite{Hulsbergen:2005pu} is performed to candidates in which the invariant mass of the \D decay products is constrained to equal the world average \D mass~\cite{PDG2014}. 
A \B mass constraint is added in the calculation of the variables that are used in the DP fit.

Candidate \B mesons with invariant mass in the range $5100$--$5800 \mev$ are retained for further analysis.
Following all selection requirements, multiple candidates are found in approximately $0.4\,\%$ of events. 
All candidates are retained and treated in the same way.

\section{Determination of signal and background yields}
\label{sec:mass-fit}

The signal and background yields are measured using an extended unbinned maximum likelihood fit to the \Dp\pim\pim invariant mass distribution.
The candidates are comprised of true signal decays and several sources of background. 
Partially reconstructed backgrounds come from \bquark~hadron decays where one or more final state particles are not reconstructed. 
Combinatorial background originates from random combinations of tracks, potentially including a real $\Dp\to\Km\pip\pip$ decay. 
Misidentified background arises from \bquark~hadron decays in which one of the final state particles is not correctly identified.
Potential residual background from charmless $B$ decays is reduced to a negligible level by the requirement that the flight distance of the $D$ candidate be greater than $1 \mm$.  

Signal candidates are modelled by the sum of two Crystal Ball (CB) functions~\cite{Skwarnicki:1986xj} with a common peak position of the Gaussian core and tails on opposite sides. 
The relative normalisation of the narrower CB shape and the ratio of widths of the CB functions are constrained, by including a Gaussian penalty term in the likelihood, to the values found in fits to simulated samples.
The tail parameters of the CB shapes are fixed to those found in simulation. 

The main source of partially reconstructed background is the $\Btodstarpipi$ channel with subsequent $\Dstarp\to\Dp\gamma$ or $\Dstarp\to\Dp\piz$ decay, where the neutral particle is not reconstructed. 
A non-parametric shape derived from simulation is used to model this contribution.
The shape is characterised by an edge around $100 \mev$ below the \B peak, where the exact position of the edge depends on properties of the decay, including the $\Dstarp$ polarisation.
As in previous studies of similar processes~\cite{LHCb-PAPER-2015-007,LHCb-PAPER-2015-012}, the fit quality improves when the shape is allowed to be offset by a small shift ($\approx 3.5 \mev$) that is determined from the data.

The combinatorial background is modelled with a linear function, where the slope is free to vary.
Many sources of misidentified background have broad $\dpipi$ invariant mass distributions that can be absorbed into the combinatorial background component. 
The exceptions are $\Btodanddstarkpi$ decays that produce distinctive shapes in the $\B$ candidate invariant mass distribution. 
These backgrounds are combined into a single non-parametric shape determined from simulated samples that are weighted to account for the known DP distribution for $\Btodkpi$ decays~\cite{LHCb-PAPER-2015-007}.
The ratio of $\Dp$ and $\Dstarp$ components in the \Btodanddstarkpi background shape is fixed from the measured values of the \Btodpipi and \Btodstarpipi branching fractions~\cite{Abe:2003zm,PDG2014} since $\mathcal{B}(\Btodstarkpi)$ is unknown.

\begin{figure}[!tb]
\centering
\includegraphics[scale=0.38]{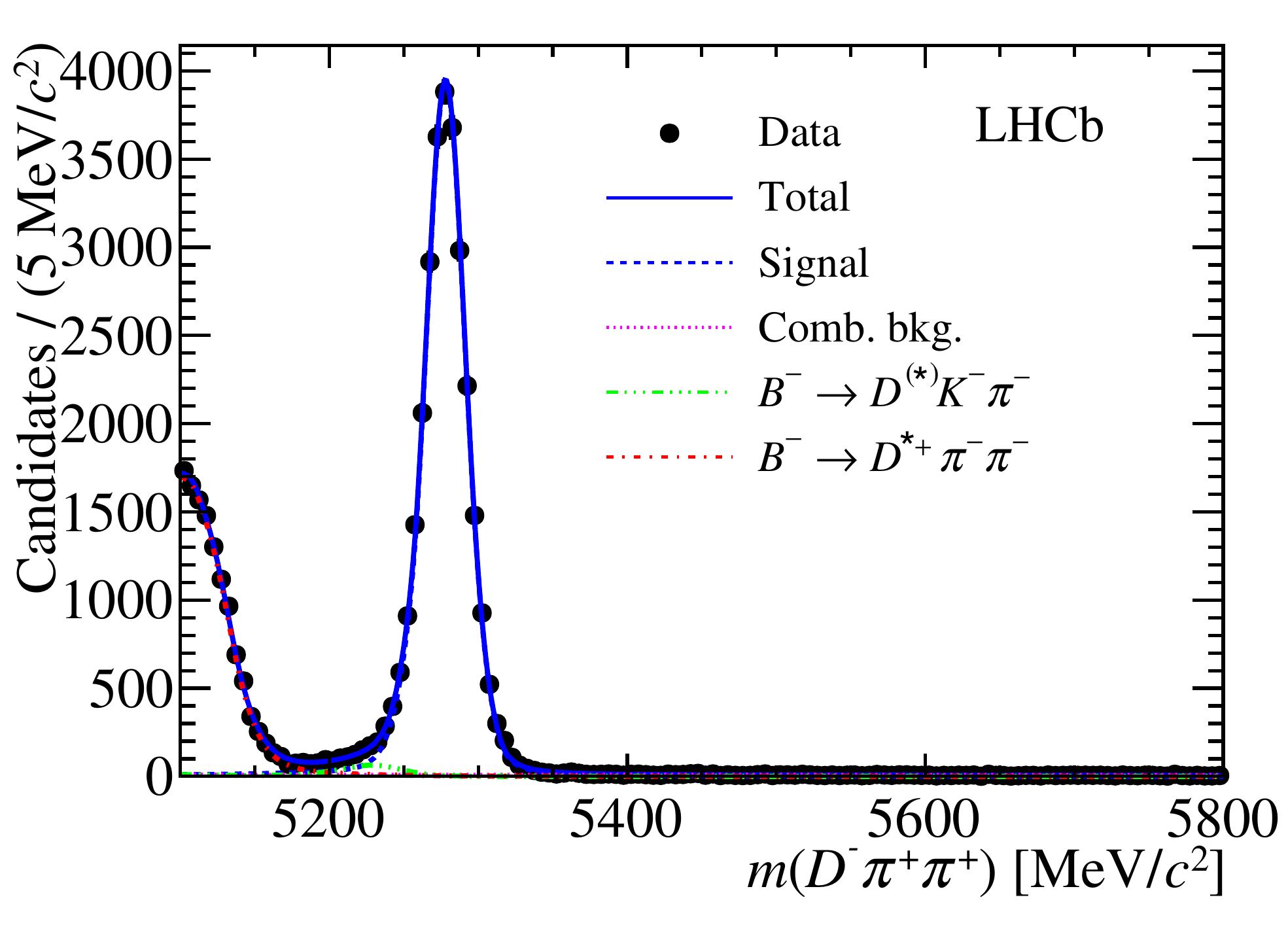}
\includegraphics[scale=0.38]{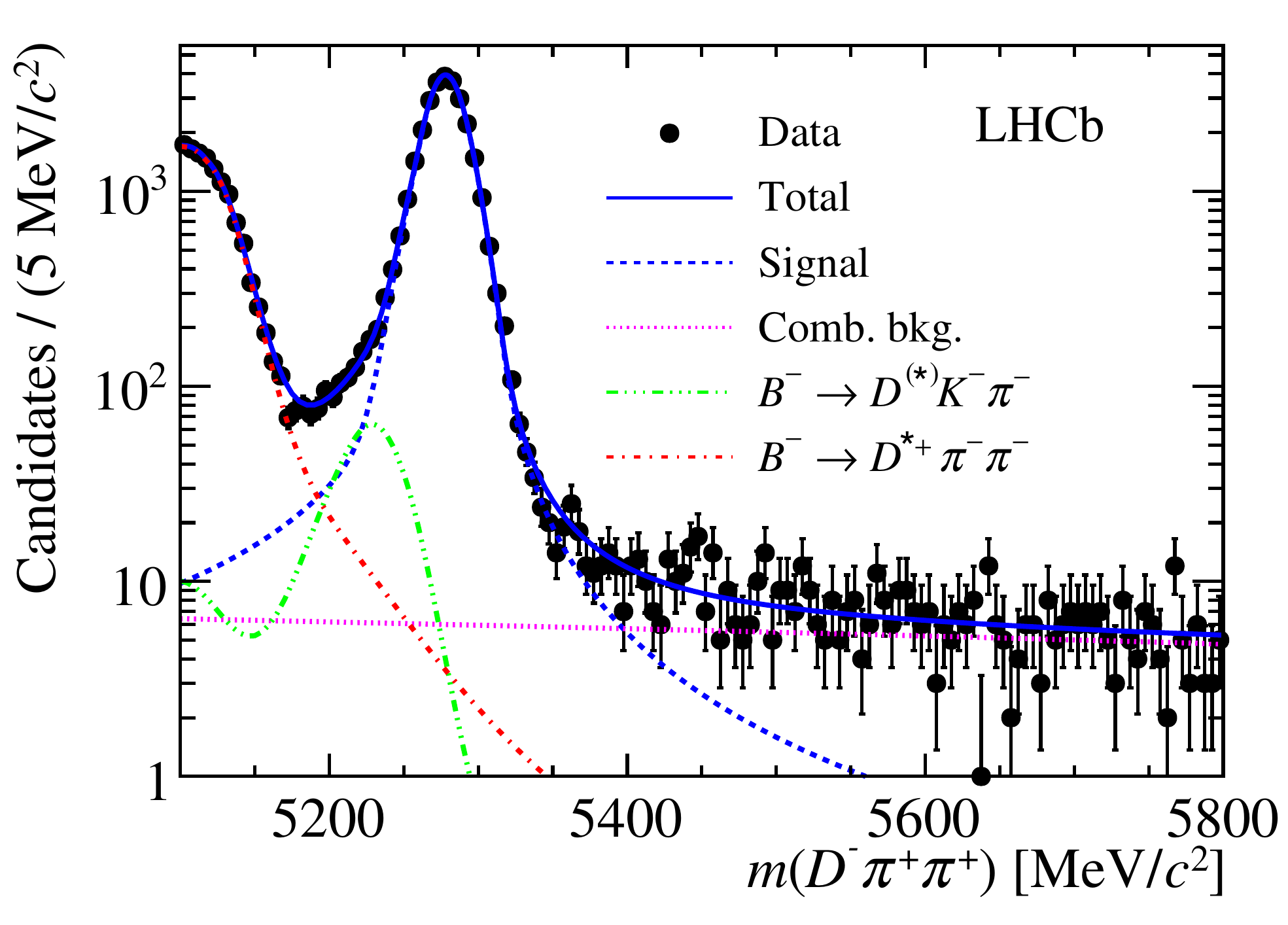}
\caption{\small 
  Results of the fit to the $B$ candidate invariant mass distribution shown with (left) linear and (right) logarithmic $y$-axis scales.
  Contributions are as described in the legend.}
\label{fig:dpimi-fit}
\end{figure}

\begin{table}[!tb] 
\centering
  \caption{\small
    Yields of the various components in the fit to $\Btodpipi$ candidate invariant mass distribution. 
    Note that the yields in the signal region are scaled from the full mass range. }
  \centering 
  \begin{tabular}{lcc} 
    \hline 
 Component & Full mass range & Signal region \\
    \hline 
 $N(\Btodpipi)$     & $29\,190 \pm 204$ & $27\,956 \pm 195$ \\
 $N(\Btodanddstarkpi)$   & $\phantom{29\,}807 \pm 123$ & $\phantom{27\,}243 \pm  37\phantom{1}$ \\
 $N(\Btodstarpipi)$ 	& $12\,120 \pm 115$ & $\phantom{27\,9}70 \pm  1\phantom{11}$ \\	      
 $N(\rm{comb. \ bkg.})$  & $\phantom{29\,}784 \pm 54\phantom{4}$ & $\phantom{27\,}103 \pm  7\phantom{11}$ \\
\hline 
  \end{tabular} 
\label{tab:DpipiFit_yields}
\end{table} 

There are 10 parameters in the fit that are free to vary: 
the yields for signal and combinatorial, $\Btodanddstarkpi$ and $\Btodstarpipi$ backgrounds;
the combinatorial background slope; 
the shared mean of the double CB shape, the width and relative normalisation of the narrower CB and the ratio of CB widths; 
and the shift parameter of the $\Btodstarpipi$ shape.
The result of the fit is shown in Fig.~\ref{fig:dpimi-fit} and gives a signal yield of approximately 29\,000 decays.
The $\chi^2$ per degree of freedom for this projection of the fit is $1.16$, calculated with statistical uncertainties only.
Component yields are shown in Table~\ref{tab:DpipiFit_yields} for both the full fit range and the signal region defined as $\pm2.5\,\sigma$ around the \B peak, where $\sigma$ is the width parameter of the dominant CB function in the signal shape; this corresponds to $5235.3 < m(\dpipi) < 5320.8 \mev$. 

\begin{figure}[!tb]
\centering
\includegraphics[scale=0.38]{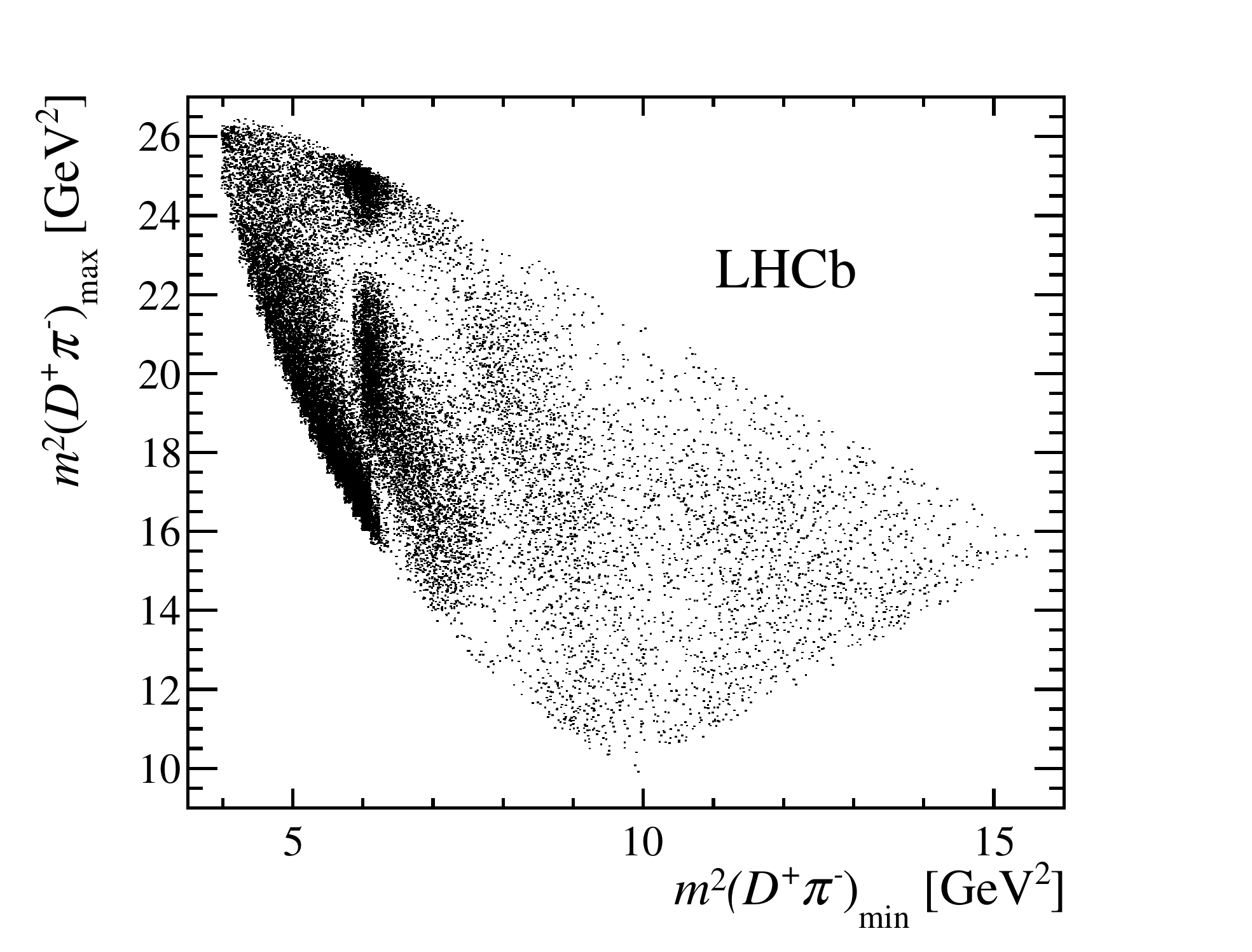}
\includegraphics[scale=0.38]{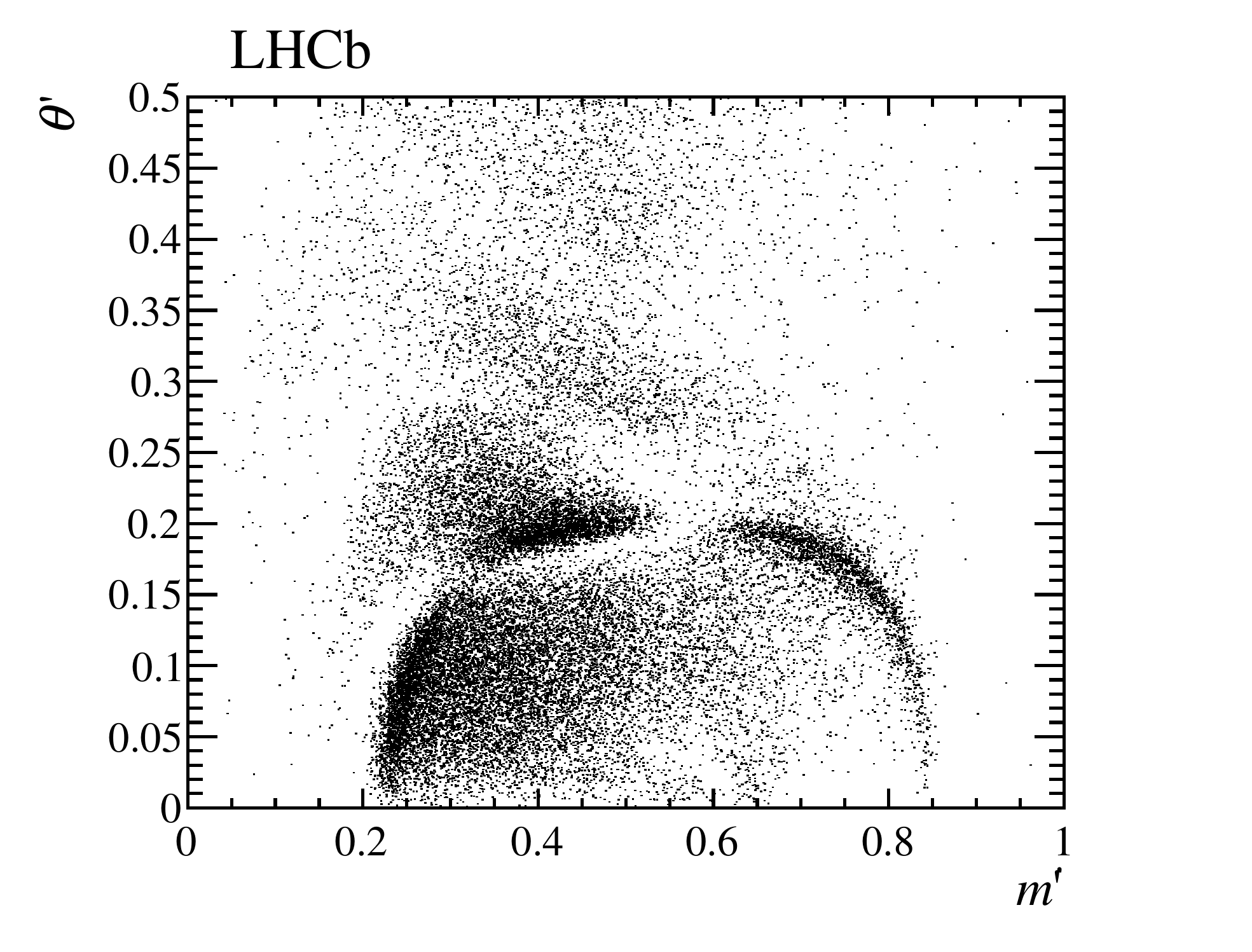}
\caption{\small Distribution of $\Btodpipi$ candidates in the signal
  region over (left) the DP and (right) the SDP.
}
\label{fig:canddp}
\end{figure}

A Dalitz plot~\cite{Dalitz:1953cp} is a two-dimensional representation of the phase space for a three-body decay in terms of two of the three possible two-body invariant mass squared combinations.
In $\Btodpipi$ decays there are two indistinguishable pions in the final state, so the two $m^2(\Dpi)$ combinations are ordered by value and the DP axes are defined as \msqdpimin and \msqdpimax.
The ordering causes a ``folding'' of the DP from the minimum value of $m^2(\Dp\pim)_{\rm max}$, which is $m^{}_\Bm m^{}_\Dp + m_\pim^2$, to the maximum value of $m^2(\Dp\pim)_{\rm min}$ at $\left( m_\Bm^2 + m_\Dp^2 - 2m_\pim^2 \right)/2$.
The DP distribution of the candidates in the signal region that are used in the DP fit is shown in Fig.~\ref{fig:canddp}~(left).
The same data are shown in the square Dalitz plot (SDP) in Fig.~\ref{fig:canddp}~(right). 
The SDP is defined by the variables \mpr\ and \thpr, which are given by 
\begin{equation}
\label{eq:sqdp-vars}
\mpr \equiv \frac{1}{\pi} \arccos\left(2\frac{m(\pim \pim) - m^{\rm min}_{\pim\pim}}{m^{\rm max}_{\pim\pim} - m^{\rm min}_{\pim\pim}} - 1 \right)
\hspace{10mm}{\rm and}\hspace{10mm}
\thpr \equiv \frac{1}{\pi}\theta(\pim\pim)\,,
\end{equation}
where $m^{\rm max}_{\pim\pim} = m_{\Bm} - m_{\Dp}$ and $m^{\rm min}_{\pim\pim} =
2m_{\pim}$ are the kinematic boundaries of $m(\pim\pim)$ and $\theta(\pim\pim)$ is the helicity angle of the $\pim\pim$ system (the angle between the momenta of the $\D$ meson and one of the pions, evaluated in the $\pim\pim$ rest frame). 
With $\mpr$ and $\thpr$ defined in terms of the $\pim\pim$ mass and helicity angle in this way, only the region of the SDP with $\thpr \leq 0.5$ is populated due to the symmetry of the two pions in the final state.
The SDP is used to describe the signal efficiency variation and distribution of background candidates, as described in Sec.~\ref{sec:dalitz}.

\section{Study of angular moments}
\label{sec:moments}

The angular moments of the \Btodpipi decays are studied to investigate which amplitudes to include in the DP fit model.
Angular moments are determined by weighting the data by the Legendre polynomial $P_L\left(\cos\theta(\Dpi)\right)$, where $\theta(\Dpi)$ is the helicity angle of the $\Dpi$ system, \ie\ the angle between the momenta of the pion in the $\Dpi$ system and the other pion from the \Bm\ decay, evaluated in the $\Dpi$ rest frame.
The moment $\left\langle P_{L}\right\rangle$ is the sum of the weighted data in a bin of $\Dpi$ mass with background contributions subtracted using sideband data and efficiency corrections, determined as in Sec.~\ref{sec:efficiency}, applied.
Each of the moments contains contributions from certain partial waves and interference terms. 
For the S-, P-, D- and F-wave amplitudes denoted by $h_j e^{i \delta_j}$ ($j=0,1,2,3$ respectively),
\begin{align}
  \left\langle P_{0}\right\rangle \propto\, &
  \left|h_0\right|{}^2+\left|h_1\right|{}^2+\left|h_2\right|{}^2+\left|h_3\right|{}^2\,, \label{eq:p0} \\ 
  \left\langle P_1\right\rangle \propto\, &
  \frac{2}{\sqrt{3}} \left|h_0\right| \left|h_1\right| \cos \left(\delta _0-\delta_1\right)+\frac{4}{\sqrt{15}} \left|h_1\right| \left|h_2\right| \cos \left(\delta_1-\delta _2\right)+ \nonumber \\ 
  &\qquad\qquad\qquad\qquad\qquad\qquad\qquad\qquad 
  \frac{6}{\sqrt{35}} \left|h_2\right| \left|h_3\right| \cos \left(\delta _2-\delta _3\right)\,, \label{eq:p1} \\ 
   \left\langle P_{2}\right\rangle \propto\, &
   \frac{6}{5} \sqrt{\frac{3}{7}} \left|h_1\right| \left|h_3\right| \cos \left(\delta_1-\delta _3\right)+\frac{2 \left|h_0\right| \left|h_2\right| \cos \left(\delta_0-\delta _2\right)}{\sqrt{5}}+\nonumber \\
   &\qquad\qquad\qquad\qquad\qquad\qquad\qquad\qquad 
   \frac{2 \left|h_1\right|{}^2}{5}+\frac{2 \left|h_2\right|{}^2}{7}+ \frac{4
   \left|h_3\right|{}^2}{15}\,, \label{eq:p2} \\ 
   \left\langle P_{3}\right\rangle \propto\, &
   \frac{6}{7} \sqrt{\frac{3}{5}} \left|h_1\right| \left|h_2\right| \cos \left(\delta_1-\delta _2\right)+\frac{2 \left|h_0\right| \left|h_3\right| \cos \left(\delta_0-\delta _3\right)}{\sqrt{7}}+\nonumber \\ 
   &\qquad\qquad\qquad\qquad\qquad\qquad\qquad\qquad 
   \frac{8 \left|h_2\right| \left|h_3\right| \cos \left(\delta _2-\delta _3\right)}{3 \sqrt{35}}\,, \label{eq:p3} \\ 
   \left\langle P_{4}\right\rangle \propto\, &
   \frac{8 \left|h_1\right| \left|h_3\right| \cos \left(\delta _1-\delta_3\right)}{3 \sqrt{21}}+ \frac{2\left|h_2\right|{}^2}{7}+\frac{2 \left|h_3\right|{}^2}{11}\,, \label{eq:p4} \\ 
  \left\langle P_5\right\rangle \propto\, &
  \frac{20}{33} \sqrt{\frac{5}{7}} \left|h_2\right| \left|h_3\right| \cos \left(\delta_2-\delta _3\right)\,, \label{eq:p5} \\ 
 \left\langle P_6\right\rangle \propto\, & 
 \frac{100 \left|h_3\right|{}^2}{429}\,. \label{eq:p6}
\end{align}
These expressions assume that there are no contributions from partial waves higher than F-wave.
Thus, they are valid only in regions of the DP unaffected by the folding, \ie\ for $m(\Dpi) \lesssim 3.2\gev$, where the full range of the $\Dpi$ helicity angle distribution is available.
Above this mass, the orthogonality of the Legendre polynomials does not hold and a straightforward interpretation of the angular moments in terms of the contributing partial waves is not possible.
Nevertheless, the angular moments provide a useful way to judge the agreement of the fit result with the data, complementary to the projections onto the invariant masses.

\begin{figure}[!htbp]
\centering
\includegraphics[scale=0.37]{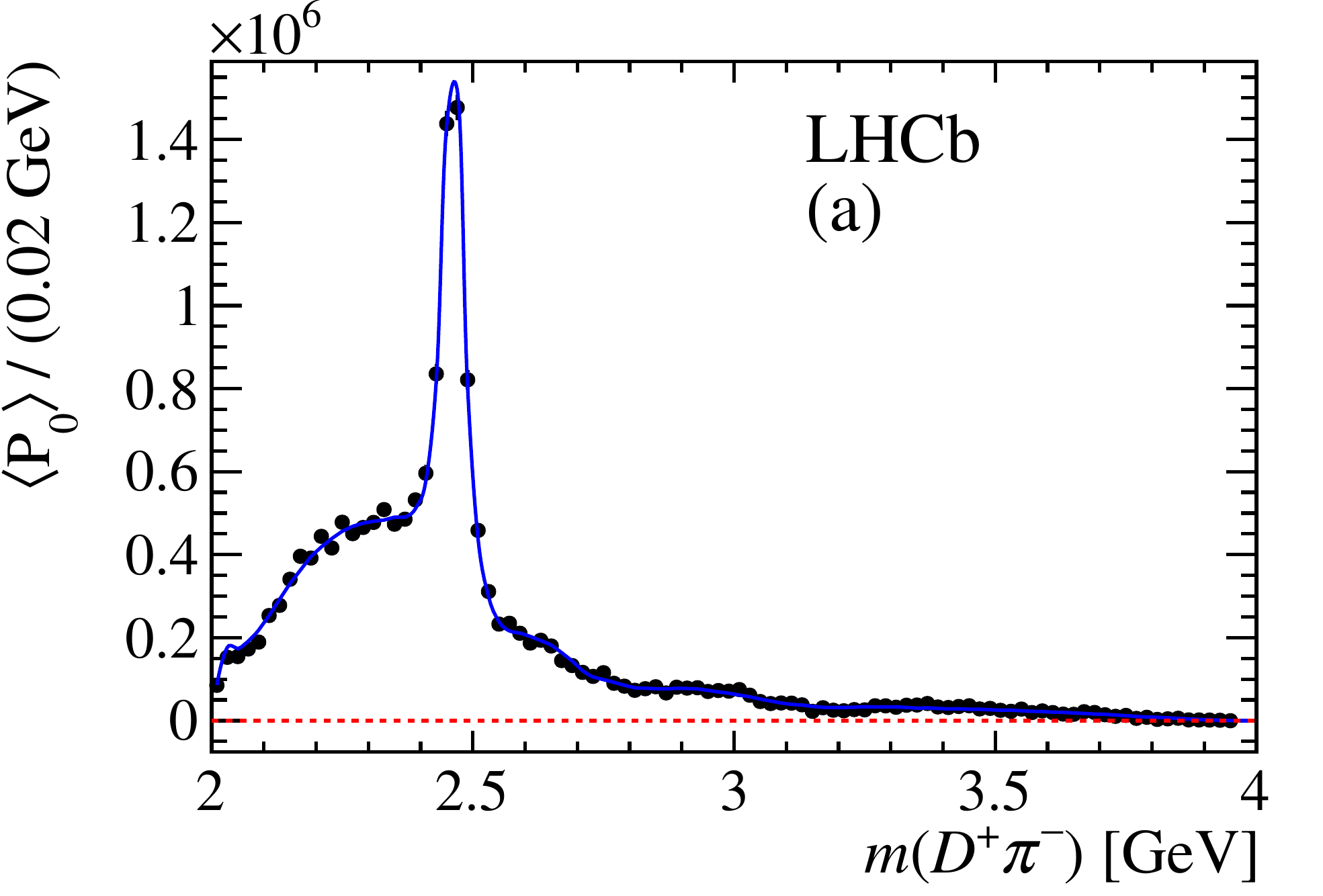}
\includegraphics[scale=0.37]{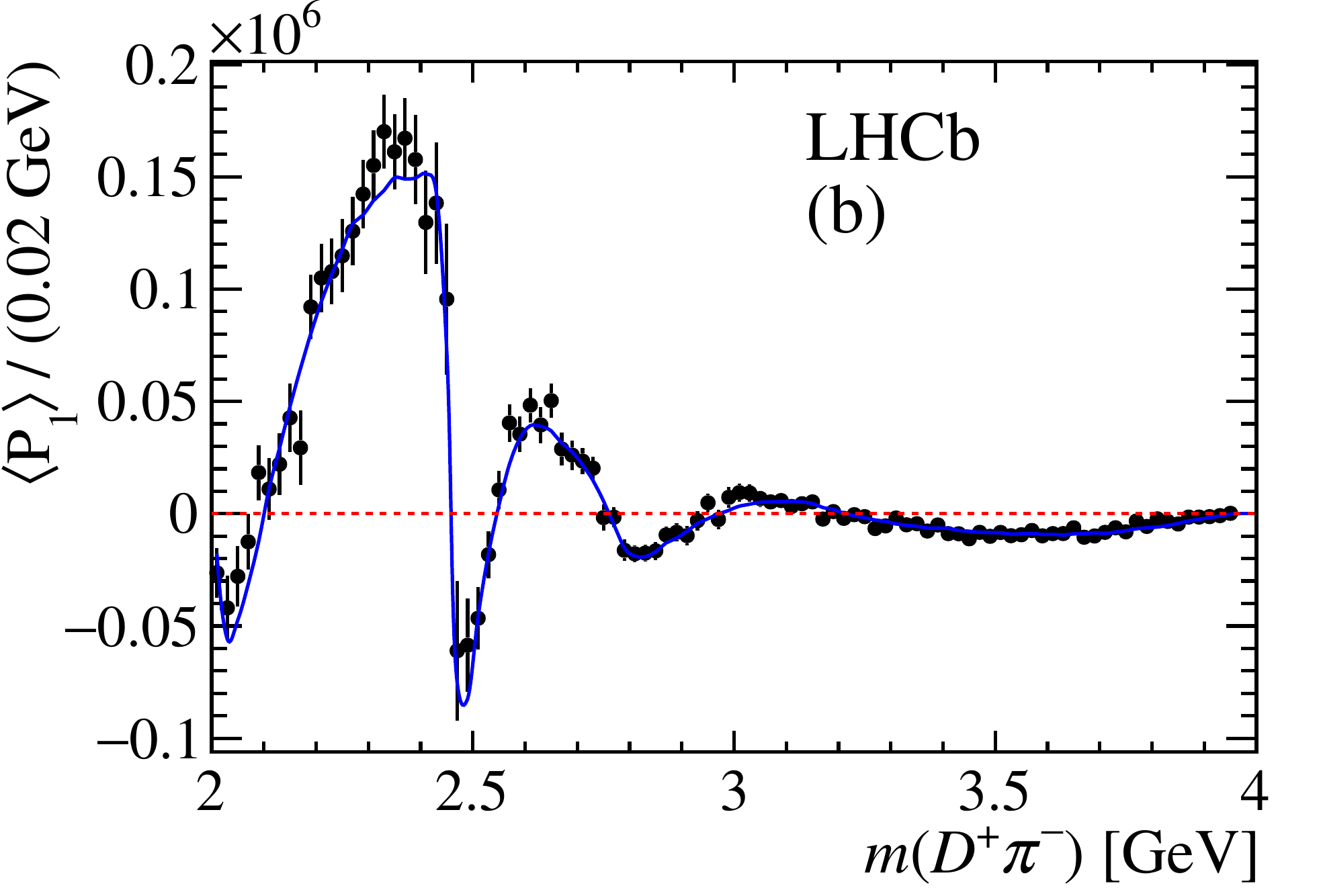}
\includegraphics[scale=0.37]{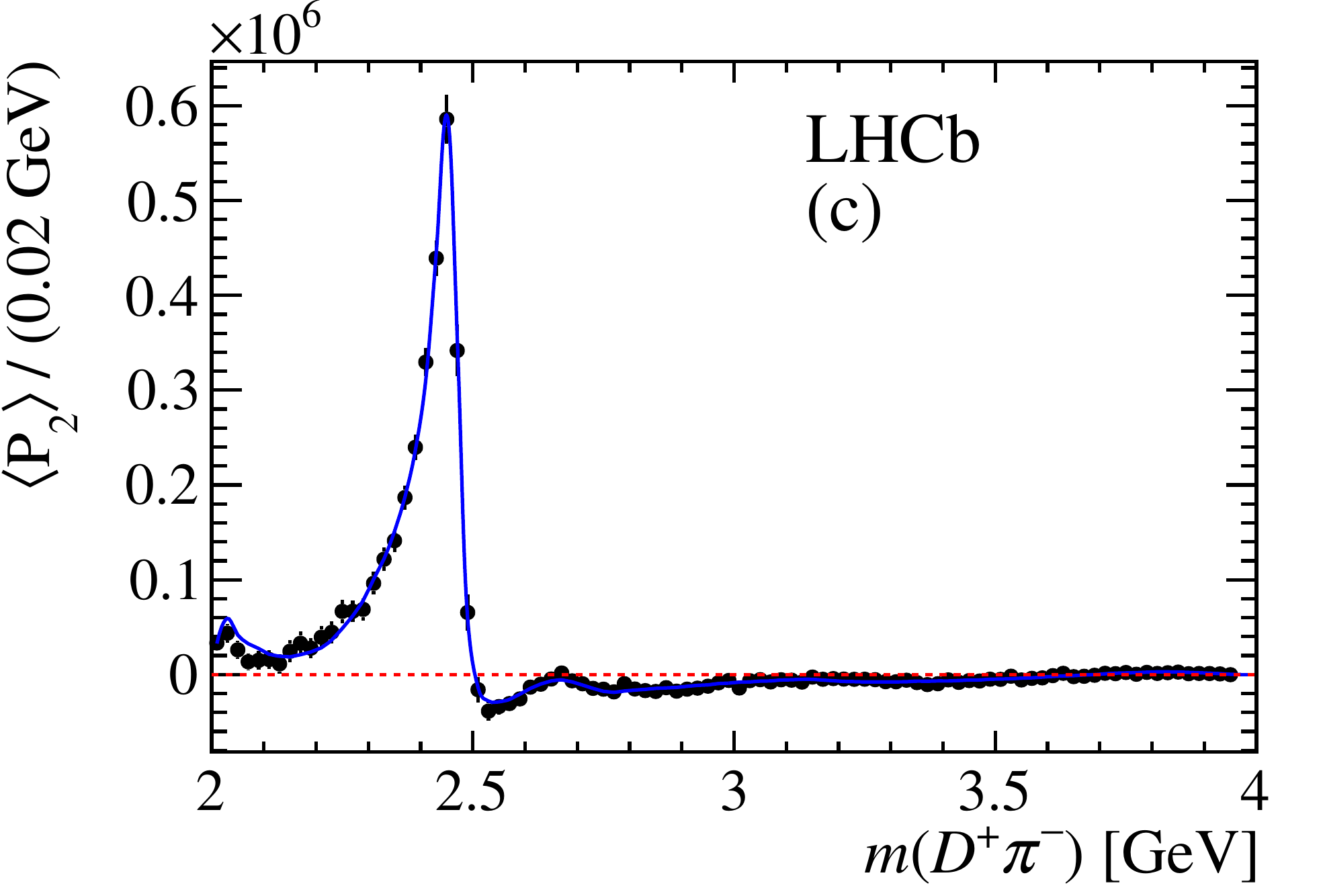}
\includegraphics[scale=0.37]{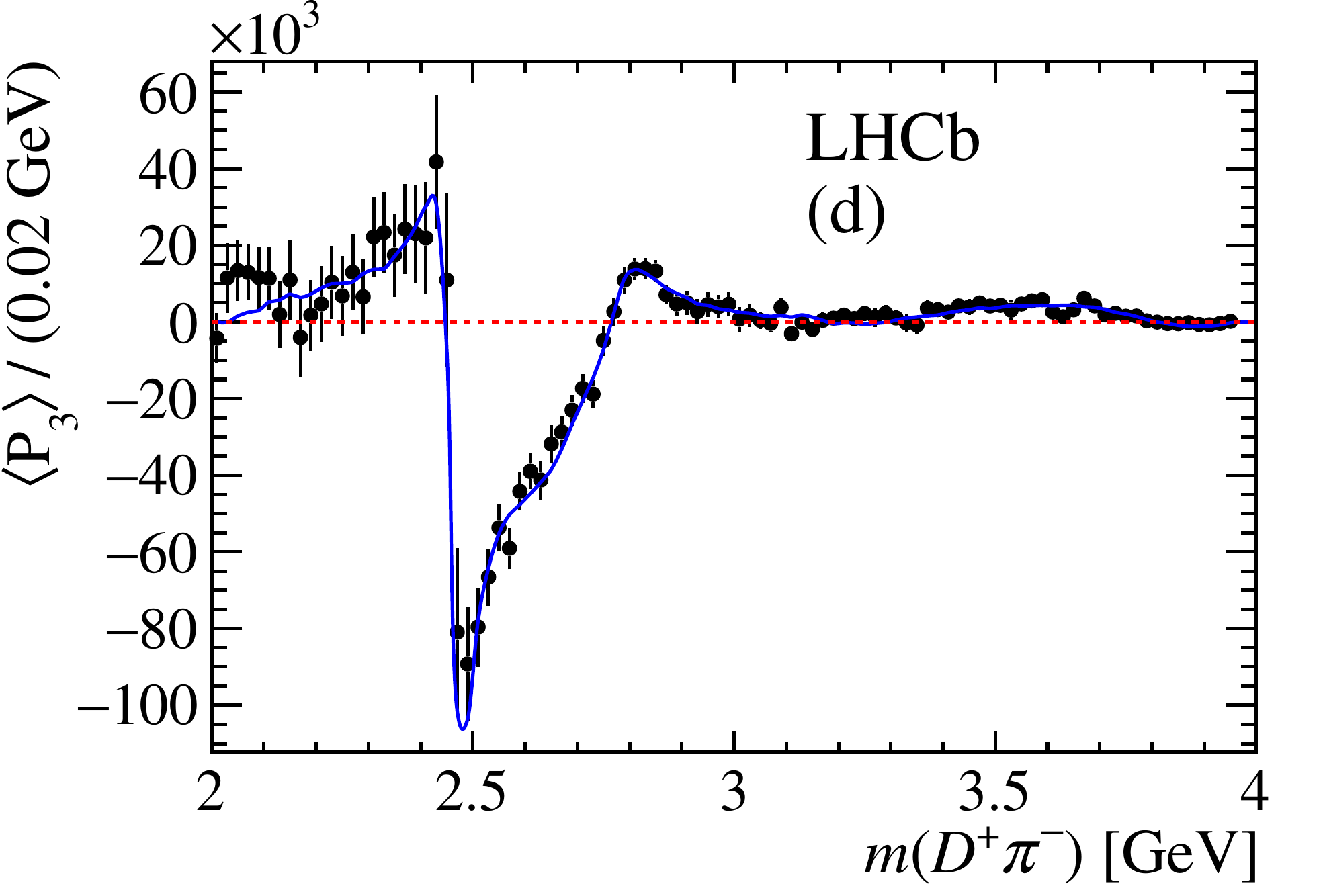}
\includegraphics[scale=0.37]{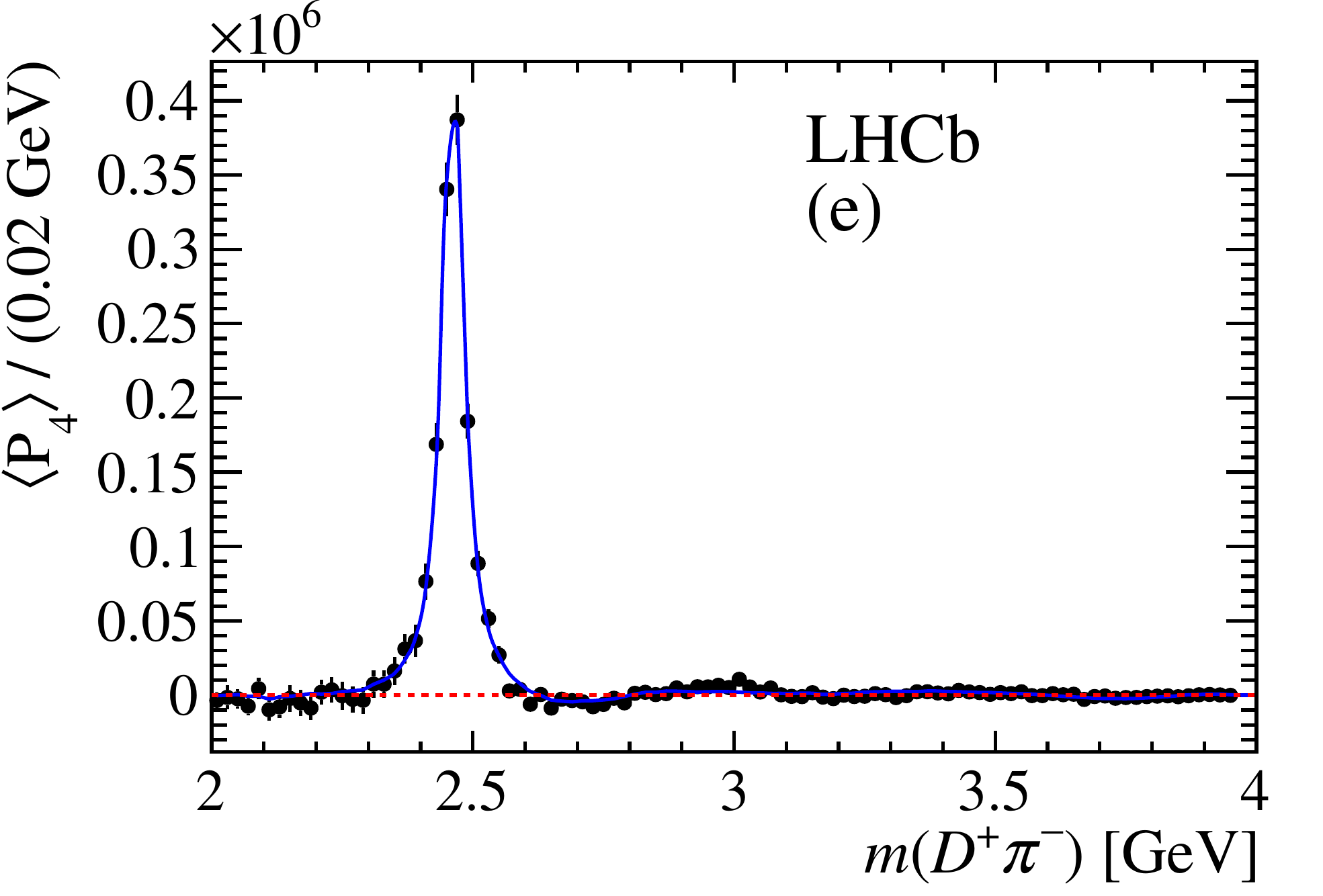}
\includegraphics[scale=0.37]{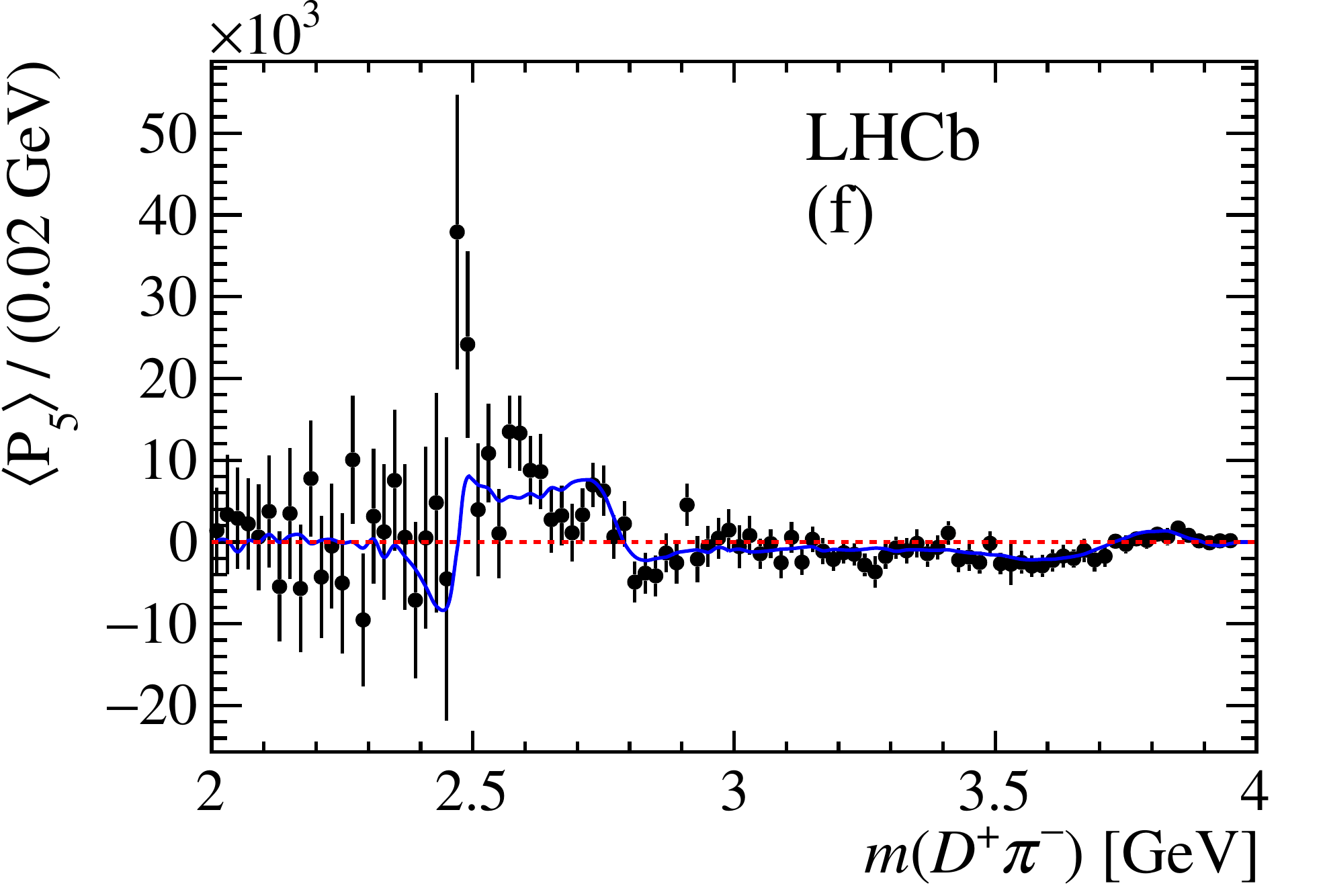}
\includegraphics[scale=0.37]{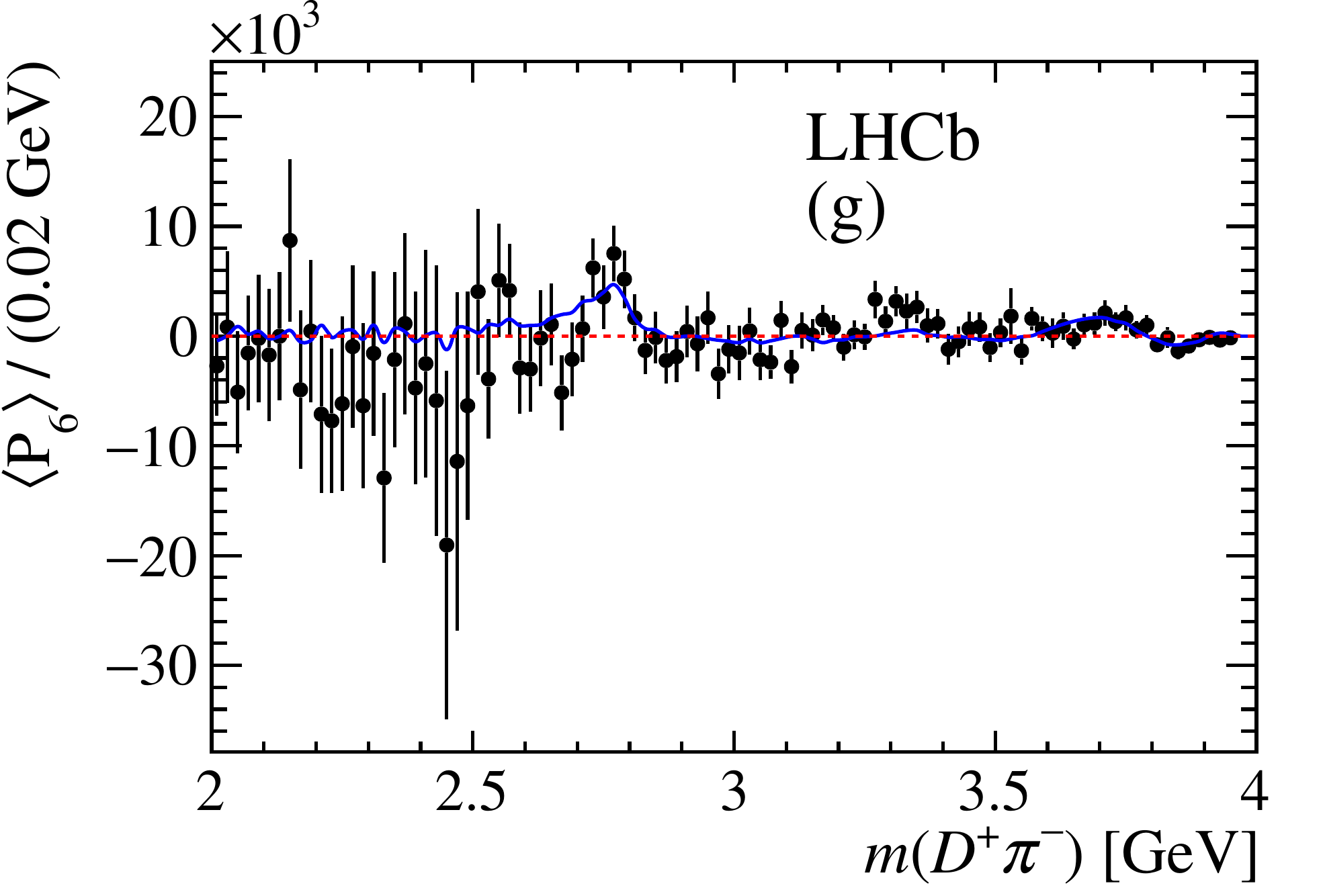}
\caption{\small
The first seven unnormalised angular moments for background-subtracted and efficiency-corrected data (black points) as a function of $m(\Dpi)$ in the range $2.0$--$4.0\gev$. 
The blue line shows the result of the DP fit described in Sec.~\ref{sec:dalitz}.}
\label{fig:moments2}
\end{figure}

\begin{figure}[!htbp]
\centering
\includegraphics[scale=0.37]{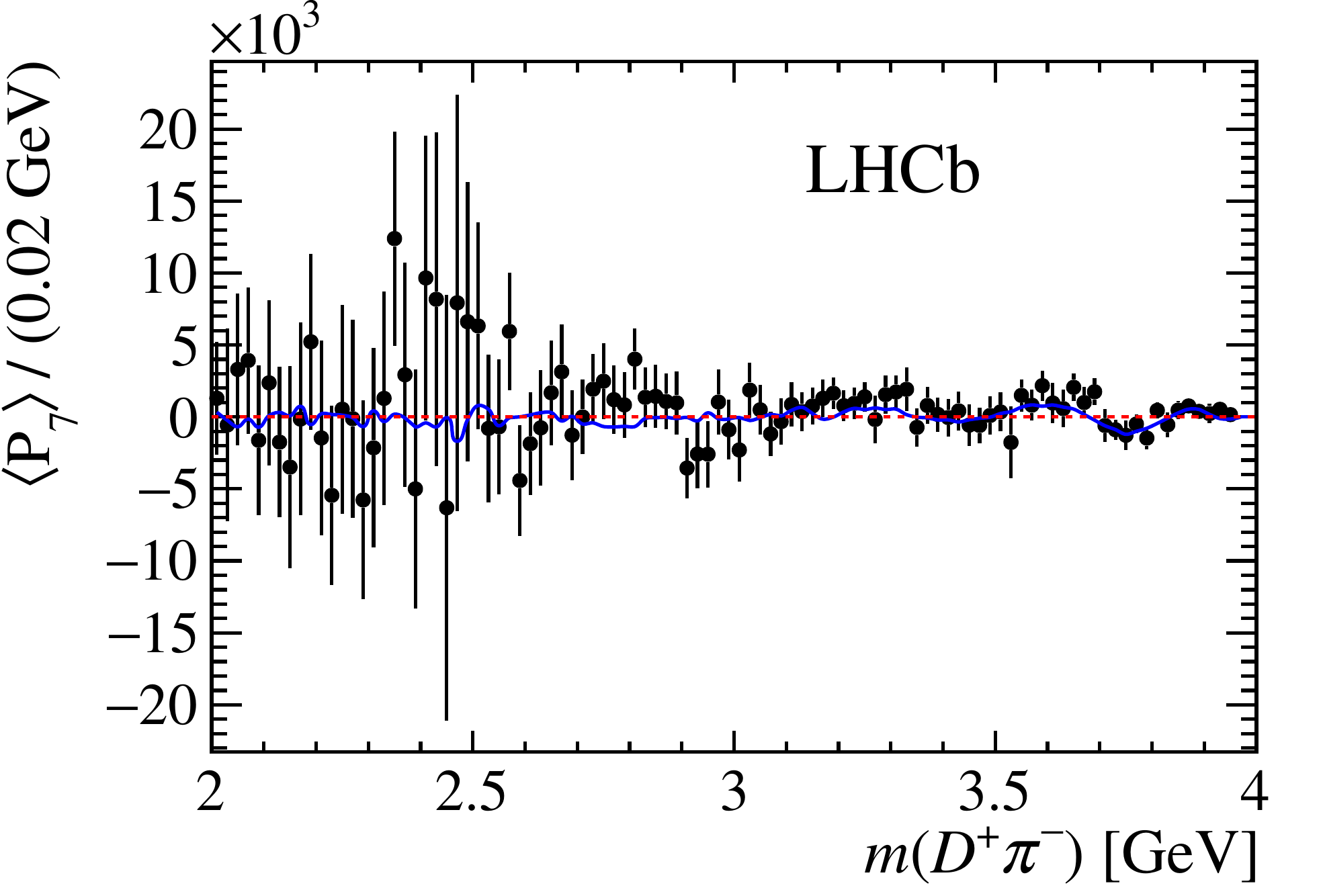}
\includegraphics[scale=0.37]{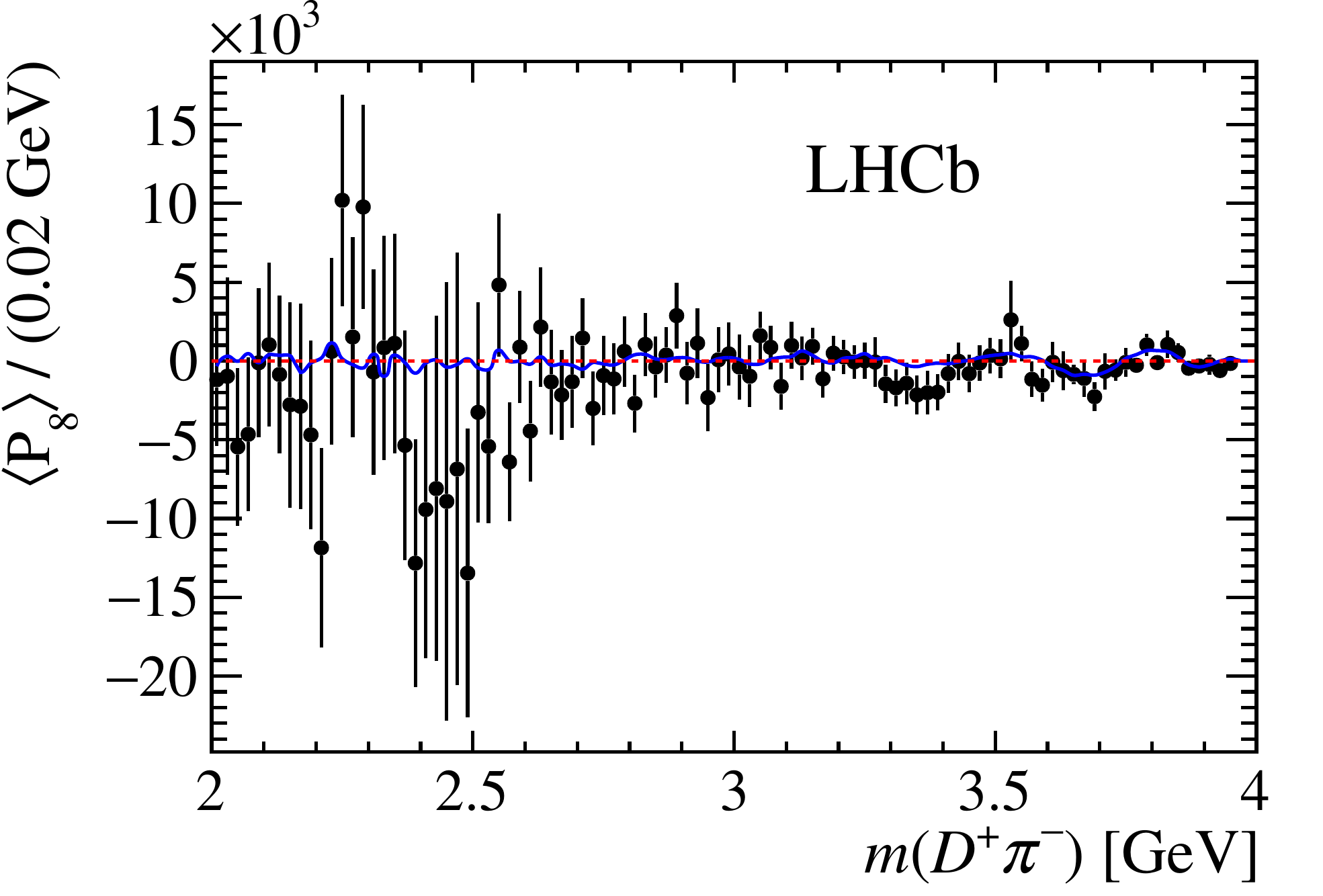}
\caption{\small
  Unnormalised angular moments $\left\langle P_7 \right\rangle$ and $\left\langle P_8 \right\rangle$ for background-subtracted and efficiency-corrected data (black points) as a function of $m(\Dpi)$ in the range $2.0$--$4.0\gev$.
  The blue line shows the result of the DP fit described in Sec.~\ref{sec:dalitz}.}
\label{fig:highmoments}
\end{figure}

The unnormalised angular moments $\left\langle P_0\right\rangle$--$\left\langle P_6\right\rangle$ are shown in Fig.~\ref{fig:moments2} for the $\Dpi$ invariant mass range $2.0$--$4.0\gev$.
The \olddstartwo\ resonance is clearly seen in the $\left\langle P_4 \right\rangle$ distribution of Fig.~\ref{fig:moments2}(e).
From Eqs.~(\ref{eq:p1}) and~(\ref{eq:p3}) it can be inferred that the structures in the distributions of $\left\langle P_1\right\rangle$ and $\left\langle P_3 \right\rangle$ below $3\gev$ suggest that there is interference both between S- and P-wave amplitudes and between P- and D-wave amplitudes. 
Therefore broad spin~0 and spin~1 components are required in the DP model.
In addition, structure in $\left\langle P_2 \right\rangle$ around $2.76 \gev$ implies the possible presence of a spin 1 resonance in that region.
The angular moments $\left\langle P_7\right\rangle$ and $\left\langle P_8\right\rangle$, shown in Fig.~\ref{fig:highmoments}, show no structure, consistent with the assumption that contributions from higher partial waves and from the isospin-2 dipion channel are small.

\begin{figure}[!tb]
\centering
\includegraphics[scale=0.37]{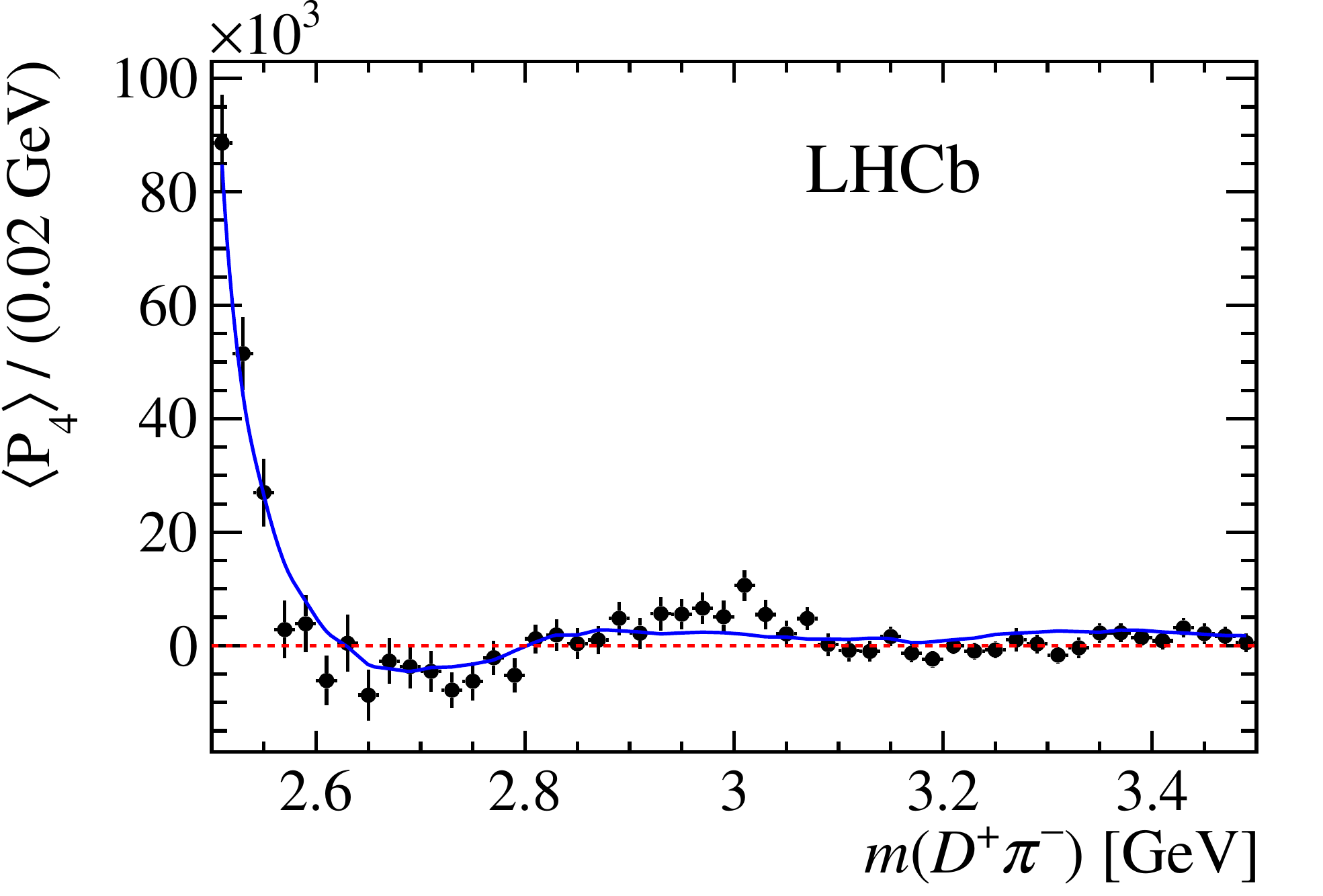}
\includegraphics[scale=0.37]{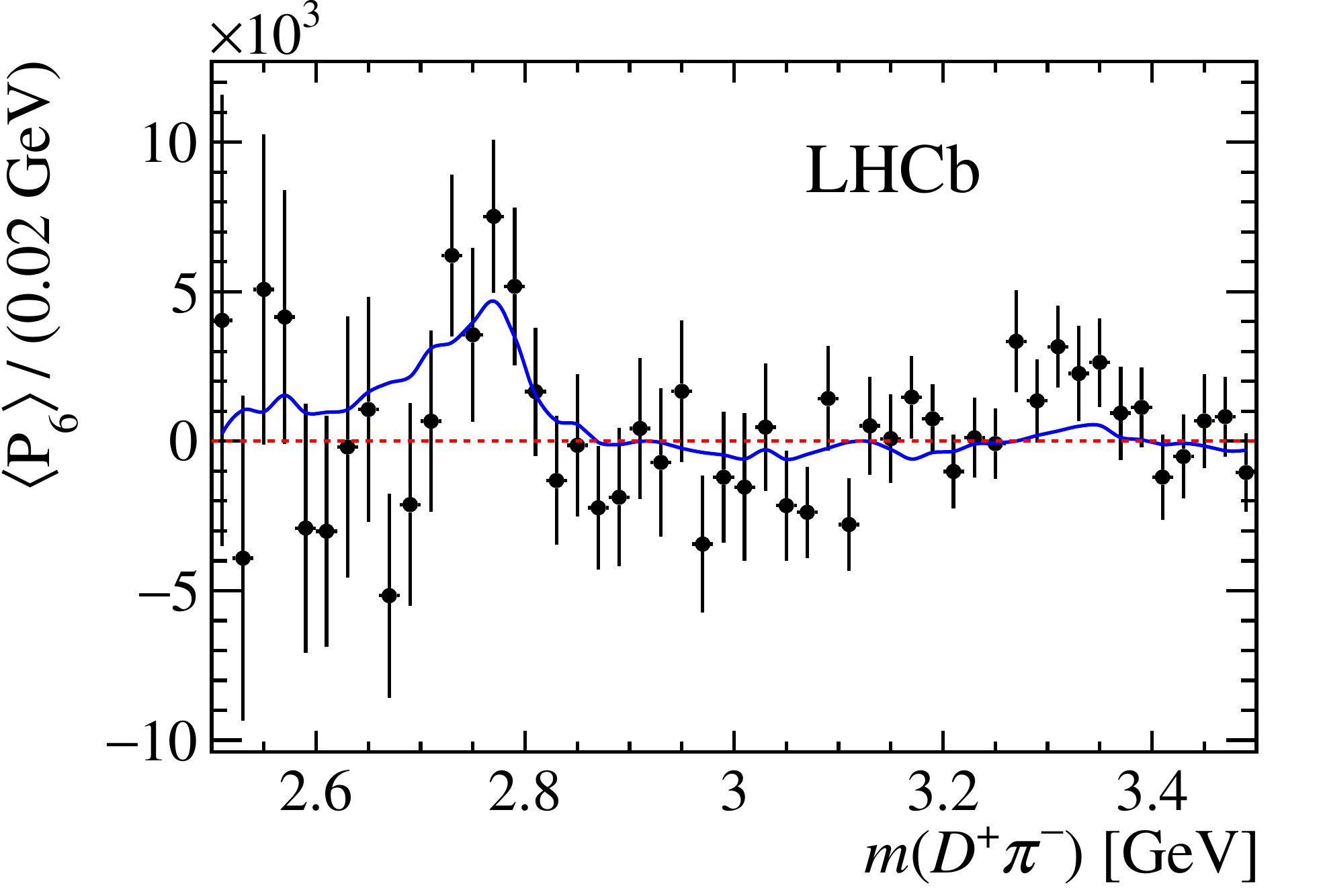}
\caption{\small
Zoomed views of the fourth and sixth unnormalised angular moments for background-subtracted and efficiency-corrected data (black points) as a function of $m(\Dpi)$.
The blue line shows the result of the DP fit described in Sec.~\ref{sec:dalitz}.}
\label{fig:moments3}
\end{figure}

Zoomed views of the fourth and sixth moments in the region around $m(\Dpi)=3\gev$ are shown in Fig.~\ref{fig:moments3}.
A wide bump is visible in the distribution of $\left\langle P_4 \right\rangle$ at $m(\Dpi) \approx 3 \gev$. 
Although close to the point where the DP folding affects the interpretation of the moments, this enhancement suggests that an additional spin~2 resonance could be contributing in this region.
A peak is also seen at $m(\Dpi) \approx 2.76 \gev$ in the $\left\langle P_6 \right\rangle$ distribution, suggesting that a spin~3 resonance should be included in the DP model.
As discussed in Sec.~\ref{sec:introduction}, other recent analyses~\cite{delAmoSanchez:2010vq,LHCb-PAPER-2013-026,LHCb-PAPER-2014-035,LHCb-PAPER-2014-036,LHCb-PAPER-2014-070,LHCb-PAPER-2015-007} suggest that both spin~1 and spin~3 states could be expected in this region.

\section{Dalitz plot analysis formalism}
\label{sec:dalitz-generalities}

The isobar approach~\cite{Fleming:1964zz,Morgan:1968zza,Herndon:1973yn} is used to describe the complex decay amplitude as the coherent sum of amplitudes for intermediate resonant and nonresonant decays. 
The total amplitude is given by
\begin{equation}\label{eqn:amp}
	{\cal A}\left(s, t\right) = \sum_{j=1}^{N} c_j F_j\left(s, t\right) \,,
\end{equation}
where the complex coefficients $c_j$ describe the relative contribution of each intermediate process. 
Here, and for the remainder of this section, \msqdpimin and \msqdpimax are referred to as $s$ and $t$, respectively.

The resonant dynamics are encoded in the $F_j\left(s, t\right)$ terms, each of which is normalised such that the integral of the magnitude squared across the DP is unity.
The amplitude is explicitly symmetrised to take account of the Bose symmetry of the final state due to the identical pions, \ie\ 
\begin{equation}
  {\cal A}\left(s,t\right) \mapsto {\cal A}\left(s,t\right) + {\cal A}\left(t,s\right) \, .
\end{equation}
This substitution is implied throughout this section.

For a $\Dpi$ resonance
\begin{equation}
  \label{eq:ResDynEqn}
  F\left(s, t\right) = 
  R\left(s\right) \times X(|\vec{p}\,|\,r_{\rm BW}) \times X(|\vec{q}\,|\,r_{\rm BW}) 
  \times T(\vec{p},\vec{q}\,) \, ,
\end{equation}
where $\vec{p}$ and $\vec{q}$ are the momenta, calculated in the \Dpi rest frame, of the particle not involved in the resonance and one of the resonance decay products, respectively. 
The functions $X$, $T$ and $R$ are described below. 

The $X(z)$ terms are Blatt--Weisskopf barrier factors~\cite{blatt-weisskopf}, where $z=|\vec{q}\,|\,r_{\rm BW}$ or $|\vec{p}\,|\,r_{\rm BW}$ and $r_{\rm BW}$ 
is the barrier radius, and are given by
\begin{equation}\begin{array}{rcl}
L = 0 \ : \ X(z) & = & \displaystyle{1}\,, \\
L = 1 \ : \ X(z) & = & \displaystyle{\sqrt{\frac{1 + z_0^2}{1 + z^2}}}\,, \\
L = 2 \ : \ X(z) & = & \displaystyle{\sqrt{\frac{z_0^4 + 3z_0^2 + 9}{z^4 + 3z^2 + 9}}}\,,\\
L = 3 \ : \ X(z) & = & \displaystyle{\sqrt{\frac{z_0^6 + 6z_0^4 + 45z_0^2 + 225}{z^6 + 6z^4 + 45z^2 + 225}}}\,,
\end{array}\label{eq:BWFormFactors}\end{equation}
where $L$ is the spin of the resonance and $z_0$ is defined as the value of $z$ where the invariant mass is equal to the mass of the resonance.
Since the \Bm meson has zero spin, $L$ is also the orbital angular momentum between the resonance and the other pion.
The barrier radius, $r_{\rm BW}$, is taken to be $4.0\gev^{-1}\approx 0.8\fm$~\cite{Aubert:2005ce,LHCb-PAPER-2014-036} for all resonances.

The $T(\vec{p},\vec{q})$ functions describe the angular distribution and are given in the Zemach tensor formalism~\cite{Zemach:1963bc,Zemach:1968zz},
\begin{equation}\begin{array}{rcl}
L = 0 \ : \ T(\vec{p},\vec{q}) & = & \displaystyle{1}\,,\\
L = 1 \ : \ T(\vec{p},\vec{q}) & = & \displaystyle{-\,2\,\vec{p}\cdot\vec{q}}\,,\\
L = 2 \ : \ T(\vec{p},\vec{q}) & = & \displaystyle{\frac{4}{3} \left[3(\vec{p}\cdot\vec{q}\,)^2 - (|\vec{p}\,||\vec{q}\,|)^2\right]}\,,\\
L = 3 \ : \ T(\vec{p},\vec{q}) & = & \displaystyle{-\,\frac{24}{15} \left[5(\vec{p}\cdot\vec{q}\,)^3 - 3(\vec{p}\cdot\vec{q}\,)(|\vec{p}\,||\vec{q}\,|)^2\right]}\,.
\end{array}\label{eq:ZTFactors}\end{equation}
These are proportional to the Legendre polynomials, $P_L(x)$, where $x$ is the cosine of the helicity angle between $\vec{p}$ and $\vec{q}$. 

The function $R\left(s\right)$ of Eq.~(\ref{eq:ResDynEqn}) describes the resonance lineshape.
Resonant contributions to the total amplitude are modelled by relativistic Breit--Wigner (RBW) functions, given by
\begin{equation}
  \label{eq:RelBWEqn}
  R(s) = \frac{1}{(m_0^2 - s) - i\, m_0 \Gamma(\sqrt{s})} \,,
\end{equation}
with a mass-dependent decay width defined as
\begin{equation}
\label{eq:GammaEqn}
\Gamma(m) = \Gamma_0 \left(\frac{q}{q_0}\right)^{2L+1}
\left(\frac{m_0}{m}\right) X^2(q\,r_{\rm BW}) \,,
\end{equation}
where $q_0$ is the value of $q \equiv |\vec{q}\,|$ when $m = m_0$ and $\Gamma_0$ is the full width. 
Virtual contributions, from resonances with pole masses outside the kinematically allowed region, can be described by RBW functions with one modification:
the pole mass $m_0$ is replaced with an effective mass, $m_0^{\rm{eff}}$, in the allowed region of $s$, when the parameter $q_{0}$ is calculated. 
The term $m_0^{\rm{eff}}$ is given by the {\it ad hoc} formula~\cite{LHCb-PAPER-2014-036}
\begin{equation}\label{eqn:effmass}
  m_0^{\rm{eff}}(m_0) = m^{\rm{min}} + (m^{\rm{max}} - m^{\rm{min}}) \left( 1 + \tanh\left( \frac{m_0 - \frac{m^{\rm{min}}+m^{\rm{max}}}{2}}{m^{\rm{max}}-m^{\rm{min}}} \right) \right)\, ,
\end{equation}
where $m^{\rm{max}}$ and $m^{\rm{min}}$ are the upper and lower thresholds of $s$.
Note that $m_0^{\rm{eff}}$ is only used in the calculation of $q_{0}$, so only the tail of such virtual contributions enters the DP.

A quasi-model-independent approach is used to describe the entire $\Dpi$ spin~0 partial wave.
The total $\Dpi$ \swave\ is fitted using cubic splines to describe the magnitude and phase variation of the spin~0 amplitude. 
Knots are defined at fixed values of $m(\Dpi)$ and splines give a smooth interpolation of the magnitude and phase of the \swave\ between these points.
The \swave\ magnitude and phase are both fixed to zero at the highest mass knot in order to ensure sensible behaviour at the kinematic limit.
For the knot at $m(\Dpi) = 2.4$\gev, close to the peak of the \olddstarzero\ resonance, the magnitude and phase values are fixed to 0.5 and 0, respectively, as a reference.
The magnitude and phase values at every other knot position are determined from the fit.

The folding of the Dalitz plot has implications for the choice of knot positions.
Since the \swave\ amplitude varies with $m(\Dpi)$, its reflection onto the other DP axis gives a helicity angle distribution that corresponds to higher partial waves.  
Equally, if knots are included at high $m(\Dpi)$, the quasi-model-independent \Dpiswave\ amplitude can absorb resonant contributions with non-zero spin due to their reflections.  
To avoid this problem, only a single knot with floated parameters is used above the minimum value of $m^2(\Dp\pim)_{\rm max}$, specifically at $4.1 \gev$ (as mentioned above, the amplitude is fixed to zero at the highest mass knot at $5.1 \gev$).  
At lower $m(\Dpi)$, knots are spaced every $0.1 \gev$ from $2.0 \gev$ up to $3.1 \gev$, except that the knot at $3.0 \gev$ is removed in order to stabilise the fit.

Neglecting reconstruction effects, the DP probability density function would be
\begin{equation}
\label{eq:SigDPLike}
{\cal{P}}_{\rm phys}\left(s, t\right) =
\frac
{|{\cal A}\left(s, t\right)|^2}
{\int\!\!\int_{\rm DP}~{|{\cal A}\left(s,t\right)|^2}~ds\,dt} \, .
\end{equation}
The effects of nonuniform signal efficiency and of background contributions are accounted for as described in Sec.~\ref{sec:dalitz}.
The probability density function depends on the complex coefficients, introduced in Eq.~(\ref{eqn:amp}), as well as the masses and widths of the resonant contributions and the parameters describing the \Dpiswave. 
These parameters are allowed to vary freely in the fit.
Results for the complex coefficients are dependent on the amplitude formalism, normalisation and phase convention, and consequently may be difficult to compare between different analyses.
It is therefore useful to define fit fractions and interference fit fractions to provide convention-independent results. 
Fit fractions are defined as the integral over the DP for a single contributing amplitude squared divided by that of the total amplitude squared, 
\begin{equation}
{\it FF}_j =
\frac
{\int\!\!\int_{\rm DP}\left|c_j F_j\left(s, t\right)\right|^2~ds\,dt}
{\int\!\!\int_{\rm DP}\left|{\cal A}\left(s, t\right)\right|^2~ds\,dt} \, .
\label{eq:fitfraction}
\end{equation}
The sum of fit fractions is not required to be unity due to the potential presence of net constructive or destructive interference. 
Interference fit fractions are defined, for $i<j$ only, as
\begin{equation}
  {\it FF}_{ij} =
  \frac
  {\int\!\!\int_{\rm DP} 2 \, \Real\left[c_ic_j^* F_i\left(s, t\right)F_j^*\left(s, t\right)\right]~ds\,dt}
  {\int\!\!\int_{\rm DP}\left|{\cal A}\left(s, t\right)\right|^2~ds\,dt} \, .
  \label{eq:intfitfraction}
\end{equation}

\section{Dalitz plot fit}
\label{sec:dalitz}

\subsection{Signal efficiency}
\label{sec:efficiency}

\begin{figure}[!tb]
 \centering
 \includegraphics[scale=0.38]{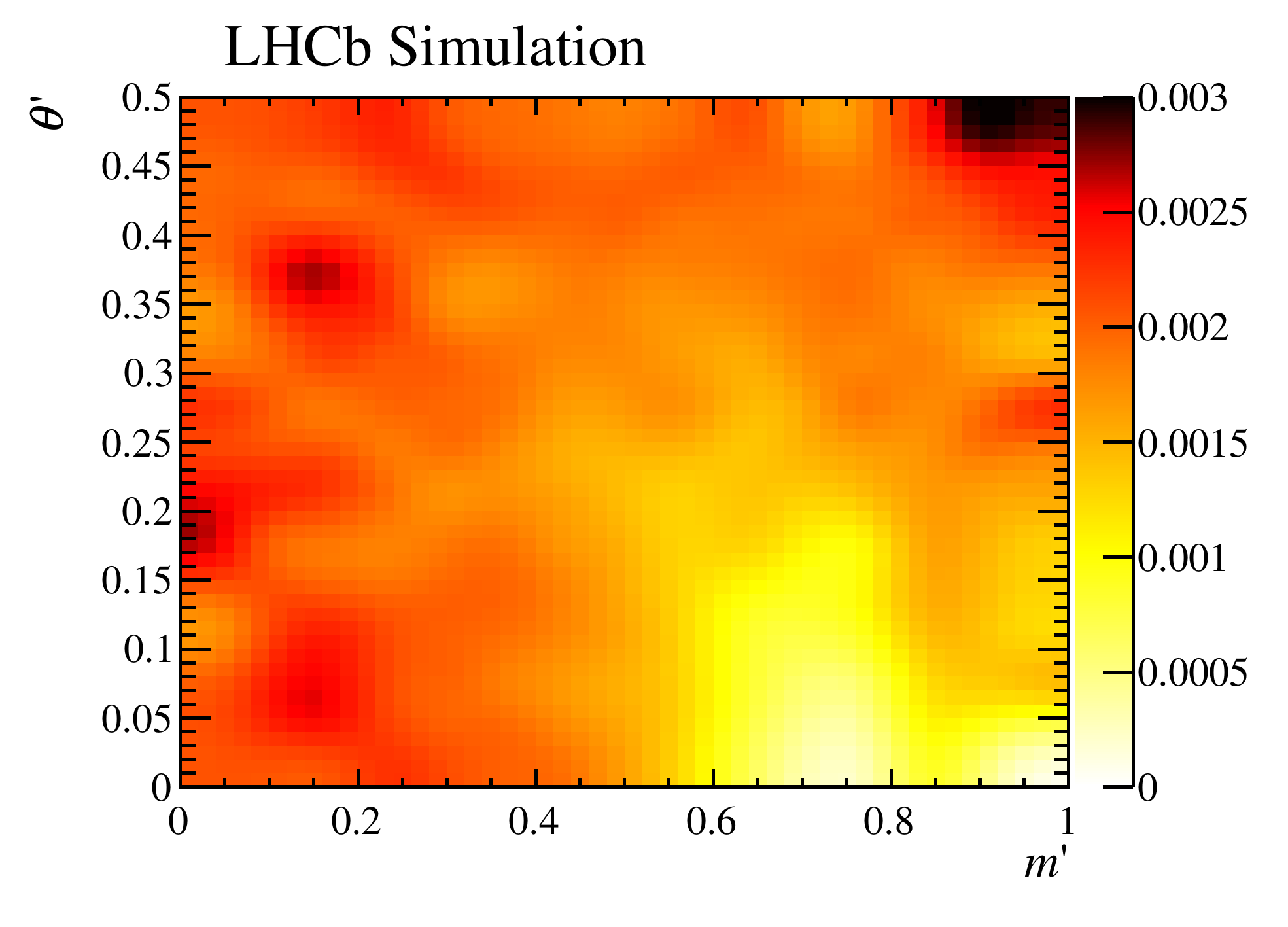}
 \caption{\small 
   Signal efficiency across the SDP for \Btodpipi decays.
   The relative uncertainty at each point is typically $5\,\%$.
}
 \label{fig:eff}
\end{figure}

Variation of the efficiency across the phase space of \Btodpipi decays is studied in terms of the SDP, since the efficiency variation is typically greatest close to the kinematic boundaries of the conventional DP.
The causes of efficiency variation across the SDP are the detector acceptance and trigger, selection and PID requirements.
Simulated samples, generated uniformly over the SDP, are used to evaluate the efficiency variation.
Data-driven corrections are applied to correct the simulation for known discrepancies with data, for the tracking, trigger and PID efficiencies, 
using identical methods to those described in Ref.~\cite{LHCb-PAPER-2014-036}.
The efficiency distributions are fitted with two-dimensional cubic splines to smooth out statistical fluctuations due to limited sample size.
Figure~\ref{fig:eff} shows the efficiency variation over the SDP.

\subsection{Background studies}
\label{sec:DPbkgs}

\begin{figure}[!tb]
\centering
\includegraphics[scale=0.35]{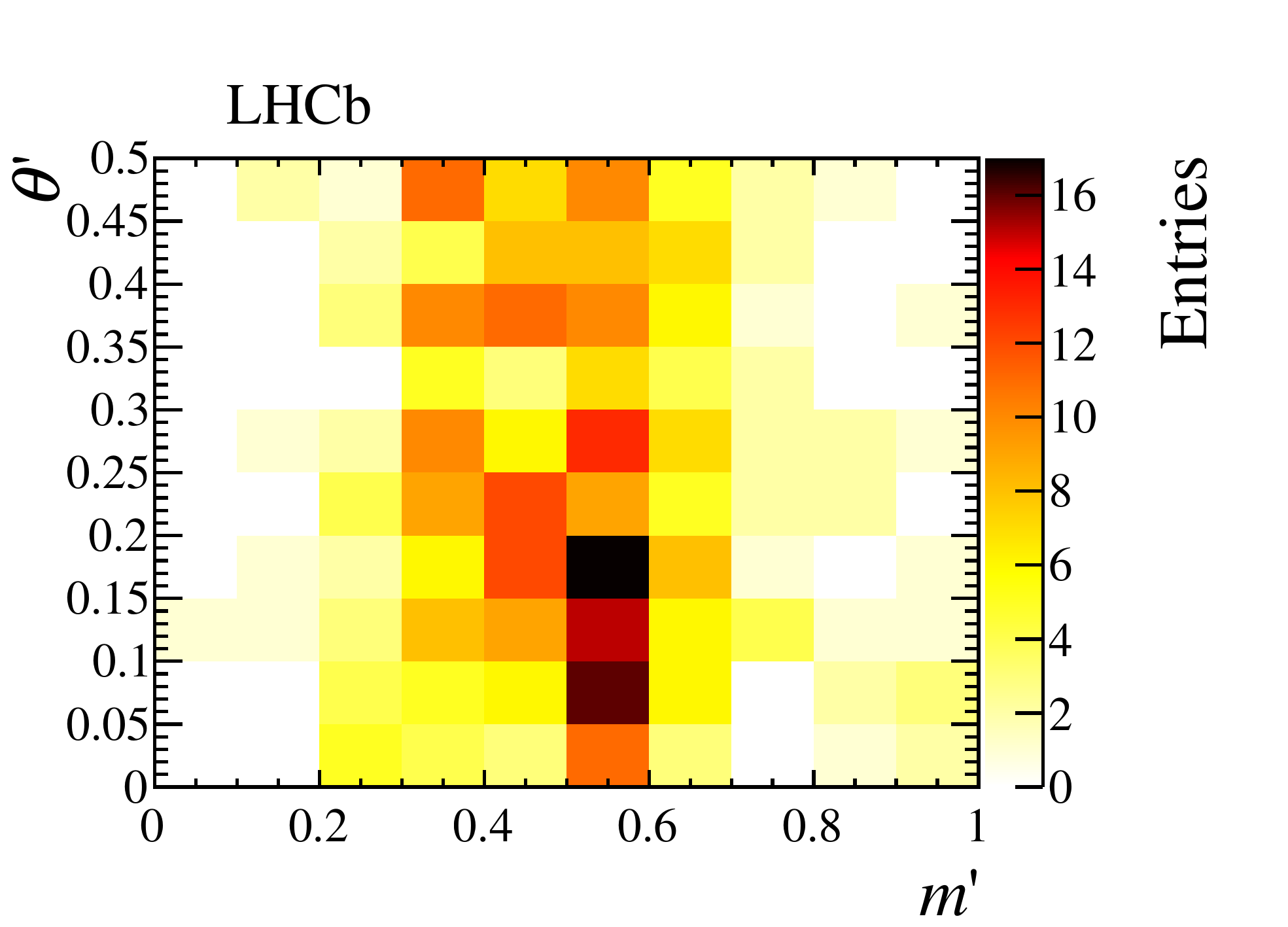}
\includegraphics[scale=0.35]{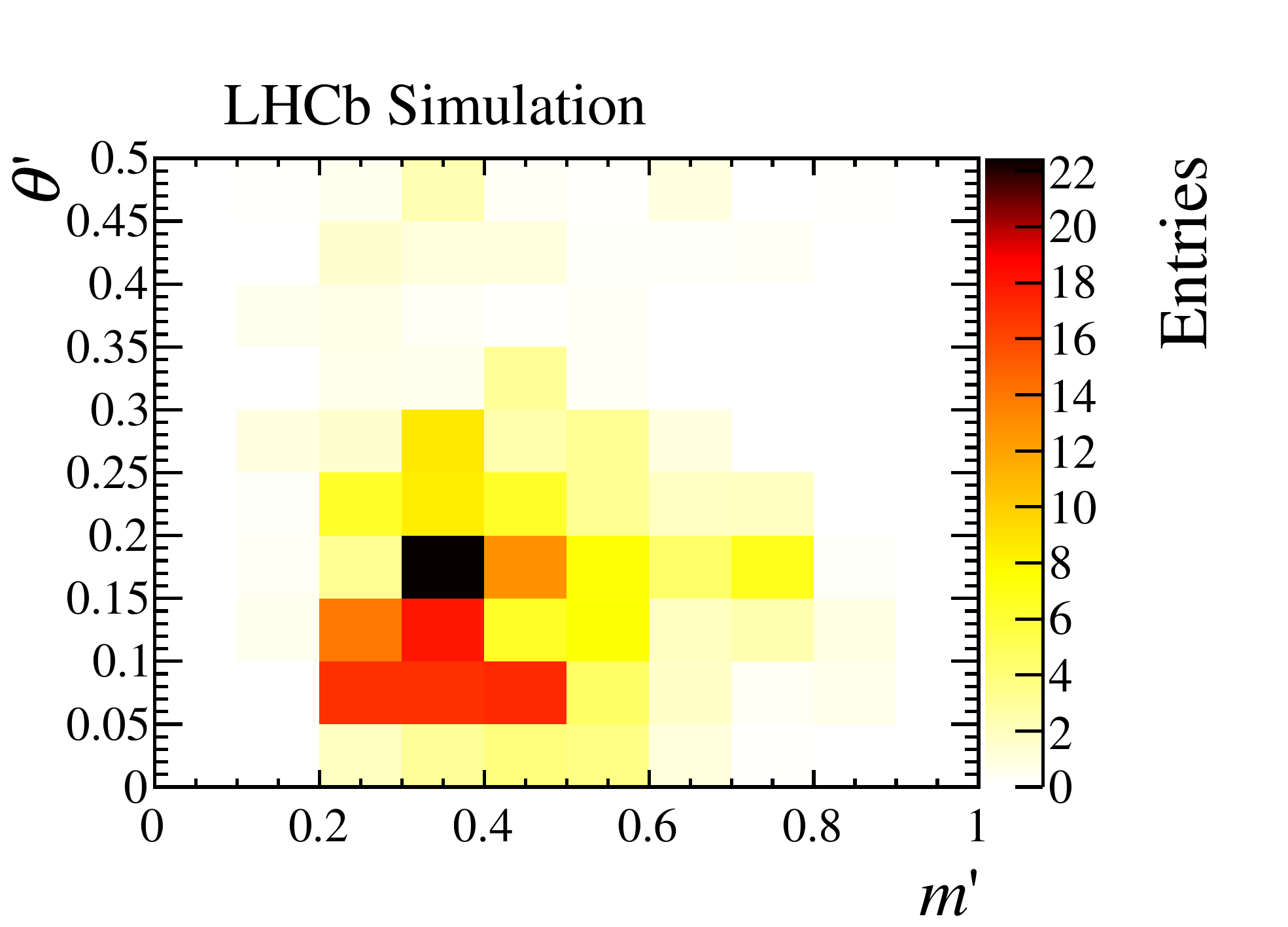}
\caption{\small Square Dalitz plot distributions for (left)~combinatorial background and (right)~\mbox{$\Btodanddstarkpi$} decays. }
\label{fig:bkgs}
\end{figure}

The yields presented in Table~\ref{tab:DpipiFit_yields} show that the important background components in the signal region are from combinatorial background and \Btodanddstarkpi decays. 
The SDP distribution of \Btodanddstarkpi decays is obtained from simulated samples using the same procedures as described in Sec.~\ref{sec:mass-fit} 
to apply weights and combine the $\Dp$ and $D^{*+}$ contributions.  
The distribution of combinatorial background events is obtained from \dpipi candidates in the high-mass sideband, defined to be $5500$--$5800\mev$. 
Figure~\ref{fig:bkgs} shows the SDP distributions of these backgrounds, which are used in the Dalitz plot fit.

\subsection{Amplitude model for \Btodpipi decays}
\label{sec:DPmodel}
The DP fit is performed using the {\sc Laura++}~\cite{Laura++} package, and the likelihood function is given by
\begin{equation}
 {\cal L} =
 \prod_i^{n_c}
 \Bigg[
 \sum_k N_k {\cal P}_k\left(s_i,t_i\right)
 \Bigg] \,,
\end{equation}
where the index $i$ runs over $n_c$ candidates, while $k$ sums over the probability density functions ${\cal P}_k$ with a yield of $N_k$ candidates in each component.
For signal events ${\cal P}_k \equiv {\cal P}_{\rm sig}$ is similar to Eq.~(\ref{eq:SigDPLike}), but is modified such that the $|{\cal A}\left(s,t\right)|^2$ 
terms are multiplied by the efficiency function described in Sec.~\ref{sec:efficiency}.
The mass resolution is approximately $2.4\mev$, which is much less than the width of the narrowest contribution to the Dalitz plot ($\sim 50 \mev$); therefore, this has negligible effect on the likelihood. 
Its effect on the measurement of masses and widths of resonances is, however, considered as a systematic uncertainty. 

Using the results of the moments analysis presented in Sec.~\ref{sec:moments} as a guide, a \mbox{\Btodpipi} DP model is constructed by including various resonant, nonresonant and virtual amplitudes. 
Only intermediate states with natural spin-parity are included because unnatural spin-parity states do not decay to two pseudoscalars.
Amplitudes that do not contribute significantly and cause the fit to become unstable are discarded. 
Alternative and additional contributions that have been considered include: an isobar description of the \Dpiswave\ including the \olddstarzero\ resonance and a nonresonant amplitude; a nonresonant P-wave component; an isospin-2 $\pi\pi$ interaction described by a unitary model as in Refs.~\cite{Achasov:2003xn,Bonvicini:2008jw} (see also Refs.~\cite{Pelaez:2004vs,Kaminski:2006qe,GarciaMartin:2011cn}); quasi-model-independent descriptions of partial waves other than the \Dpiswave.

The resulting baseline signal model consists of the seven components listed in Table~\ref{tab:resonances}: four resonances, two virtual resonances and a quasi-model-independent description of the \Dpiswave.
There are 42 free parameters in this model.
The broad P-wave structure indicated by the angular moments is adequately described by the virtual $D^*(2007)^0$ and $B^{*0}$ amplitudes.
The peaks seen in various moments are described by the \olddstartwo, \newdstarone, \newdstarthree and \newdstartwo resonances.
Here, and throughout the paper, these states are labelled as such since it is not clear if the \newdstarone\ state corresponds to one of the previously observed peaks (see Table~\ref{tab:PDG}), while the parameters of the \newdstarthree\ resonance seem to be consistent with earlier measurements.
An excess at $m(\Dpi) \approx 3000 \mev$ was reported in Ref.~\cite{LHCb-PAPER-2013-026}, but the parameters of this state were not reported with systematic uncertainties.
The baseline model provides a better quality fit than the alternative models that are discussed in Sec.~\ref{sec:systematics}. 
The inclusion of all components of the model is necessary to obtain a good description of the data, as described in Sec.~\ref{sec:results}. 

\begin{table}[!tb]
\centering
\caption{\small
  Signal contributions to the fit model, where parameters and uncertainties are taken from Ref.~\cite{PDG2014}. 
  States labelled with subscript $v$ are virtual contributions.
  The model ``MIPW'' refers to the quasi-model-independent partial wave approach.
}
\label{tab:resonances}
\begin{tabular}{lccc}
\hline
  \noalign{\vskip 1mm}  
Resonance & Spin & Model & Parameters \\
\hline
  \noalign{\vskip 1mm}  
\olddstartwo &2& RBW & \multirow{4}{*}{Determined from data (see Table~\ref{tab:masswidth})} \\
\newdstarone &1& RBW & \\ 
\newdstarthree &3& RBW & \\ 
\newdstartwo &2& RBW & \\ 
  \noalign{\vskip 1mm}  
\hline
  \noalign{\vskip 1mm}  
\dstarv &1& RBW & $m = 2006.98 \pm 0.15 \mev$, $\Gamma = 2.1 \mev$ \\
$\B^{*0}_{v}$ &1& RBW & $m = 5325.2 \pm 0.4 \mev$, $\Gamma = 0.0 \mev$  \\
  \noalign{\vskip 1mm}  
\hline
  \noalign{\vskip 1mm}  
Total S-wave &0& MIPW & See text\\
  \noalign{\vskip 1mm}  
\hline
\end{tabular}
\end{table}

The real and imaginary parts of the complex coefficients for each of the components are free parameters of the fit, except for the \olddstartwo\ contribution that is taken to be a reference amplitude with real and imaginary parts of its complex coefficient $c_k$ fixed to 1 and 0, respectively.
Parameters such as magnitudes and phases for each amplitude, the fit fractions and interference fit fractions are calculated from these quantities. 
The statistical uncertainties are determined using large samples of pseudoexperiments to ensure that correlations between parameters are accounted for.

\subsection{Dalitz plot fit results}
\label{sec:DPresults}
\begin{figure}[!tb]
\centering
\includegraphics[scale=0.45]{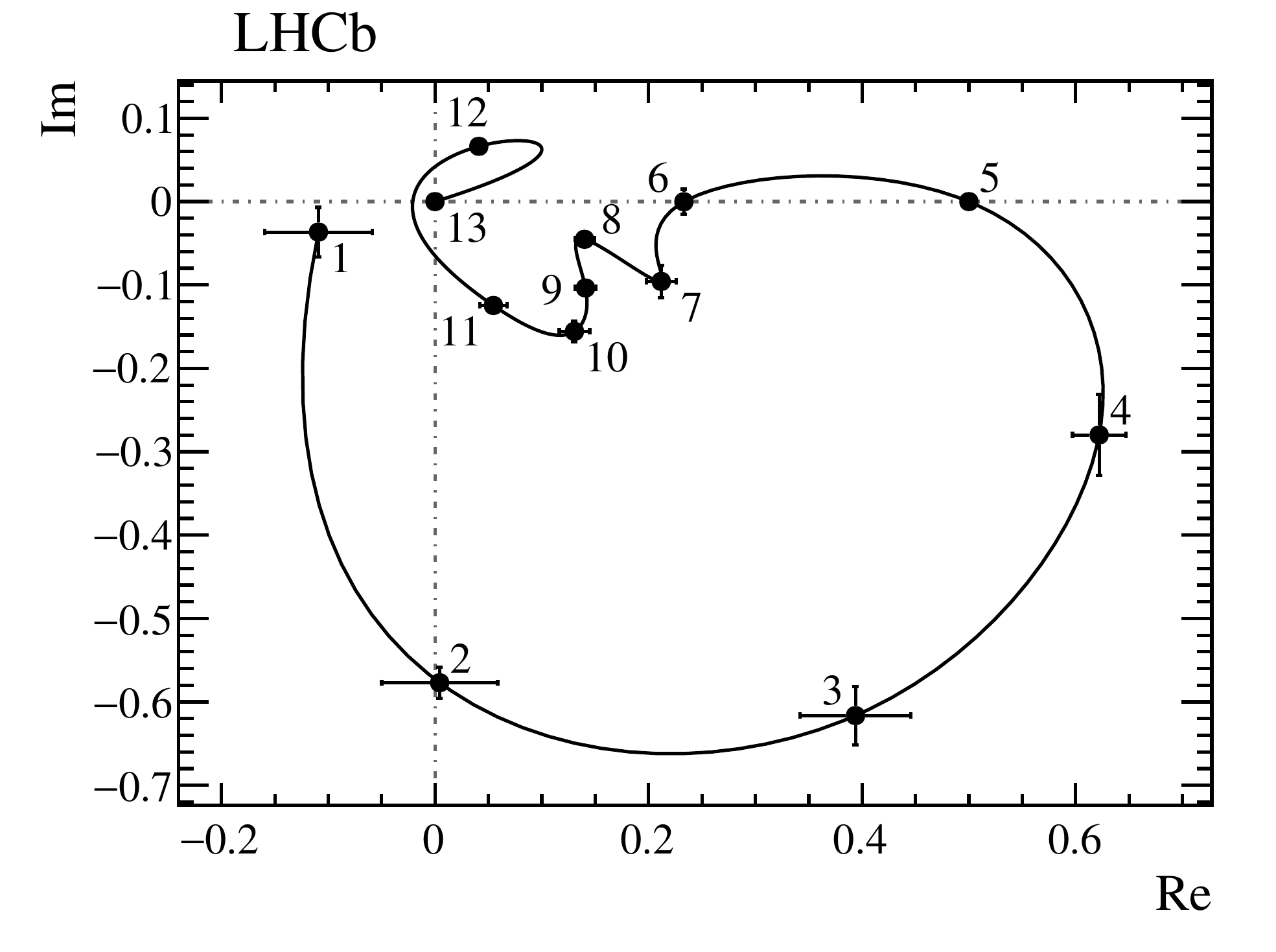} 
\caption{\small 
  Real and imaginary parts of the S-wave amplitude, shown in an Argand diagram.
  The knots are shown with statistical uncertainties only, connected by the cubic spline interpolation used in the fit.
  The leftmost point is that at the lowest value of $m(\Dpi)$, with mass increasing along the connected points.
Each point, labelled 1--13, corresponds to the position of a knot in the spline, at values of $m(\Dpi) = \{ 2.01, 2.10, 2.20, 2.30, 2.40, 2.50, 2.60, 2.70, 2.80, 2.90, 3.10, 4.10, 5.14 \} \,\gev$. 
  The points at $(0.5,0.0)$ and $(0.0,0.0)$ are fixed.
  The anticlockwise rotation of the phase at low $m(\Dpi)$ is as expected due to the presence of the \olddstarzero\ resonance.
}
\label{fig:Argand_Swave}
\end{figure}

The masses and widths of the \olddstartwo, \newdstarone, \newdstarthree\ and \newdstartwo\ resonances are determined from the fit and are given in Table~\ref{tab:masswidth}.
\begin{table}[!tb]
\centering
\caption{\small Masses and widths determined in the fit to data, with statistical uncertainties only.}
\label{tab:masswidth}
\begin{tabular}{lcc}
\hline
Contribution & Mass (MeV) & Width (MeV) \\ 
\hline 
  \noalign{\vskip 1mm}  
\olddstartwo &   $2463.7 \pm 0.4$ & $\phantom{1}47.0 \pm 0.8$\\ 
\newdstarone &   $2681.1 \pm 5.6$ & $186.7 \pm 8.5$\\ 
\newdstarthree &   $2775.5 \pm 4.5$ & $\phantom{1}95.3 \pm 9.6$\\ 
\newdstartwo & $\phantom{1}3214 \pm 29$ & $\phantom{1}186 \pm 38$\\ 
\hline
\end{tabular}
\end{table}
The floated complex coefficients at each knot position and the splines describing the total \Dpiswave\ are shown in Fig.~\ref{fig:Argand_Swave}.
The phase motion at low $m(\Dpi)$ is consistent with that expected due to the presence of the \olddstarzero\ state.
There is, however, an ambiguous solution with the opposite phase motion in this region, which occurs since there are significant contributions only from S- and P-waves and thus only $\cos(\delta_0 - \delta_1)$ can be determined as seen in Eq.~(\ref{eq:p1}).
Since the P-wave in this region is described by the \dstarv\ amplitude, and hence has slowly varying phase, the entire \Dpiswave\ has a sign ambiguity.
Similar ambiguities have been observed previously~\cite{Aitala:2005yh}.
Only results consistent with the expected phase motion are reported.

Table~\ref{tab:ffstat} shows the values of the complex coefficients and fit fractions for each amplitude.
The interference fit fractions are given in Appendix~\ref{app:iffstat}. 
\begin{table}[!tb]
\centering
\caption{\small Complex coefficients and fit fractions determined from the
  Dalitz plot fit. Uncertainties are statistical only.
}
\label{tab:ffstat}
\resizebox{\textwidth}{!}{ 
\begin{tabular}{lccccc} 
\hline 
  \noalign{\vskip 1mm}  
 \multirow{2}{*}{Contribution}& & \multicolumn{4}{c}{Isobar model coefficients} \\ 
 & Fit fraction (\%) & Real part & Imaginary part & Magnitude & Phase (rad) \\ 
\hline 
  \noalign{\vskip 1mm}  
 \olddstartwo & 	$35.7 \pm 0.6$ & 	 $1.00$ & 	 $0.00$ & 	$1.00$ & 	 $0.00$ \\ 
 \newdstarone & 	$\phantom{1}8.3 \pm 0.6$ & 	 $-0.38 \pm 0.02$ & 		 $\phantom{-}0.30 \pm 0.02$ & 	$0.48 \pm 0.02$ & 	 $\phantom{-}2.47 \pm 0.09$\\ 
 \newdstarthree &	$\phantom{1}1.0 \pm 0.1$ & 	 $\phantom{-}0.17 \pm 0.01$ & 	 $\phantom{-}0.00 \pm 0.01$ & 	$0.17 \pm 0.01$ & 	 $\phantom{-}0.01 \pm 0.20$\\ 
 \newdstartwo &	$\phantom{1}0.23 \pm 0.07$ & 	 $\phantom{-}0.05 \pm 0.02$ & 	 $-0.06 \pm 0.02$ & 		$0.08 \pm 0.01$ & 	 $-0.84 \pm 0.28$\\ 
  \noalign{\vskip 1mm}  
\hline 
  \noalign{\vskip 1mm}  
 \dstarv & 	$10.8 \pm 0.7$ & 	 $\phantom{-}0.51 \pm 0.03$ & 	 $-0.20 \pm 0.05$ & 		$0.55 \pm 0.02$ & 	 $-0.38 \pm 0.19$\\ 
$\B^{*0}_{v}$ 	& 	$\phantom{1}2.7 \pm 1.0$ & 	 $\phantom{-}0.27 \pm 0.03$ & 	 $\phantom{-}0.04 \pm 0.04$ & 	$0.27 \pm 0.05$ & 	 $\phantom{-}0.14 \pm 0.38$\\ 
  \noalign{\vskip 1mm}  
\hline 
  \noalign{\vskip 1mm}  
Total S-wave 		& 	$57.0 \pm 0.8$ & 	 $\phantom{-}1.21 \pm 0.02$ & 	 $-0.35 \pm 0.04$ & 		$1.26 \pm 0.01$ & 	 $-0.28 \pm 0.05$\\ 
\hline 
Total fit fraction & 	115.7\\ 
\hline 
\end{tabular} 
}
\end{table}

Given the complexity of the DP fit, the minimisation procedure may find local minima in the likelihood function.
To try to ensure that the global minimum is found, the fit is performed many times with randomised initial values for the $c_j$ terms.
No other minima are found with negative log-likelihood values close to that of the global minimum so they are not considered further.

\begin{figure}[!tb]
\centering
\includegraphics[scale=0.56]{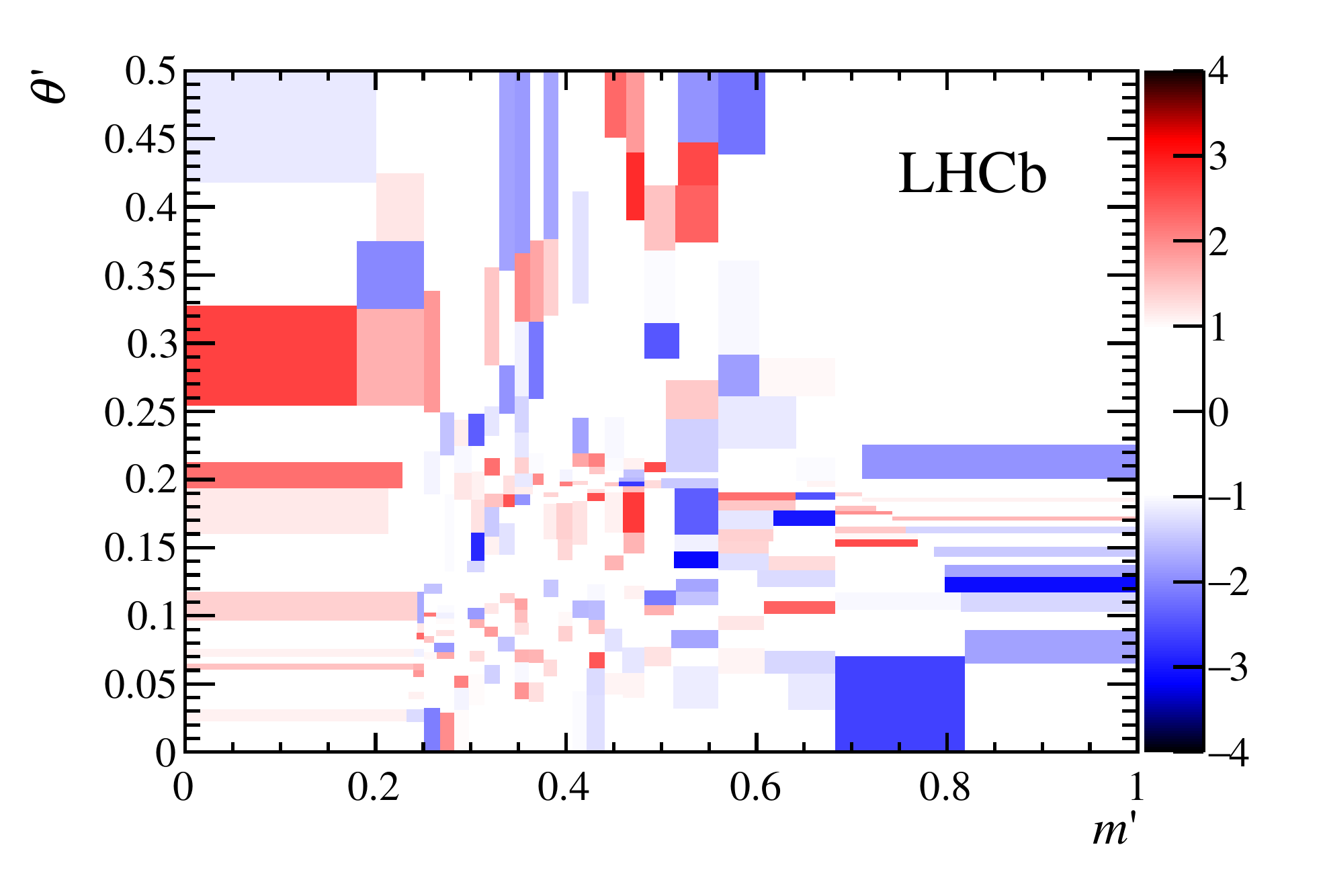}
\caption{\small 
  Differences between the SDP distribution of the data and fit model, in terms of the normalised residual in each bin.
  No bin lies outside the $z$-axis limits.
}
\label{fig:sdppull}
\end{figure}

The consistency of the fit model and the data is evaluated in several ways.
Numerous one-dimensional projections comparing the data and fit model (including several shown below and those from the moments study in Sec.~\ref{sec:moments}) show good agreement.
Additionally, a two-dimensional $\chi^2$ value is calculated by comparing the data and the fit model distributions across the SDP in $484$ equally populated bins. 
Figure~\ref{fig:sdppull} shows the normalised residual in each bin.
The distribution of the $z$-axis values from Fig.~\ref{fig:sdppull} is consistent with a unit Gaussian centered on zero.
Further checks using unbinned fit quality tests~\cite{Williams:2010vh} show satisfactory agreement between the data and the fit model.

\begin{figure}[!tb]
\centering
\includegraphics[scale=0.35]{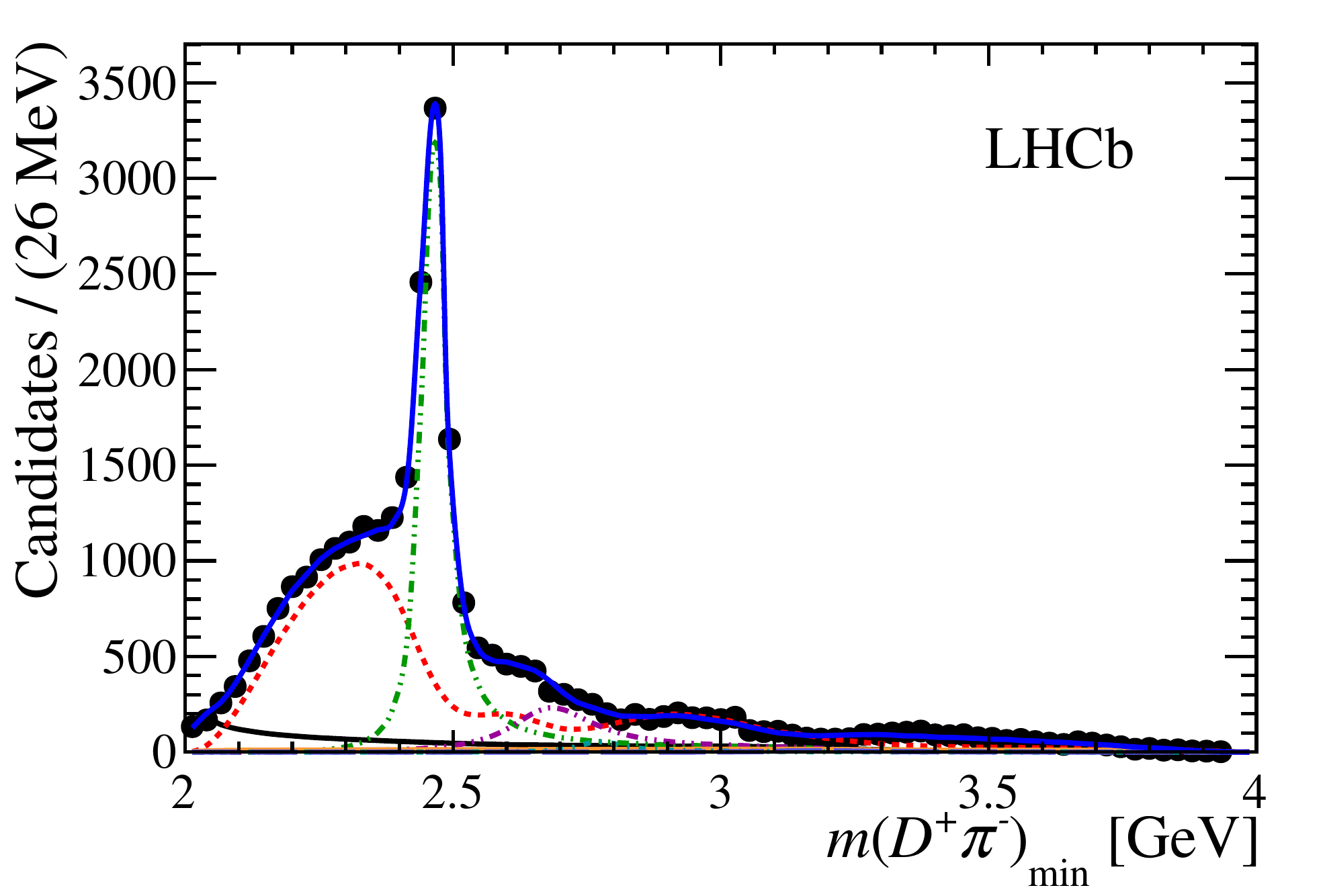}
\includegraphics[scale=0.35]{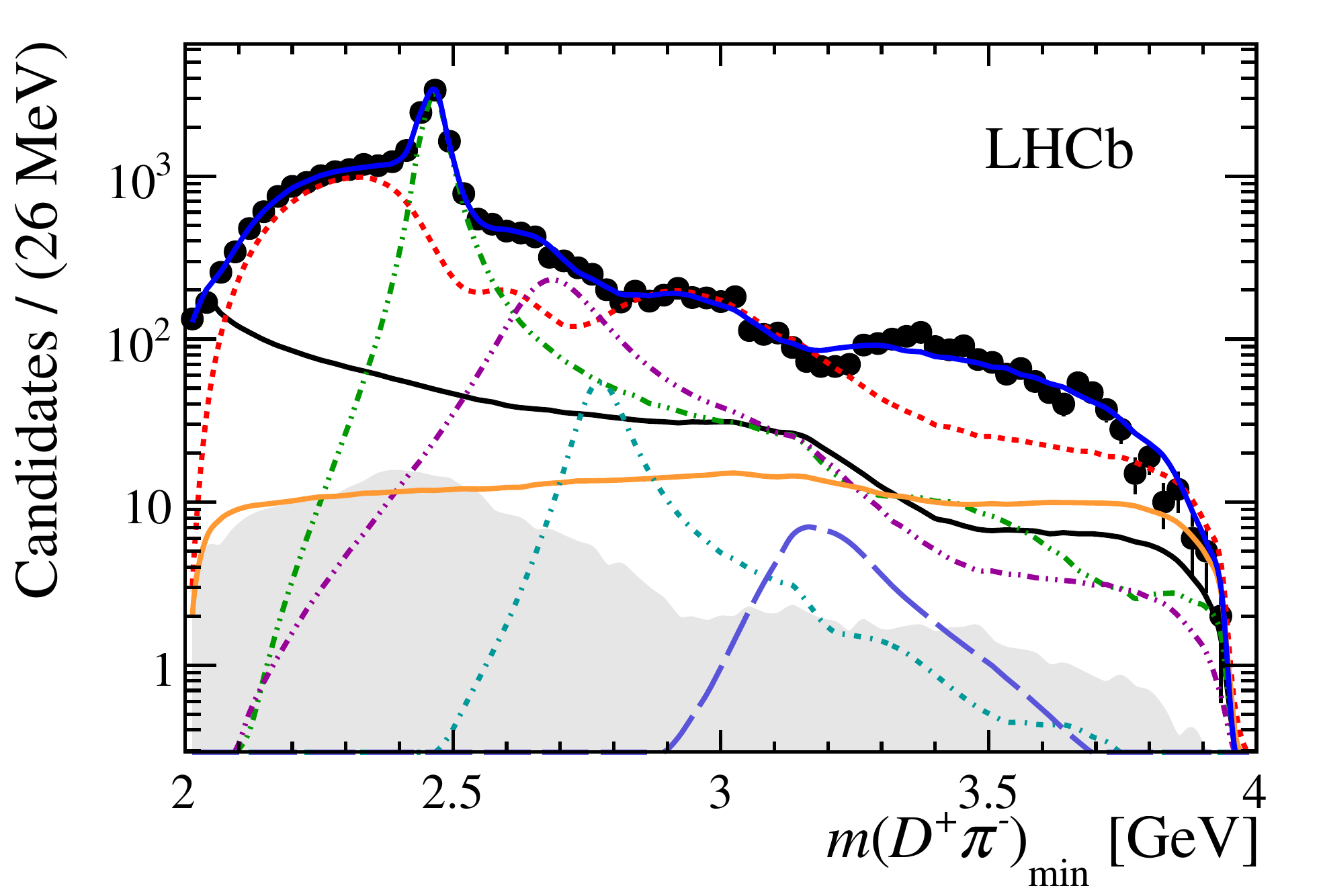}
\includegraphics[scale=0.35]{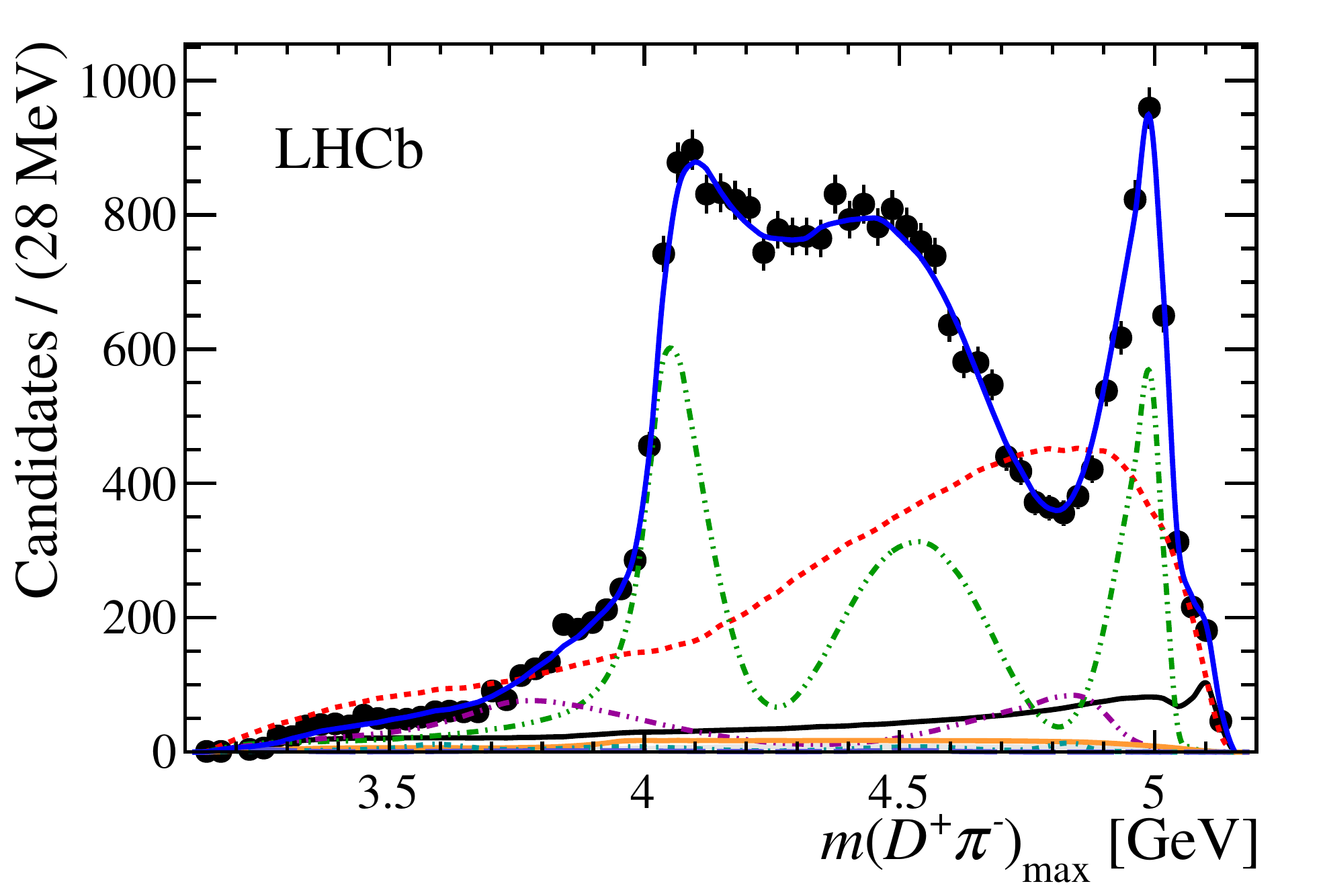}
\includegraphics[scale=0.35]{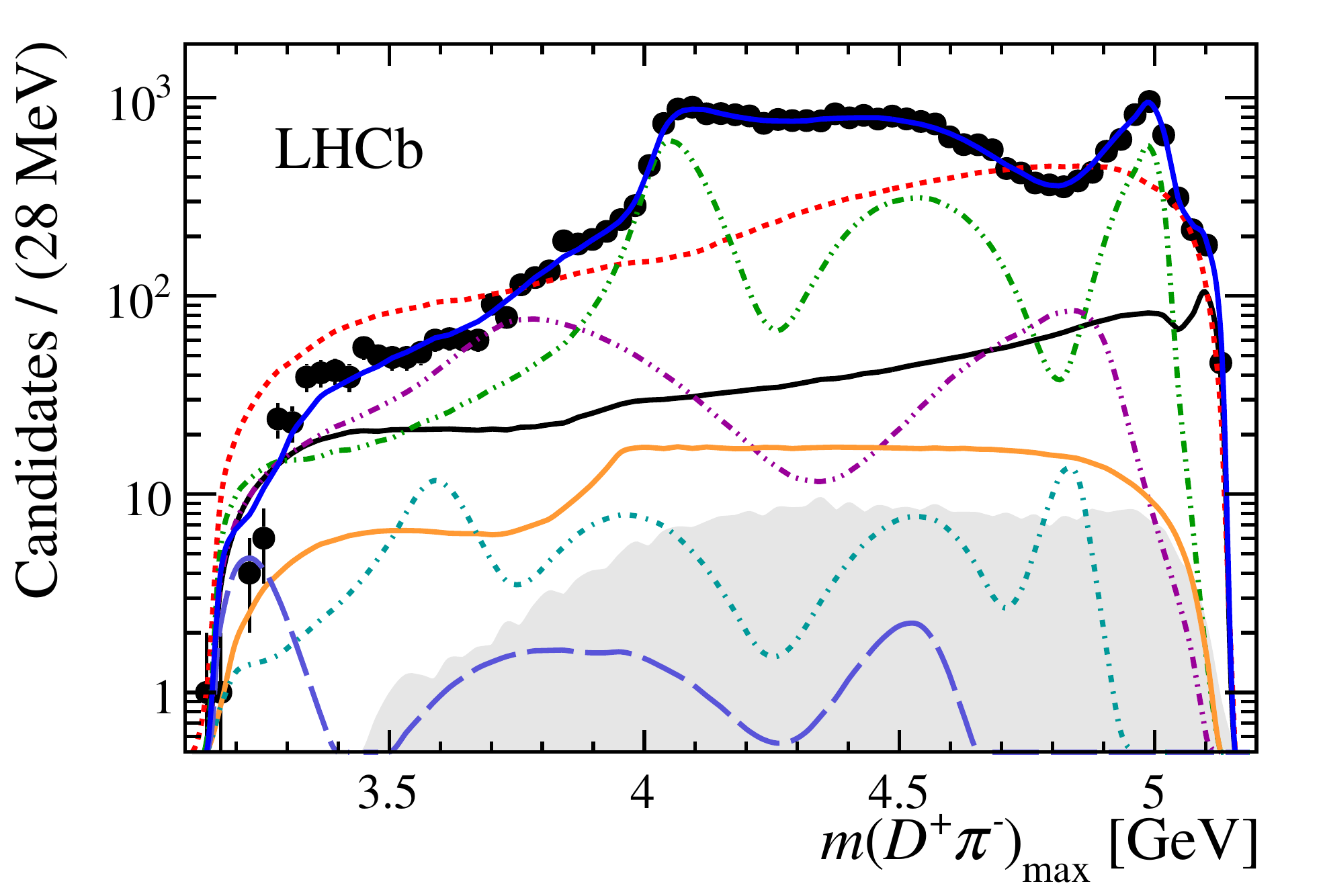}
\includegraphics[scale=0.35]{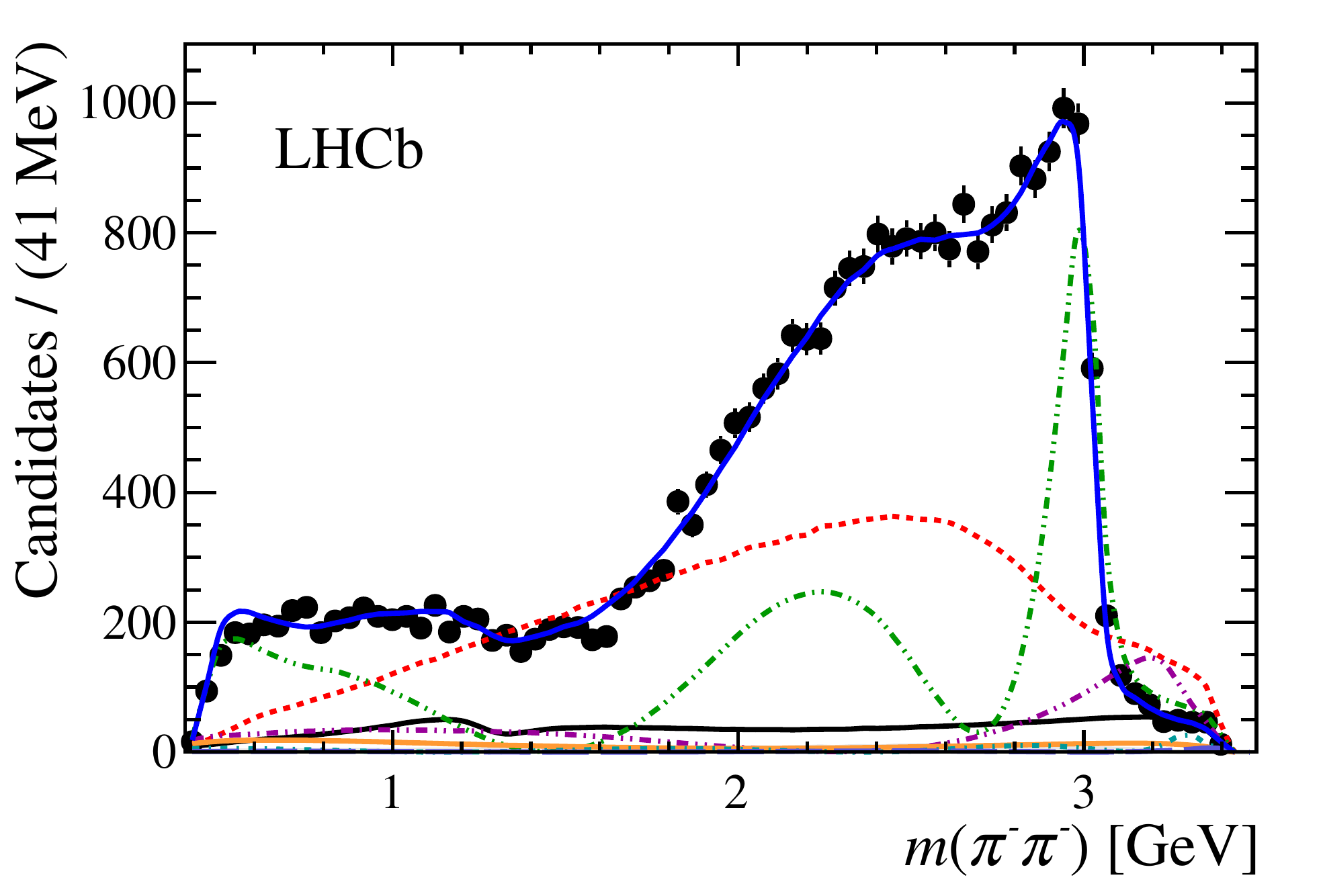}
\includegraphics[scale=0.35]{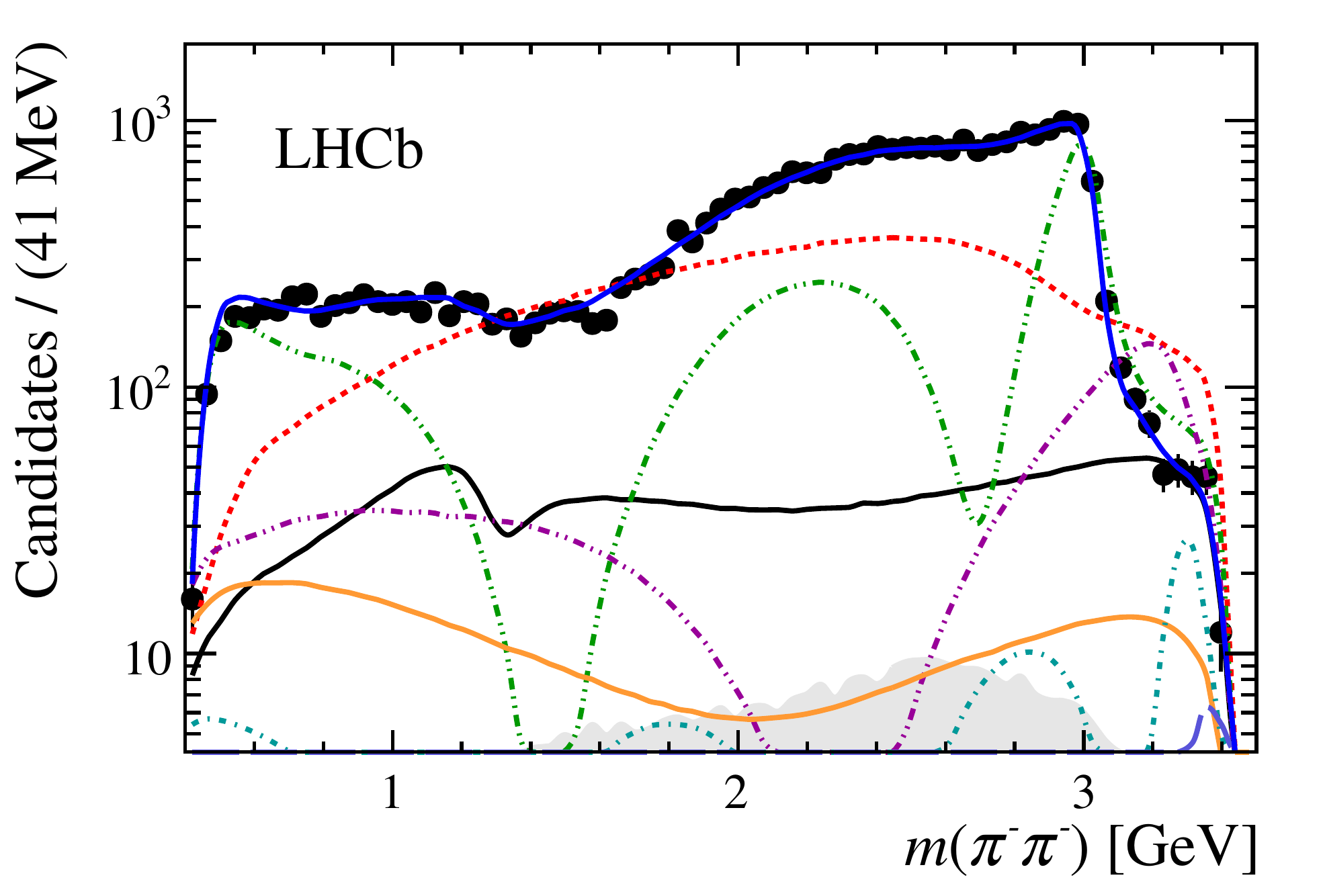} 
\includegraphics[scale=0.50]{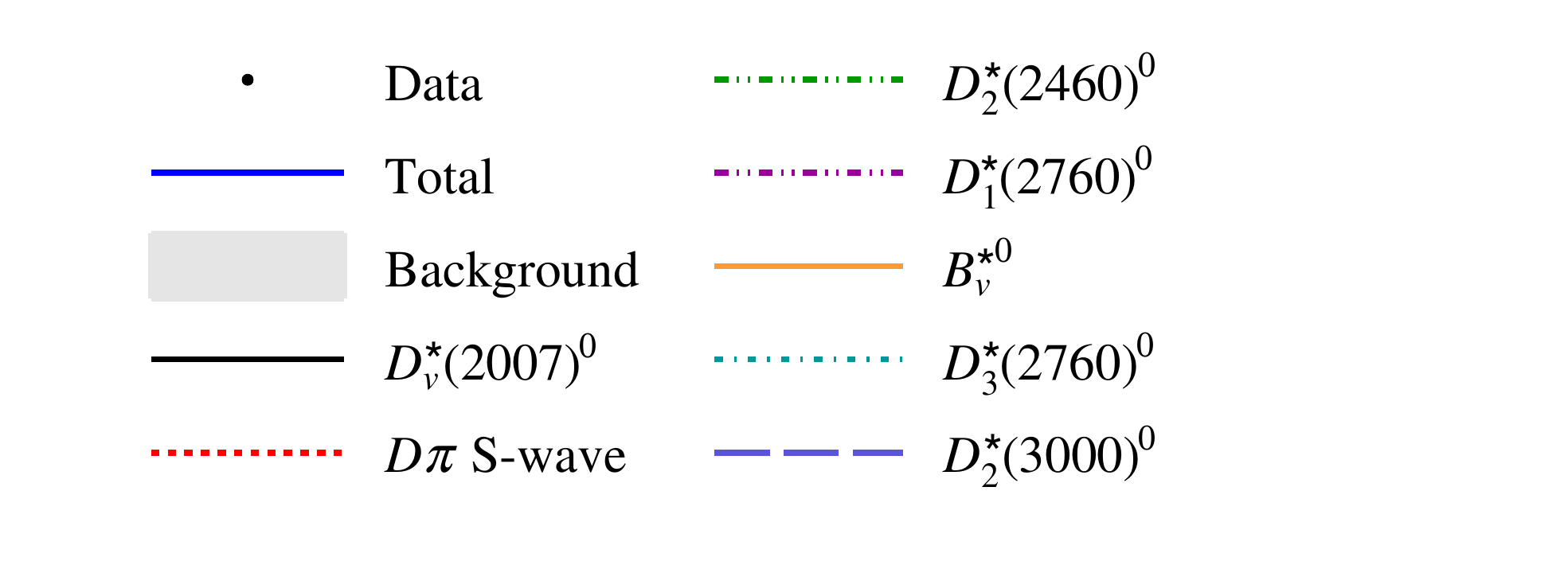}
\caption{\small Projections of the data and amplitude fit onto (top)~$m(\Dpi)_{\rm min}$, (middle)~$m(\Dpi)_{\rm max}$ and (bottom)~$m(\pim\pim)$, with the same 
projections shown (right) with a logarithmic $y$-axis scale. Components are described in the legend.}
\label{fig:cfit}
\end{figure}

\begin{figure}[!tb]
\centering
\includegraphics[scale=0.35]{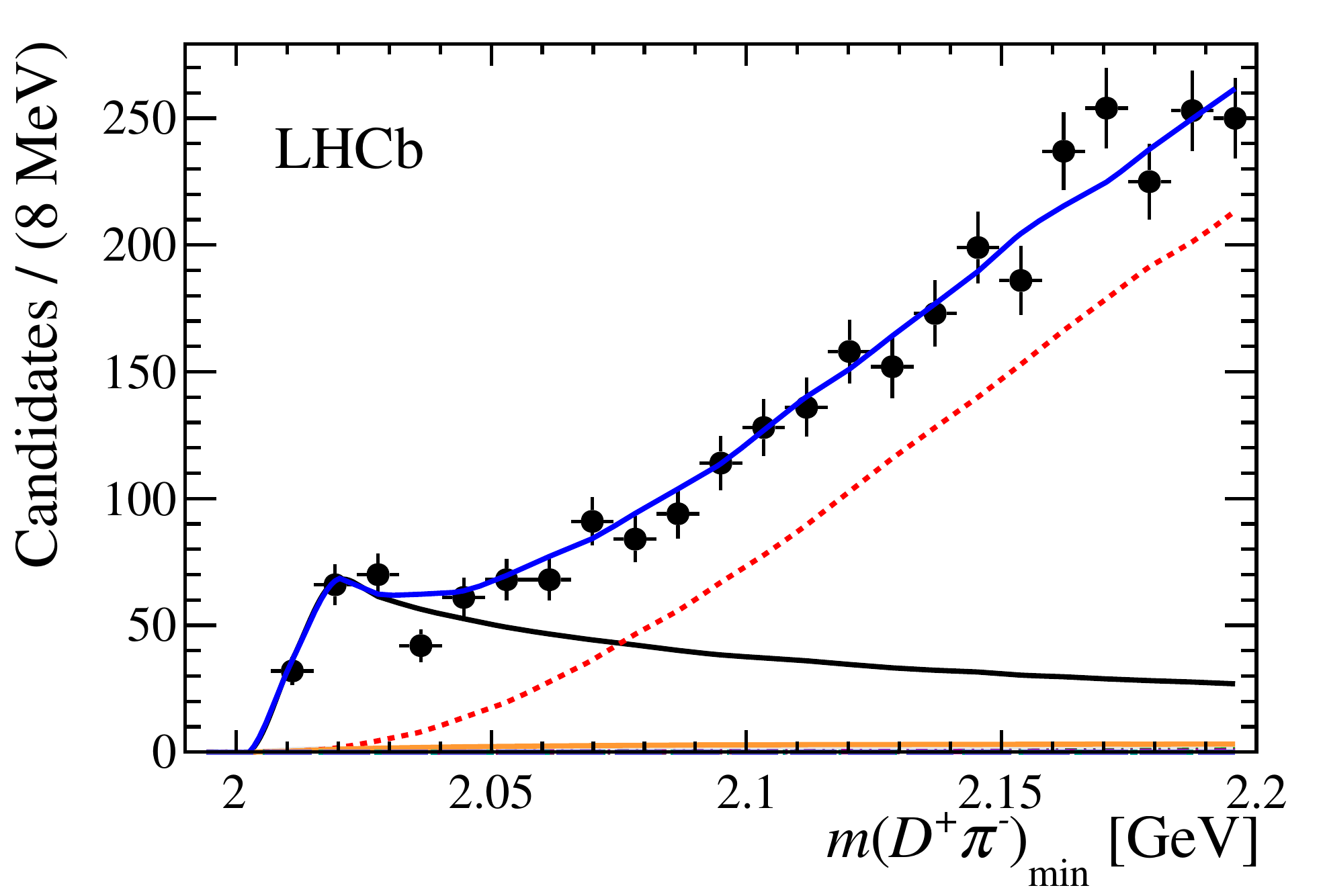}
\includegraphics[scale=0.35]{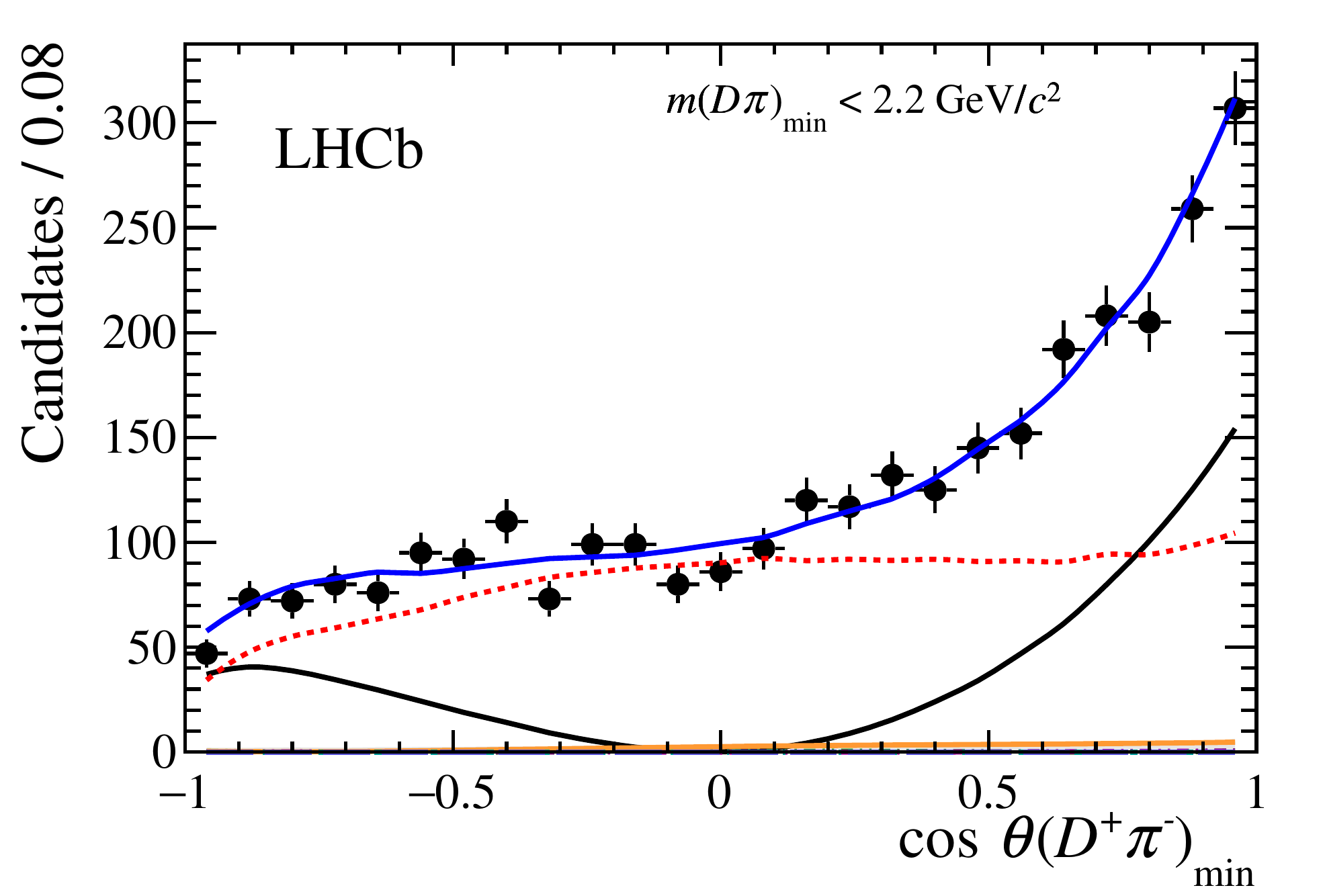}
\includegraphics[scale=0.35]{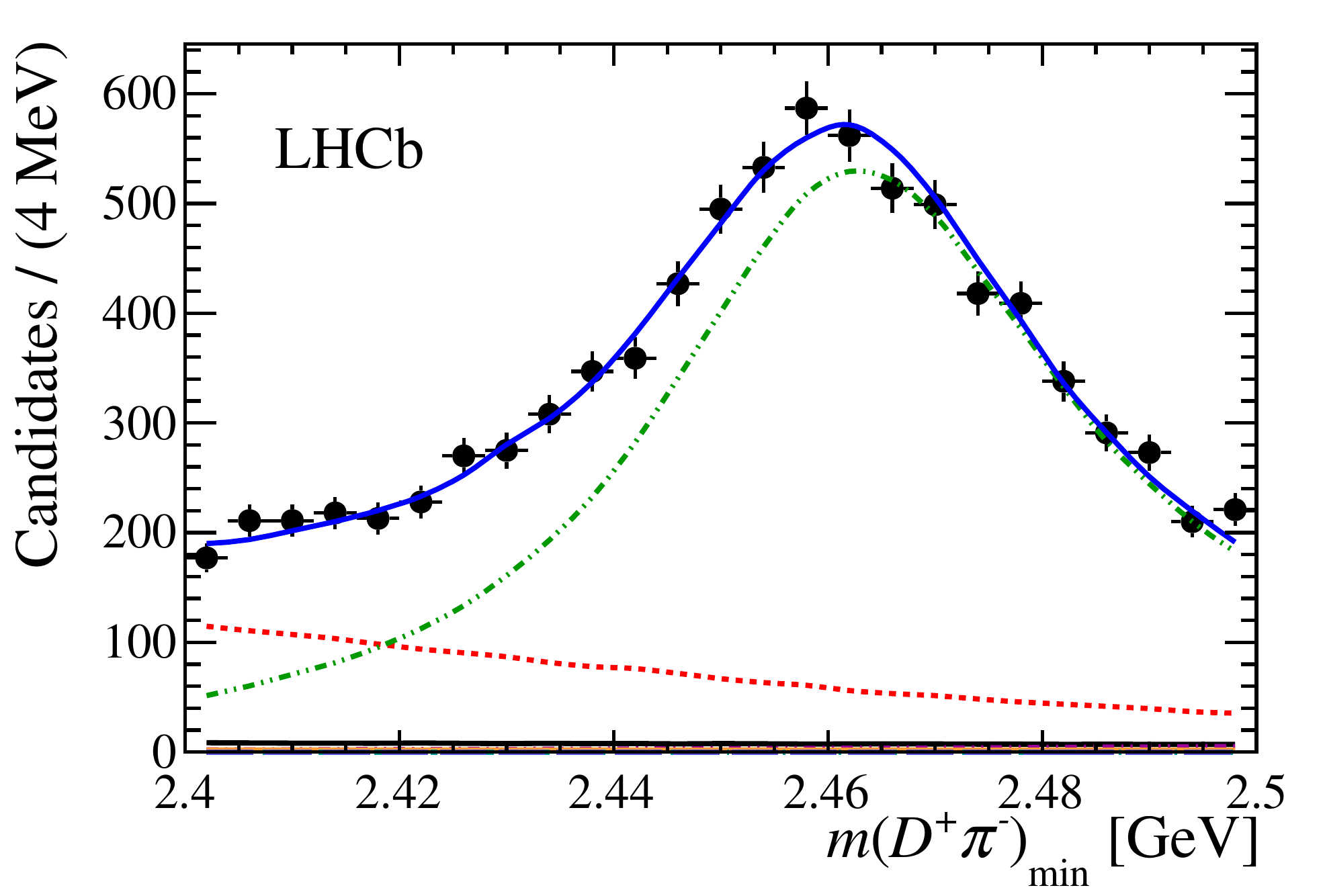}
\includegraphics[scale=0.35]{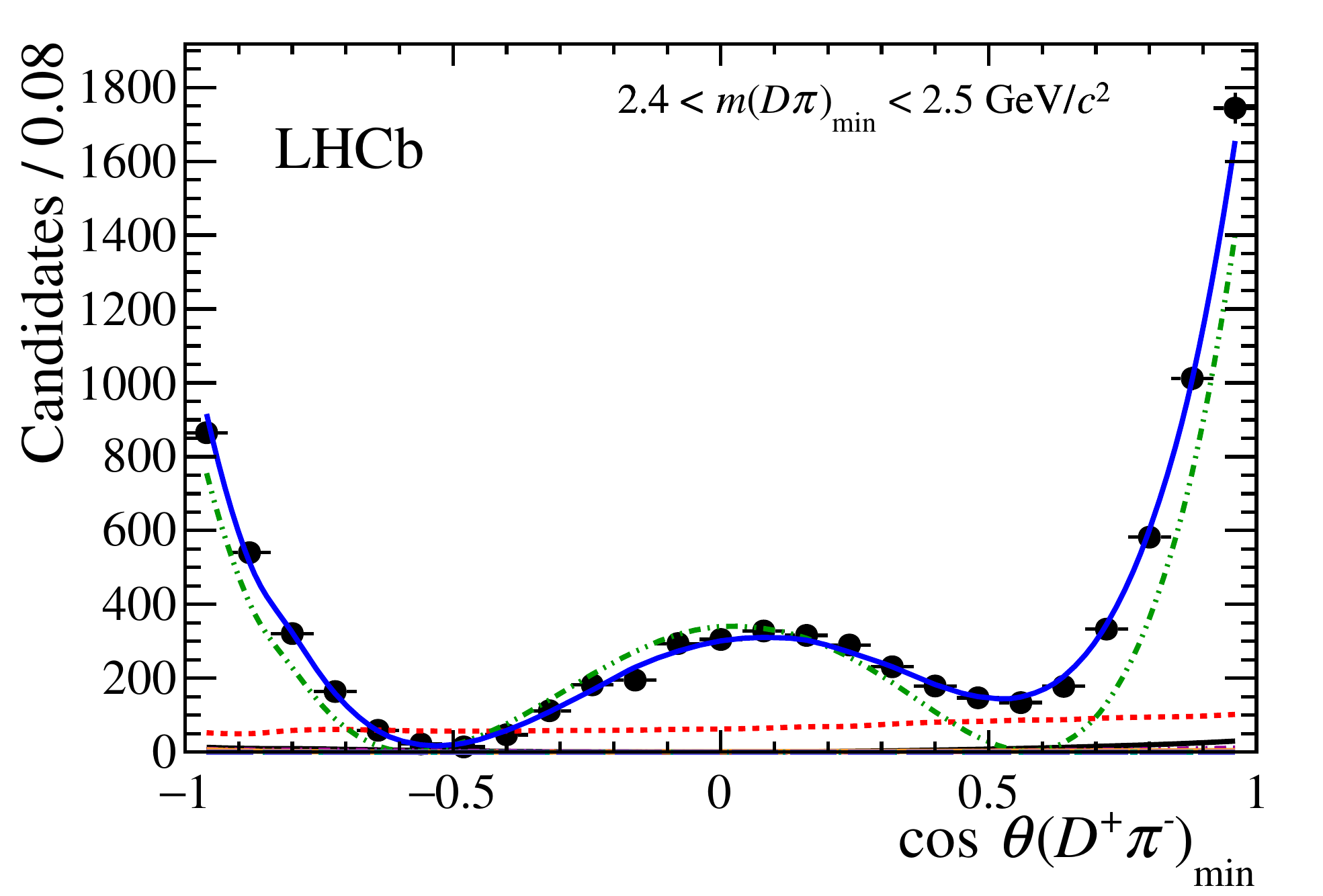}
\includegraphics[scale=0.35]{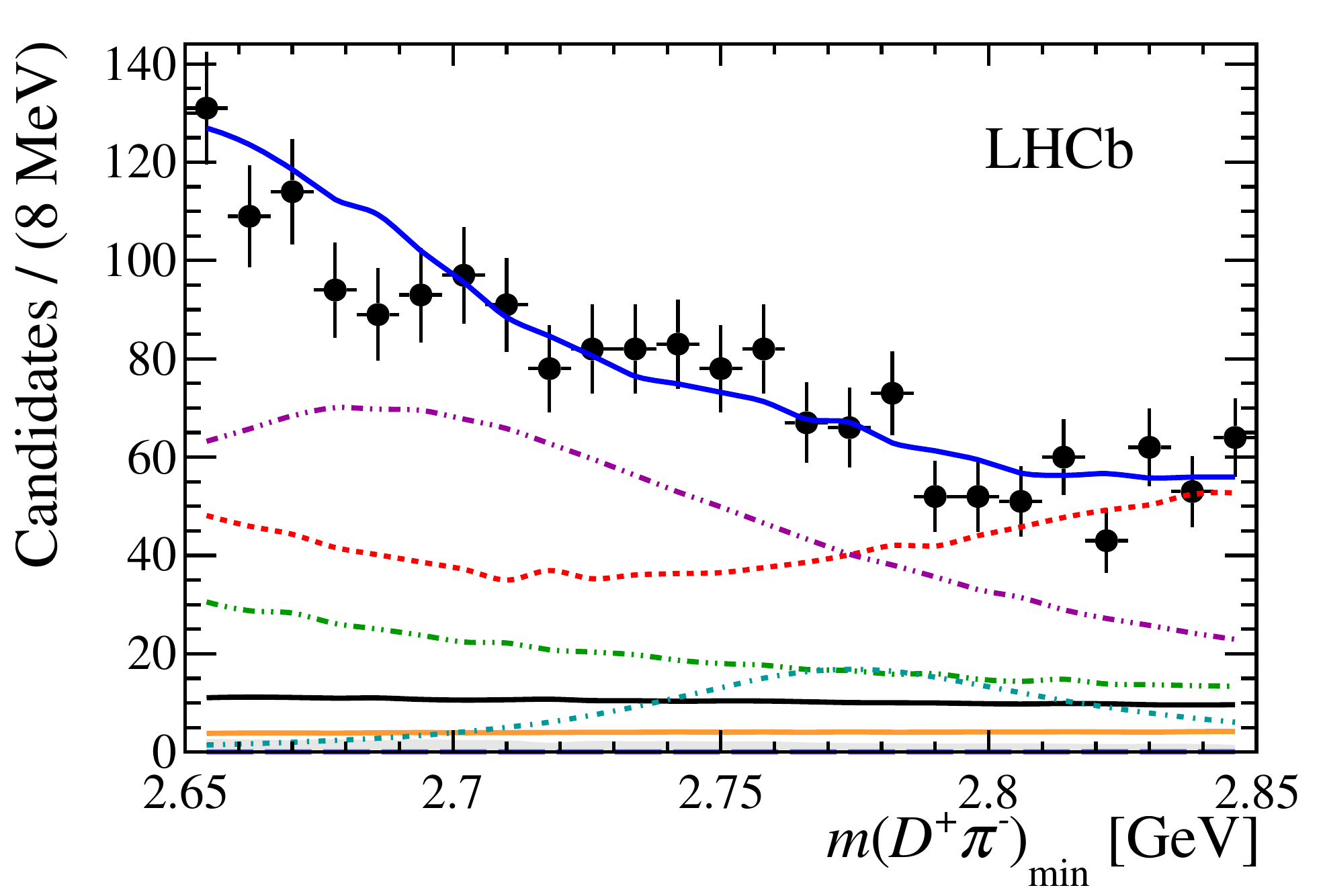}
\includegraphics[scale=0.35]{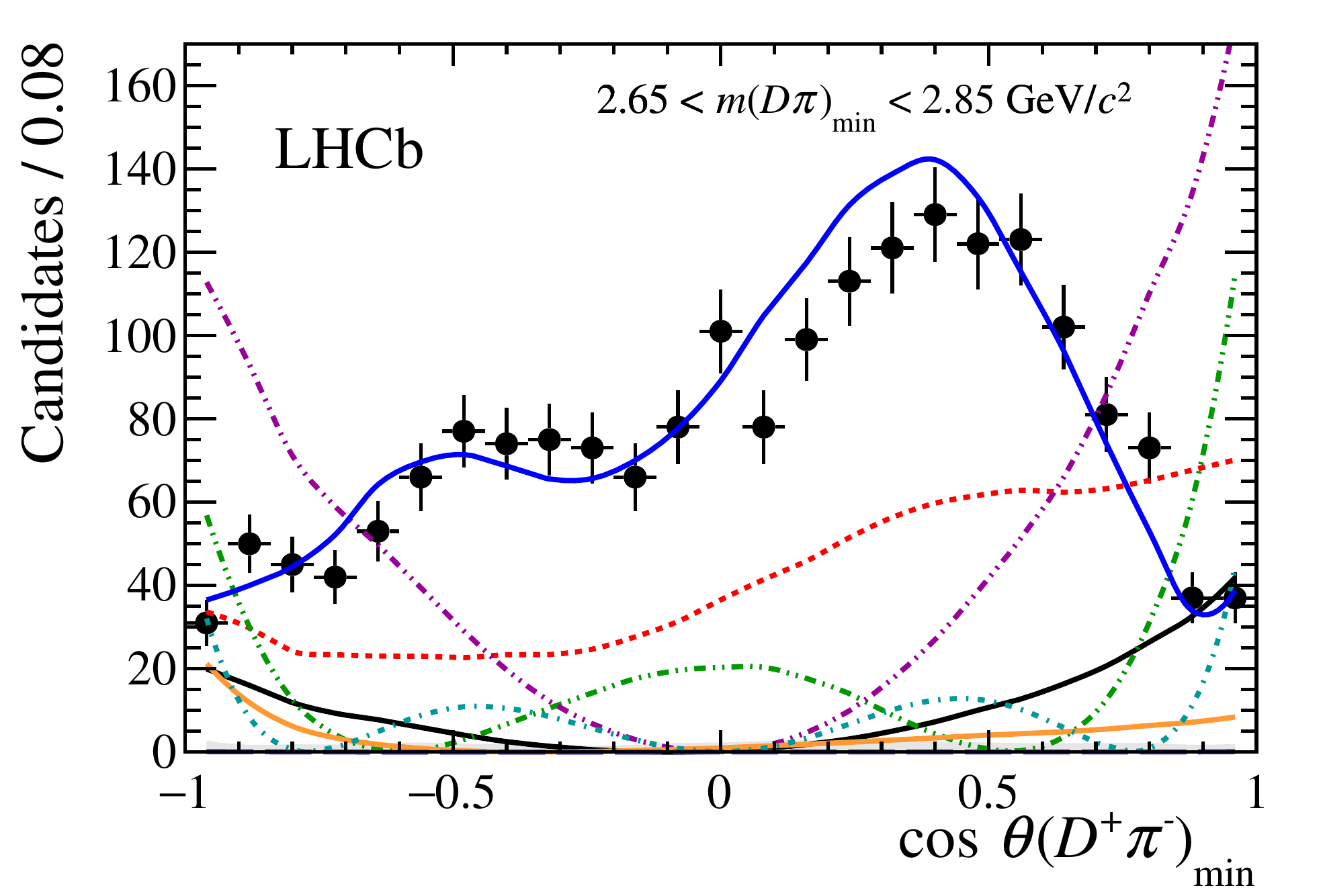}
\includegraphics[scale=0.35]{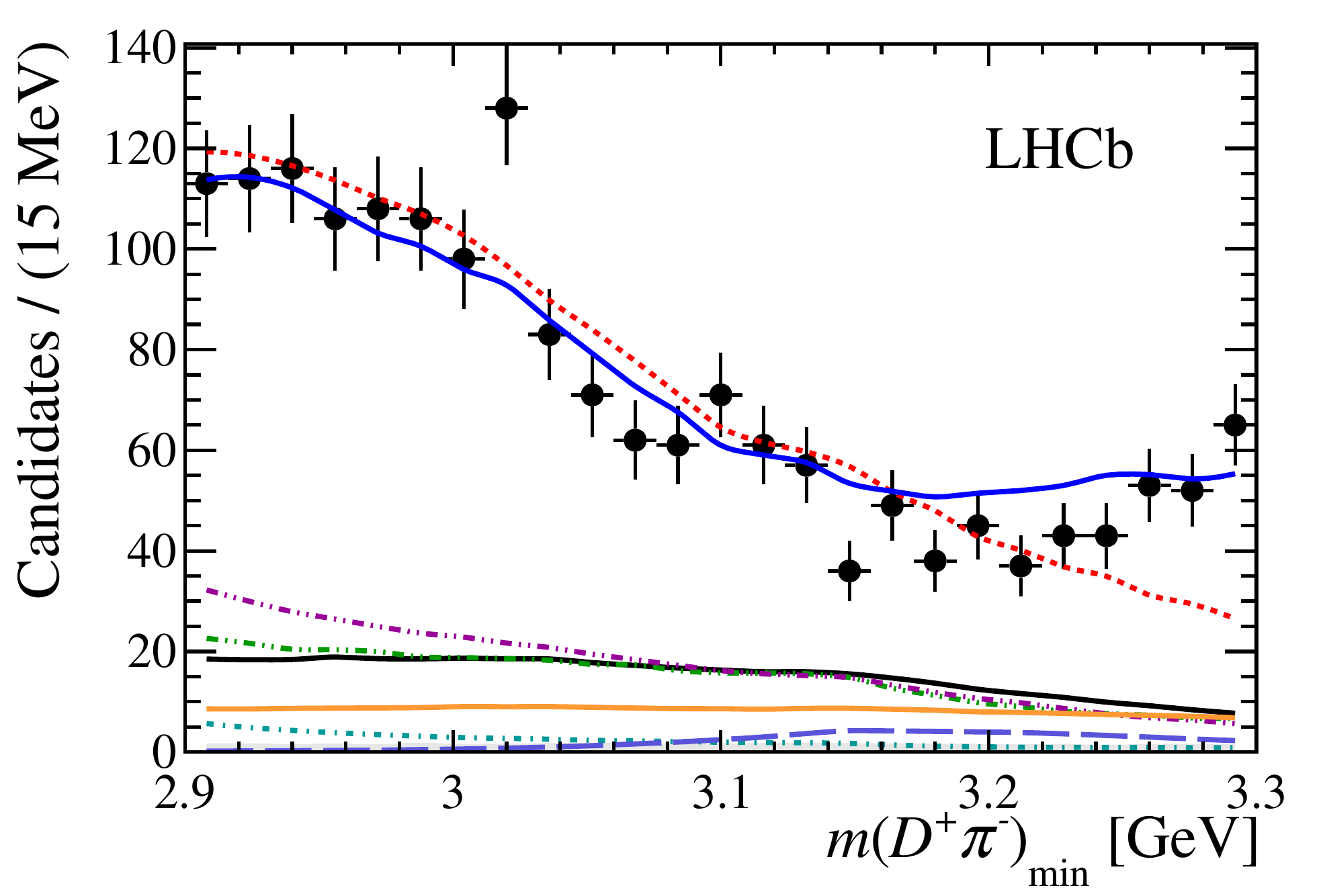}
\includegraphics[scale=0.35]{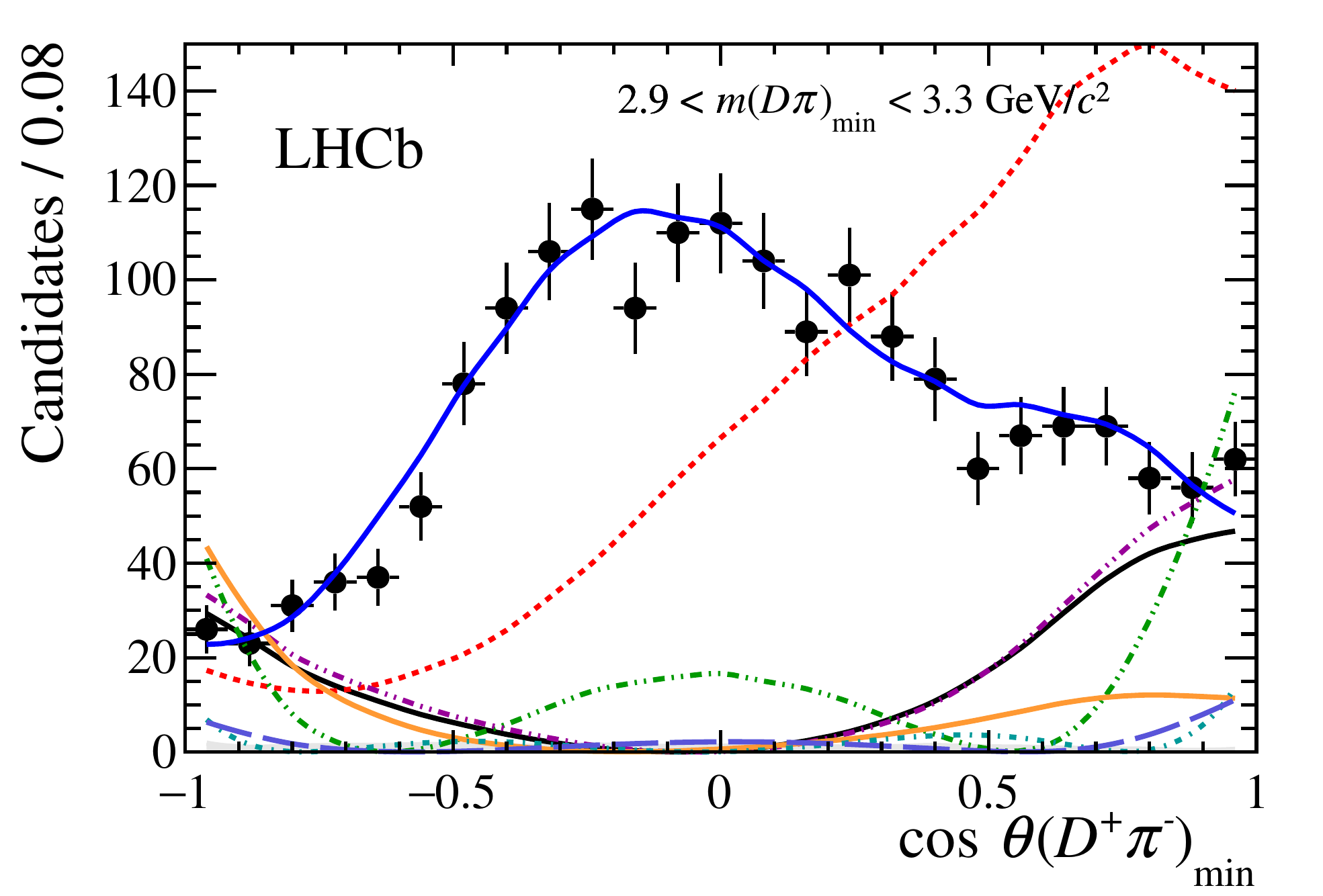}
\caption{\small Projections of the data and amplitude fit onto (left)~$m(\Dpi)$ and (right)~the cosine of the helicity angle for the $\Dpi$ system in (top to bottom)~the low mass threshold region, the \olddstartwo region, the \newdstarone--\newdstarthree\ region and the \newdstartwo region. Components are as shown in Fig.~\ref{fig:cfit}.}
\label{fig:pipi_fitprojzoom}
\end{figure}

One-dimensional projections of the baseline fit model and data onto $m(\Dpi)_{\rm min}$, $m(\Dpi)_{\rm max}$ and $m(\pim\pim)$ are shown in Fig.~\ref{fig:cfit}.
The model is seen to give a good description of the data sample, with the most evident discrepancy at low values of $m(\Dpi)_{\rm max}$, a region of the DP (that corresponds to high values of $m(\pim\pim)$ and $m(\Dpi)_{\rm min} \approx 3.2 \gev$) in which many different amplitudes contribute.
In Fig.~\ref{fig:pipi_fitprojzoom}, zoomed views of the $m(\Dpi)_{\rm min}$ invariant mass projection are provided for regions at threshold and around the \olddstartwo, \newdstarone--\newdstarthree\ and \newdstartwo resonances.
Projections of the cosine of the $\Dpi$ helicity angle in the same regions of $m(\Dpi)_{\rm min}$ are also shown in Fig.~\ref{fig:pipi_fitprojzoom}.
Good agreement is seen in all these projections, suggesting that the model gives an acceptable description of the data and the spin assignments of the \newdstarone, \newdstarthree\ and \newdstartwo\ states are correct.

\afterpage{\clearpage}

\section{Systematic uncertainties}
\label{sec:systematics}

Sources of systematic uncertainty are divided into two categories: experimental and model uncertainties. 
The sources of experimental systematic uncertainty are the signal and background yields in the signal region, the SDP distributions of the background components, the efficiency variation across the SDP, and possible fit bias. 
Model uncertainties arise due to the fixed parameters in the amplitude model, the addition of amplitudes not included in the baseline fit, the modelling of the amplitudes from virtual resonances, and the effect of removing the least well modelled part of the phase space.
The systematic uncertainties from each source are combined in quadrature.

The signal and background yields in the signal region are determined from the fit to the $B$ candidate invariant mass distribution, as described in Sec.~\ref{sec:mass-fit}. 
The total uncertainty on each yield, including systematic effects due to the modelling of the components in the \B candidate mass fit, is calculated, and the yields varied accordingly in the DP fit. 
The deviations from the baseline DP fit result are assigned as systematic uncertainties.

The effect of imperfect knowledge of the background distributions over the SDP is tested by varying the bin contents of the histograms used to model the shapes within their statistical uncertainties. 
For $\Btodanddstarkpi$ decays the ratio of the $\Dstarp$ and $\Dp$ contributions is varied. 
Where applicable, the reweighting of the SDP distribution of the simulated samples is removed.
Changes in the results compared to the baseline DP fit result are again assigned as systematic uncertainties. 

The uncertainty related to the knowledge of the variation of efficiency across the SDP is determined by varying the efficiency histograms before the spline fit is performed. 
The central bin in each $3 \times 3$ cluster is varied by its statistical uncertainty and the surrounding bins in the cluster are varied by interpolation. 
This procedure accounts for possible correlations between the bins, since a systematic effect on a given bin is likely also to affect neighbouring bins.
An ensemble of DP fits are performed, each with a unique efficiency histogram, and the effects on the results are assigned as systematic uncertainties. 
An additional systematic uncertainty is assigned by varying the binning scheme of the control sample used to determine the PID efficiencies.

Systematic uncertainties related to possible intrinsic fit bias are investigated using an ensemble of pseudoexperiments. 
Differences between the input and fitted values from the ensemble for the fit parameters are found to be small. 
Systematic uncertainties are assigned as the sum in quadrature of the difference between the input and output values and the uncertainty on the mean of the output value determined from a fit to the ensemble.

The only fixed parameter in the lineshapes of resonant amplitudes is the \mbox{Blatt--Weisskopf} barrier radius, $r_{\rm BW}$.
To account for potential systematic effects, this is varied between 3 and 5\,$\gev^{-1}$~\cite{LHCb-PAPER-2014-036}, and the difference compared to the baseline fit model is assigned as an uncertainty.
The choice of knot positions in the quasi-model-independent description of the \Dpiswave is another source of possible systematic uncertainty.
This is evaluated from the change in the fit results when more knots are added at low $m(\Dpi)$.
As discussed in Sec.~\ref{sec:dalitz-generalities}, it is not possible to add more knots at high $m(\Dpi)$ without destabilising the fit.

As discussed in Sec.~\ref{sec:introduction}, it is possible that there is more than one spin 1 resonance in the range $2.6 < m(\Dpi) < 2.8 \gev$.
The measured parameters of the \newdstarone\ resonance are most consistent with those given for the $D^*(2650)$ state in Table~\ref{tab:PDG}, therefore the effect of including an additional $D^*(2760)$ contribution is considered as a source of systematic uncertainty. 
Separate fits are performed with the parameters of the $D^*(2760)$ state fixed to the values determined by \babar~\cite{delAmoSanchez:2010vq} and LHCb~\cite{LHCb-PAPER-2013-026} and the larger of the deviations from the baseline results is taken as the associated uncertainty.  
Additional fits are performed with the value of the \dstarv\ width given in Table~\ref{tab:resonances}, which corresponds to the current experimental upper limit~\cite{PDG2014}, replaced by the measured central value for the $\Dstar(2010)^+$ ($83.4 \kev$); the associated systematic uncertainty is negligible. 
The dependence of the results on the effective pole mass description of Eq.~(\ref{eqn:effmass}) that is used for the virtual resonance contributions is found by using a fixed width in Eq.~(\ref{eq:RelBWEqn}), removing the dependence on $m^{\rm eff}_0$.

A discrepancy between the model and the data is seen in the low $m(\Dpi)_{\rm max}$ region, as discussed in Sec.~\ref{sec:DPresults}.
Since this may not be accounted for by the other sources of systematic uncertainty, the effect on the results is determined by performing fits where this region of the DP is vetoed by removing separately candidates with either $m(\Dpi)_{\rm max} < 3.3\gev$ or $m(\pim\pim) > 3.05 \gev$.  
Systematic uncertainties are assigned as the difference in the fitted parameters compared to the baseline fit.

Contributions to the experimental and model systematic uncertainties for the fit fractions, masses and widths are broken down in Tables~\ref{tab:exptsystbreak} and~\ref{tab:modsystbreak}.
The largest source of experimental systematic uncertainty for many parameters is the knowledge of the efficiency variation across the Dalitz plot.
The various parameters are affected differently by the sources of model uncertainty, with some being affected by the variation of fixed parameters in the model, others (notably the parameters associated with the \newdstarone\ amplitude) by the introduction of an additional $D_1^*(2760)^0$ resonance, and some changing when the poorly-modelled region of phase space is vetoed.
The effect of the finite mass resolution, described in Sec.~\ref{sec:DPmodel}, on the measurements of the masses and widths of resonances is found to be negligible.

\begin{table}[!tb]
\centering
\caption{\small Breakdown of experimental systematic uncertainties on the fit fractions (\%) and masses and widths $(\mevnsp)$.}
\label{tab:exptsystbreak}
\begin{tabular}{lr@{$\,\pm\,$}lccccc} 
\hline 
 & \multicolumn{2}{c}{Nominal} & S/B frac. & Eff. & Bkgd. & Fit bias & Total \\ 
\hline 
\olddstartwo & $35.7$ & $0.6$ & 0.1 & 1.3 & 0.0 & 0.2 & 1.4 \\ 
\newdstarone & $8.3$ & $0.6$ & 0.0 & 0.7 & 0.1 & 0.1 & 0.7 \\ 
\newdstarthree & $1.0$ & $0.1$ & 0.0 & 0.1 & 0.0 & 0.0 & 0.1 \\ 
\newdstartwo & $0.2$ & $0.1$ & 0.0 & 0.1 & 0.0 & 0.0 & 0.1 \\ 
\hline 
\dstarv & $10.8$ & $0.7$ & 0.0 & 0.7 & 0.1 & 0.1 & 0.7 \\ 
$B^{*}_{v}$ & $2.7$ & $1.0$ & 0.0 & 1.4 & 0.1 & 0.2 & 1.4 \\ 
\hline 
Total S-wave & $57.0$ & $0.8$ & 0.1 & 0.6 & 0.1 & 0.1 & 0.6 \\ 
\hline 
$m\left( \olddstartwo\right)$ & $2463.7$ & $0.4$ &0.0 & 0.3 & 0.1 & 0.1 & 0.3 \\ 
$\Gamma\left( \olddstartwo\right)$ & $47.0$ & $0.8$ &0.1 & 0.9 & 0.1 & 0.0 & 0.9 \\ 
\hline 
$m\left( \newdstarone\right)$ & $2681.1$ & $5.6$ &0.1 & 4.8 & 0.9 & 0.2 & 4.9 \\ 
$\Gamma\left( \newdstarone\right)$ & $186.7$ & $8.5$ &0.5 & 8.4 & 1.0 & 1.2 & 8.6 \\ 
\hline 
$m\left( \newdstarthree\right)$ & $2775.5$ & $4.5$ &0.4 & 4.4 & 0.6 & 0.4 & 4.5 \\ 
$\Gamma\left( \newdstarthree\right)$ & $95.3$ & $9.6$ &0.9 & 5.9 & 1.5 & 4.9 & 7.9 \\ 
\hline 
$m\left( \newdstartwo\right)$ & $3214$ & $29$ & 3 & 29 & 13 & 9 & 33 \\ 
$\Gamma\left( \newdstartwo\right)$ & $186$ & $38$ & 2 & 31 & 8 & 12 & 34 \\ 
\hline 
\end{tabular} 
\end{table}

\begin{table}[!tb]
\centering
\caption{\small Breakdown of model uncertainties on the fit fractions (\%) and masses and widths $(\mevnsp)$.}
\label{tab:modsystbreak}
\begin{tabular}{lr@{$\,\pm\,$}lccccc} 
\hline 
 & \multicolumn{2}{c}{Nominal} & Fixed & Add & Alternative & DP veto & Total \\ 
 & \multicolumn{2}{c}{} &  params. & $D_1^*(2760)^0$ & models \\
\hline 
\olddstartwo & $35.7$ & $0.6$ & 0.9 & 0.0 & 0.0 & 0.1 & 0.9 \\ 
\newdstarone & $8.3$ & $0.6$ & 0.2 & 0.9 & 0.0 & 1.5 & 1.8 \\ 
\newdstarthree & $1.0$ & $0.1$ & 0.0 & 0.0 & 0.0 & 0.2 & 0.2 \\ 
\newdstartwo & $0.2$ & $0.1$ & 0.0 & 0.0 & 0.0 & 0.1 & 0.1 \\ 
\hline 
\dstarv & $10.8$ & $0.7$ & 2.3 & 0.1 & 0.0 & 0.2 & 2.3 \\ 
$B^{*}_{v}$ & $2.7$ & $1.0$ & 1.2 & 0.2 & 0.0 & 1.0 & 1.6 \\ 
\hline 
Total S-wave & $57.0$ & $0.8$ & 0.8 & 0.4 & 0.0 & 0.1 & 0.9 \\ 
\hline 
$m\left( \olddstartwo\right)$ & $2463.7$ & $0.4$ &0.4 & 0.1 & 0.0 & 0.4 & 0.6 \\ 
$\Gamma\left( \olddstartwo\right)$ & $47.0$ & $0.8$ &0.2 & 0.0 & 0.0 & 0.1 & 0.3 \\ 
\hline 
$m\left( \newdstarone\right)$ & $2681.1$ & $5.6$ &4.7 & 11.8 & 0.1 & 3.0 & 13.1 \\ 
$\Gamma\left( \newdstarone\right)$ & $186.7$ & $8.5$ &3.2 & 4.5 & 0.3 & 6.0 & 8.2 \\ 
\hline 
$m\left( \newdstarthree\right)$ & $2775.5$ & $4.5$ &3.4 & 0.4 & 0.0 & 3.3 & 4.7 \\ 
$\Gamma\left( \newdstarthree\right)$ & $95.3$ & $9.6$ &2.8 & 3.2 & 0.0 & 32.9 & 33.1 \\ 
\hline 
$m\left( \newdstartwo\right)$ & $3214$ & $29$ & 25 & 1 & 1 & 26 & 36 \\ 
$\Gamma\left( \newdstartwo\right)$ & $186$ & $38$ & 7 & 19 & 0 & 60 & 63 \\ 
\hline 
\end{tabular} 
\end{table}

Several cross-checks are performed to confirm the stability of the results. 
The data sample is divided into two parts depending on the charge of the \B candidate, the polarity of the magnet and the year of data taking. 
All fits give consistent results.

\section{Results and summary}
\label{sec:results}

Results for the complex coefficients multiplying each amplitude are reported in Table~\ref{tab:cf-results}, and those that describe the \Dpiswave\ amplitude are shown in Table~\ref{tab:splines-results}. 
These complex numbers are reported in terms of real and imaginary parts and also in terms of magnitude and phase as, due to correlations, the propagation of uncertainties from one form to the other may not be trivial.
Results for the interference fit fractions are given in Appendix~\ref{app:iffstat}.

The fit fractions, summarised in Table~\ref{tab:cfitfrac-results}, for resonant contributions are converted into quasi-two-body product branching fractions by multiplying by the $\Btodpipi$ branching fraction.
This value is taken from the world average after a correction for the relative branching fractions of $\Bp\Bm$ and $\Bz\Bzb$ pairs at the $\FourS$ resonance, $\Gamma(\FourS\to\Bp\Bm)/\Gamma(\FourS\to\Bz\Bzb) = 1.055 \pm 0.025$~\cite{PDG2014}, giving ${\cal B}\left(\Btodpipi\right) = (1.014 \pm 0.054) \times 10^{-3}$.
The product branching fractions are shown in Table~\ref{tab:BFresults}; they cannot be converted into absolute branching fractions because the branching fractions for the resonance decays to $\Dpi$ are unknown.

\begin{table}[!tb]
\centering
\caption{\small Results for the complex amplitudes. 
The three quoted errors are statistical, experimental systematic and model uncertainties.}
\label{tab:cf-results}
\begin{tabular}{lcc} 
\hline 
Resonance & \multicolumn{2}{c}{Isobar model coefficients} \\ 
& Real part & Imaginary part \\ 
\hline 
\olddstartwo   & $1.00$ & $0.00$ \\ 
\newdstarone   & $-0.38 \pm 0.02 \pm 0.05 \pm 0.08$ & $\phantom{-}0.30 \pm 0.02 \pm 0.08 \pm 0.03$ \\ 
\newdstarthree & $\phantom{-}0.17 \pm 0.01 \pm 0.01 \pm 0.02$ & $\phantom{-}0.00 \pm 0.01 \pm 0.05 \pm 0.02$ \\ 
\newdstartwo   & $\phantom{-}0.05 \pm 0.02 \pm 0.02 \pm 0.04$ & $-0.06 \pm 0.02 \pm 0.05 \pm 0.03$ \\ 
\hline 
\dstarv & $\phantom{-}0.51 \pm 0.03 \pm 0.02 \pm 0.05$ & $-0.20 \pm 0.05 \pm 0.11 \pm 0.05$ \\ 
$B^*_v$ & $\phantom{-}0.27 \pm 0.03 \pm 0.11 \pm 0.10$ & $\phantom{-}0.04 \pm 0.04 \pm 0.12 \pm 0.05$ \\ 
\hline 
Total S-wave & $\phantom{-}1.21 \pm 0.02 \pm 0.01 \pm 0.02$ & $-0.35 \pm 0.04 \pm 0.07 \pm 0.03$ \\ 
\hline \\ [-1.5ex]
\hline 
& Magnitude & Phase \\ 
\hline                 
\olddstartwo   & $1.00$ & $0.00$ \\
\newdstarone   & $0.48 \pm 0.02 \pm 0.01 \pm 0.06$ & $\phantom{-}2.47 \pm 0.09 \pm 0.18 \pm 0.12$ \\
\newdstarthree & $0.17 \pm 0.01 \pm 0.01 \pm 0.02$ & $\phantom{-}0.01 \pm 0.20 \pm 0.11 \pm 0.09$ \\
\newdstartwo   & $0.08 \pm 0.01 \pm 0.01 \pm 0.01$ & $-0.84 \pm 0.28 \pm 0.52 \pm 0.63$ \\
\hline                 
\dstarv & $0.55 \pm 0.02 \pm 0.01 \pm 0.06$ & $-0.38 \pm 0.19 \pm 0.15 \pm 0.08$ \\
$B^*_v$ & $0.27 \pm 0.05 \pm 0.13 \pm 0.09$ & $\phantom{-}0.14 \pm 0.38 \pm 0.19 \pm 0.25$ \\
\hline                 
Total S-wave & $1.26 \pm 0.01 \pm 0.02 \pm 0.02$ & $-0.28 \pm 0.05 \pm 0.05 \pm 0.03$ \\
\hline
\end{tabular} 
\end{table}

\begin{table}[!tb]
\centering
\caption{\small
Results for the \Dpiswave\ amplitude at the spline knots. 
The three quoted errors are statistical, experimental systematic and model uncertainties.}
\label{tab:splines-results}
\begin{tabular}{lcc} 
\hline 
Knot mass & \multicolumn{2}{c}{$\Dpi$ S-wave amplitude} \\
$(\gevnsp)$ & Real part & Imaginary part \\
\hline 
2.01 & $-0.11 \pm 0.05 \pm 0.07 \pm 0.09$ & $-0.04 \pm 0.03 \pm 0.05 \pm 0.11$ \\ 
2.10 & $\phantom{-}0.00 \pm 0.05 \pm 0.11 \pm 0.05$ & $-0.58 \pm 0.02 \pm 0.03 \pm 0.03$ \\ 
2.20 & $\phantom{-}0.39 \pm 0.05 \pm 0.08 \pm 0.05$ & $-0.62 \pm 0.04 \pm 0.07 \pm 0.04$ \\ 
2.30 & $\phantom{-}0.62 \pm 0.02 \pm 0.03 \pm 0.01$ & $-0.28 \pm 0.05 \pm 0.10 \pm 0.03$ \\ 
2.40 & 0.50 & 0.00 \\ 
2.50 & $\phantom{-}0.23 \pm 0.01 \pm 0.01 \pm 0.01$ & $-0.00 \pm 0.02 \pm 0.04 \pm 0.01$ \\ 
2.60 & $\phantom{-}0.21 \pm 0.01 \pm 0.01 \pm 0.01$ & $-0.10 \pm 0.02 \pm 0.03 \pm 0.06$ \\ 
2.70 & $\phantom{-}0.14 \pm 0.01 \pm 0.01 \pm 0.01$ & $-0.05 \pm 0.01 \pm 0.02 \pm 0.02$ \\ 
2.80 & $\phantom{-}0.14 \pm 0.01 \pm 0.01 \pm 0.01$ & $-0.10 \pm 0.01 \pm 0.02 \pm 0.04$ \\ 
2.90 & $\phantom{-}0.13 \pm 0.01 \pm 0.02 \pm 0.01$ & $-0.16 \pm 0.01 \pm 0.02 \pm 0.02$ \\ 
3.10 & $\phantom{-}0.05 \pm 0.01 \pm 0.02 \pm 0.02$ & $-0.12 \pm 0.01 \pm 0.01 \pm 0.01$ \\ 
4.10 & $\phantom{-}0.04 \pm 0.01 \pm 0.01 \pm 0.01$ & $\phantom{-}0.07 \pm 0.01 \pm 0.01 \pm 0.01$ \\ 
5.14 & 0.00 & 0.00 \\ 
\hline \\ [-1.5ex]
\hline
& Magnitude & Phase \\
\hline
2.01 & $\phantom{-}0.12 \pm 0.05 \pm 0.07 \pm 0.06$ & $-2.82 \pm 0.22 \pm 0.28 \pm 1.47$ \\
2.10 & $\phantom{-}0.58 \pm 0.02 \pm 0.03 \pm 0.03$ & $-1.56 \pm 0.09 \pm 0.17 \pm 0.08$ \\
2.20 & $\phantom{-}0.73 \pm 0.01 \pm 0.03 \pm 0.02$ & $-1.00 \pm 0.08 \pm 0.15 \pm 0.08$ \\
2.30 & $\phantom{-}0.68 \pm 0.01 \pm 0.03 \pm 0.01$ & $-0.42 \pm 0.08 \pm 0.14 \pm 0.05$ \\
2.40 & 0.50 & 0.00 \\
2.50 & $\phantom{-}0.23 \pm 0.01 \pm 0.01 \pm 0.01$ & $-0.00 \pm 0.06 \pm 0.07 \pm 0.05$ \\
2.60 & $\phantom{-}0.23 \pm 0.01 \pm 0.01 \pm 0.03$ & $-0.42 \pm 0.09 \pm 0.13 \pm 0.24$ \\
2.70 & $\phantom{-}0.15 \pm 0.01 \pm 0.01 \pm 0.01$ & $-0.31 \pm 0.07 \pm 0.11 \pm 0.15$ \\
2.80 & $\phantom{-}0.17 \pm 0.01 \pm 0.01 \pm 0.01$ & $-0.63 \pm 0.08 \pm 0.10 \pm 0.19$ \\
2.90 & $\phantom{-}0.20 \pm 0.01 \pm 0.01 \pm 0.01$ & $-0.87 \pm 0.09 \pm 0.12 \pm 0.10$ \\
3.10 & $\phantom{-}0.14 \pm 0.00 \pm 0.01 \pm 0.01$ & $-1.16 \pm 0.10 \pm 0.13 \pm 0.13$ \\
4.10 & $\phantom{-}0.08 \pm 0.00 \pm 0.01 \pm 0.01$ & $\phantom{-}1.02 \pm 0.12 \pm 0.20 \pm 0.16$ \\
5.14 & 0.00 & 0.00 \\
\hline
\end{tabular} 
\end{table}

\begin{table}[!tb]
\centering
\caption{\small Results for the fit fractions. 
The three quoted errors are statistical, experimental systematic and model uncertainties.}
\label{tab:cfitfrac-results}
\begin{tabular}{lc} 
\hline 
 Resonance & Fit fraction (\%) \\ 
 \hline 
\olddstartwo & $35.69 \pm 0.62 \pm 1.37 \pm 0.89$ \\ 
\newdstarone & $\phantom{3}8.32 \pm 0.62 \pm 0.69 \pm 1.79$ \\ 
\newdstarthree & $\phantom{3}1.01 \pm 0.13 \pm 0.13 \pm 0.25$ \\ 
\newdstartwo & $\phantom{3}0.23 \pm 0.07 \pm 0.07 \pm 0.08$ \\ 
\hline 
\dstarv & $10.79 \pm 0.68 \pm 0.74 \pm 2.34$ \\ 
$B^*_v$ & $\phantom{3}2.69 \pm 1.01 \pm 1.43 \pm 1.61$ \\ 
\hline 
Total S-wave & $56.96 \pm 0.78 \pm 0.62 \pm 0.87$ \\ 
\hline 
\end{tabular} 
\end{table}

\begin{table}[!tb]
\centering
\caption{\small Results for the product branching fractions ${\cal B}(\Bm \to R\pim) \times {\cal B}(R \to \Dp\pim)$. 
The four quoted errors are statistical, experimental systematic, model and inclusive branching fraction uncertainties.}
\label{tab:BFresults}
\begin{tabular}{lc} 
\hline 
 Resonance & Branching fraction ($10^{-4}$) \\ 
 \hline 
\olddstartwo & $3.62 \pm 0.06 \pm 0.14 \pm 0.09 \pm 0.25$ \\ 
\newdstarone & $0.84 \pm 0.06 \pm 0.07 \pm 0.18 \pm 0.06$ \\ 
\newdstarthree & $0.10 \pm 0.01 \pm 0.01 \pm 0.02 \pm 0.01$ \\ 
\newdstartwo & $0.02 \pm 0.01 \pm 0.01 \pm 0.01 \pm 0.00$ \\ 
\hline 
\dstarv & $1.09 \pm 0.07 \pm 0.07 \pm 0.24 \pm 0.07$ \\ 
$B^*_v$ & $0.27 \pm 0.10 \pm 0.14 \pm 0.16 \pm 0.02$ \\ 
\hline 
Total S-wave & $5.78 \pm 0.08 \pm 0.06 \pm 0.09 \pm 0.39$ \\ 
\hline 
\end{tabular} 
\end{table}

The masses and widths of the \olddstartwo, \newdstarone, \newdstarthree\ and \newdstartwo\ resonances are determined to be
\begin{eqnarray*}
m(\olddstartwo) & = & 2463.7 \pm \phantom{1}0.4  \pm \phantom{2}0.4  \pm \phantom{2}0.6 \mev\, ,\\ 
\Gamma(\olddstartwo) & = & \phantom{24}47.0 \pm \phantom{1}0.8  \pm \phantom{2}0.9  \pm \phantom{2}0.3 \mev\, ,\\
m(\newdstarone) & = & 2681.1 \pm \phantom{1}5.6  \pm \phantom{2}4.9  \pm 13.1 \mev\, ,\\ 
\Gamma(\newdstarone) & = & \phantom{2}186.7 \pm \phantom{1}8.5  \pm \phantom{2}8.6  \pm \phantom{2}8.2 \mev\, ,\\
m(\newdstarthree) & = & 2775.5 \pm \phantom{1}4.5  \pm \phantom{2}4.5  \pm \phantom{2}4.7 \mev\, ,\\ 
\Gamma(\newdstarthree) & = & \phantom{27}95.3 \pm \phantom{1}9.6  \pm \phantom{2}7.9  \pm 33.1 \mev\, ,\\
m(\newdstartwo) & = & \phantom{1.}3214 \pm \phantom{1.}29  \pm \phantom{1.}33  \pm \phantom{1.}36 \mev \, ,\\
\Gamma(\newdstartwo) & = & \phantom{13.}186 \pm \phantom{1.}38 \pm \phantom{1.}34 \pm \phantom{1.}63 \mev\, ,
\end{eqnarray*}
where the three quoted errors are statistical, experimental systematic and model uncertainties.
The results for the $D^{*}_{2}(2460)^{0}$ are consistent with the PDG averages~\cite{PDG2014} given in Table~\ref{tab:PDG}.
The \newdstarone\ state has parameters close to those measured for the $D^*(2650)$ resonance observed by LHCb in prompt production in $pp$ collisions~\cite{LHCb-PAPER-2013-026}.
As discussed in Sec.~\ref{sec:introduction}, both 2S and 1D states with spin-parity $J^P = 1^-$ are expected in this region.
Similarly, the \newdstarthree\ state has parameters close to those for the $D^*(2760)$ states reported in Refs.~\cite{LHCb-PAPER-2013-026,delAmoSanchez:2010vq} and for the charged $D_3^*(2760)^+$ state~\cite{LHCb-PAPER-2014-070}.
It appears likely to be a member of the 1D family.
The \newdstartwo\ state has parameters that are not consistent with any previously observed resonance, although due to the large uncertainties it cannot be ruled out that it has a common origin with the $D^*(3000)$ state that was reported, without evaluation of systematic uncertainties, in Ref.~\cite{LHCb-PAPER-2013-026}.
It could potentially be a member of the 2P or 1F family.

Removal of any of the \newdstarone, \newdstarthree\ and \newdstartwo\ states from the baseline fit model results in large changes of the likelihood value.
To investigate the effect of the systematic uncertainties, a similar likelihood ratio test is performed in the alternative models that give the largest uncertainties on the parameters of these resonances.  
Accounting for the four degrees of freedom associated with each resonance, the significances of the \newdstarone\ and \newdstarthree\ states including systematic uncertainties are found to be above $10\,\sigma$, while that for the \newdstartwo\ state is $6.6\,\sigma$.
Assigning alternative spin hypotheses to these states results in similarly large changes in likelihood.  

In summary, an analysis of the amplitudes contributing to $\Btodpipi$ decays has been performed using a data sample corresponding to $3.0\invfb$ of $pp$ collision data recorded by the LHCb experiment.
The Dalitz plot fit model containing resonant contributions from the \olddstartwo, \newdstarone, \newdstarthree\ and \newdstartwo\ states, virtual \dstarv\ and $B^{*0}_{v}$ resonances and a quasi-model-independent description of the full \Dpiswave\ was found to give a good description of the data.
These results constitute the first observations of the \newdstarthree\ and \newdstartwo\ resonances and may be useful to develop improved models of the dynamics in the \Dpi\ system.

\section*{Acknowledgements}

\noindent We express our gratitude to our colleagues in the CERN
accelerator departments for the excellent performance of the LHC. We
thank the technical and administrative staff at the LHCb
institutes. We acknowledge support from CERN and from the national
agencies: CAPES, CNPq, FAPERJ and FINEP (Brazil); NSFC (China);
CNRS/IN2P3 (France); BMBF, DFG and MPG (Germany); INFN (Italy); 
FOM and NWO (The Netherlands); MNiSW and NCN (Poland); MEN/IFA (Romania); 
MinES and FASO (Russia); MinECo (Spain); SNSF and SER (Switzerland); 
NASU (Ukraine); STFC (United Kingdom); NSF (USA).
We acknowledge the computing resources that are provided by CERN, IN2P3 (France), KIT and DESY (Germany), INFN (Italy), SURF (The Netherlands), PIC (Spain), GridPP (United Kingdom), RRCKI and Yandex LLC (Russia), CSCS (Switzerland), IFIN-HH (Romania), CBPF (Brazil), PL-GRID (Poland) and OSC (USA). We are indebted to the communities behind the multiple open 
source software packages on which we depend.
Individual groups or members have received support from AvH Foundation (Germany),
EPLANET, Marie Sk\l{}odowska-Curie Actions and ERC (European Union), 
Conseil G\'{e}n\'{e}ral de Haute-Savoie, Labex ENIGMASS and OCEVU, 
R\'{e}gion Auvergne (France), RFBR and Yandex LLC (Russia), GVA, XuntaGal and GENCAT (Spain), Herchel Smith Fund, The Royal Society, Royal Commission for the Exhibition of 1851 and the Leverhulme Trust (United Kingdom).

\clearpage
\appendix
\section{Results for interference fit fractions}
\label{app:iffstat}

The central values and statistical errors for the interference fit fractions are shown in Table~\ref{tab:iffstat}. 
The experimental systematic and model uncertainties are given in Tables~\ref{tab:iffexp}.  

\begin{table}[!htb]
\caption{\small Interference fit fractions (\%) and statistical uncertainties. The amplitudes are: ($A_0$) \dstarv, ($A_1$) \Dpiswave, ($A_2$) \olddstartwo, ($A_3$) \newdstarone, ($A_4$) $B^{*0}_v$, ($A_5$) \newdstarthree, ($A_6$) \newdstartwo. The diagonal elements are the same as the conventional fit fractions.
}
\centering
\vspace{1ex}
\resizebox{\textwidth}{!}{
\begin{tabular}{lccccccc}
\hline
& $A_0$ & $A_1$ & $A_2$ & $A_3$ & $A_4$ & $A_5$ & $A_6$ \\
\hline
$A_0$ & $\phantom{-}10.8\pm0.7$ & $\phantom{-}3.1\pm1.0$ & $-0.8\pm0.0$ & $\phantom{-}0.7\pm1.9$ & $-6.2\pm1.3$ & $\phantom{-}0.1\pm0.0$ & $-0.2\pm0.0$ \\
$A_1$ &  & $\phantom{-}57.0\pm0.8$ & $-2.4\pm0.2$ & $-5.5\pm0.4$ & $-1.9\pm1.4$ & $-0.0\pm0.0$ & $-0.3\pm0.1$ \\
$A_2$ &  &  & $\phantom{-}35.7\pm0.6$ & $-0.3\pm0.1$ & $-0.7\pm0.4$ & $-0.2\pm0.0$ & $-0.5\pm0.2$ \\
$A_3$ &  &  &  & $\phantom{-}8.3\pm0.6$ & $-0.9\pm1.8$ & $\phantom{-}0.1\pm0.0$ & $\phantom{-}0.1\pm0.0$ \\
$A_4$ &  &  &  &  & $\phantom{-}2.7\pm1.0$ & $-0.0\pm0.0$ & $\phantom{-}0.1\pm0.0$ \\
$A_5$ &  &  &  &  &  & $\phantom{-}1.0\pm0.1$ & $\phantom{-}0.0\pm0.0$ \\
$A_6$ &  &  &  &  &  &  & $\phantom{-}0.2\pm0.1$ \\
\hline
\end{tabular}
}
\label{tab:iffstat}
\end{table}

\begin{table}[!htb]
\centering
\caption{\small (Top) Experimental and (bottom) model systematic uncertainties on the interference fit fractions (\%). The amplitudes are: ($A_0$) \dstarv, ($A_1$) \Dpiswave, ($A_2$) \olddstartwo, ($A_3$) \newdstarone, ($A_4$) $B^{*0}_{v}$, ($A_5$) \newdstarthree, ($A_6$) \newdstartwo. The diagonal elements are the same as the conventional fit fractions.}
\label{tab:iffexp}
\begin{tabular}{lccccccc} 
\hline 
& $A_0$ & $A_1$ & $A_2$ & $A_3$ & $A_4$ & $A_5$ & $A_6$ \\ 
\hline 
$A_0$ & 0.74 & 0.42 & 0.04 & 1.46 & 1.42 & 0.01 & 0.06 \\ 
$A_1$ & & 0.62 & 0.21 & 0.34 & 0.58 & 0.03 & 0.13 \\ 
$A_2$ & & & 1.37 & 0.13 & 0.14 & 0.01 & 0.24 \\ 
$A_3$ & & & & 0.69 & 2.11 & 0.00 & 0.06 \\ 
$A_4$ & & & & & 1.43 & 0.15 & 0.05 \\ 
$A_5$ & & & & & & 0.13 & 0.01 \\ 
$A_6$ & & & & & & & 0.07 \\ 
\hline 
\end{tabular} 
\vspace{2ex}
\begin{tabular}{lccccccc} 
\hline 
& $A_0$ & $A_1$ & $A_2$ & $A_3$ & $A_4$ & $A_5$ & $A_6$ \\ 
\hline 
$A_0$ & 2.34 & 0.91 & 0.21 & 1.01 & 3.11 & 0.04 & 0.12 \\ 
$A_1$ & & 0.87 & 0.21 & 0.48 & 1.74 & 0.02 & 0.16 \\ 
$A_2$ & & & 0.89 & 0.07 & 0.53 & 0.08 & 0.34 \\ 
$A_3$ & & & & 1.79 & 0.87 & 0.02 & 0.04 \\ 
$A_4$ & & & & & 1.61 & 0.04 & 0.05 \\ 
$A_5$ & & & & & & 0.25 & 0.03 \\ 
$A_6$ & & & & & & & 0.08 \\ 
\hline 
\end{tabular} 
\end{table}

\clearpage
\addcontentsline{toc}{section}{References}

\begin{mcitethebibliography}{10}
\mciteSetBstSublistMode{n}
\mciteSetBstMaxWidthForm{subitem}{\alph{mcitesubitemcount})}
\mciteSetBstSublistLabelBeginEnd{\mcitemaxwidthsubitemform\space}
{\relax}{\relax}

\bibitem{Godfrey:1985xj}
S.~Godfrey and N.~Isgur, \ifthenelse{\boolean{articletitles}}{\emph{{Mesons in
  a relativized quark model with chromodynamics}},
  }{}\href{http://dx.doi.org/10.1103/PhysRevD.32.189}{Phys.\ Rev.\
  \textbf{D32} (1985) 189}\relax
\mciteBstWouldAddEndPuncttrue
\mciteSetBstMidEndSepPunct{\mcitedefaultmidpunct}
{\mcitedefaultendpunct}{\mcitedefaultseppunct}\relax
\EndOfBibitem
\bibitem{Isgur:1991wq}
N.~Isgur and M.~B. Wise,
  \ifthenelse{\boolean{articletitles}}{\emph{{Spectroscopy with heavy quark
  symmetry}}, }{}\href{http://dx.doi.org/10.1103/PhysRevLett.66.1130}{Phys.\
  Rev.\ Lett.\  \textbf{66} (1991) 1130}\relax
\mciteBstWouldAddEndPuncttrue
\mciteSetBstMidEndSepPunct{\mcitedefaultmidpunct}
{\mcitedefaultendpunct}{\mcitedefaultseppunct}\relax
\EndOfBibitem
\bibitem{Colangelo:2012xi}
P.~Colangelo, F.~De~Fazio, F.~Giannuzzi, and S.~Nicotri,
  \ifthenelse{\boolean{articletitles}}{\emph{{New meson spectroscopy with open
  charm and beauty}},
  }{}\href{http://dx.doi.org/10.1103/PhysRevD.86.054024}{Phys.\ Rev.\
  \textbf{D86} (2012) 054024},
  \href{http://arxiv.org/abs/1207.6940}{{\normalfont\ttfamily
  arXiv:1207.6940}}\relax
\mciteBstWouldAddEndPuncttrue
\mciteSetBstMidEndSepPunct{\mcitedefaultmidpunct}
{\mcitedefaultendpunct}{\mcitedefaultseppunct}\relax
\EndOfBibitem
\bibitem{Mohler:2012na}
D.~Mohler, S.~Prelovsek, and R.~M. Woloshyn,
  \ifthenelse{\boolean{articletitles}}{\emph{{$D \pi$ scattering and $D$ meson
  resonances from lattice QCD}},
  }{}\href{http://dx.doi.org/10.1103/PhysRevD.87.034501}{Phys.\ Rev.\
  \textbf{D87} (2013) 034501},
  \href{http://arxiv.org/abs/1208.4059}{{\normalfont\ttfamily
  arXiv:1208.4059}}\relax
\mciteBstWouldAddEndPuncttrue
\mciteSetBstMidEndSepPunct{\mcitedefaultmidpunct}
{\mcitedefaultendpunct}{\mcitedefaultseppunct}\relax
\EndOfBibitem
\bibitem{Moir:2016srx}
G.~Moir {\em et~al.},
  \ifthenelse{\boolean{articletitles}}{\emph{{Coupled-Channel $D\pi$, $D\eta$
  and $D_{s}\bar{K}$ Scattering from Lattice QCD}},
  }{}\href{http://arxiv.org/abs/1607.07093}{{\normalfont\ttfamily
  arXiv:1607.07093}}, {submitted to JHEP}\relax
\mciteBstWouldAddEndPuncttrue
\mciteSetBstMidEndSepPunct{\mcitedefaultmidpunct}
{\mcitedefaultendpunct}{\mcitedefaultseppunct}\relax
\EndOfBibitem
\bibitem{delAmoSanchez:2010vq}
\babar collaboration, P.~del Amo~Sanchez {\em et~al.},
  \ifthenelse{\boolean{articletitles}}{\emph{{Observation of new resonances
  decaying to $D\pi$ and $D^*\pi$ in inclusive $e^+e^-$ collisions near
  $\sqrt{s}=$10.58 GeV}},
  }{}\href{http://dx.doi.org/10.1103/PhysRevD.82.111101}{Phys.\ Rev.\
  \textbf{D82} (2010) 111101},
  \href{http://arxiv.org/abs/1009.2076}{{\normalfont\ttfamily
  arXiv:1009.2076}}\relax
\mciteBstWouldAddEndPuncttrue
\mciteSetBstMidEndSepPunct{\mcitedefaultmidpunct}
{\mcitedefaultendpunct}{\mcitedefaultseppunct}\relax
\EndOfBibitem
\bibitem{LHCb-PAPER-2013-026}
LHCb collaboration, R.~Aaij {\em et~al.},
  \ifthenelse{\boolean{articletitles}}{\emph{{Study of $D_J$ meson decays to
  $D^+\pi^-$, $D^0\pi^+$ and $D^{*+}\pi^-$ final states in $pp$ collisions}},
  }{}\href{http://dx.doi.org/10.1007/JHEP09(2013)145}{JHEP \textbf{09} (2013)
  145}, \href{http://arxiv.org/abs/1307.4556}{{\normalfont\ttfamily
  arXiv:1307.4556}}\relax
\mciteBstWouldAddEndPuncttrue
\mciteSetBstMidEndSepPunct{\mcitedefaultmidpunct}
{\mcitedefaultendpunct}{\mcitedefaultseppunct}\relax
\EndOfBibitem
\bibitem{Abe:2003zm}
Belle collaboration, K.~Abe {\em et~al.},
  \ifthenelse{\boolean{articletitles}}{\emph{{Study of $\Bm \to D^{**0} \pim \
  (D^{**0} \to D^{(*)+} \pim)$ decays}},
  }{}\href{http://dx.doi.org/10.1103/PhysRevD.69.112002}{Phys.\ Rev.\
  \textbf{D69} (2004) 112002},
  \href{http://arxiv.org/abs/hep-ex/0307021}{{\normalfont\ttfamily
  arXiv:hep-ex/0307021}}\relax
\mciteBstWouldAddEndPuncttrue
\mciteSetBstMidEndSepPunct{\mcitedefaultmidpunct}
{\mcitedefaultendpunct}{\mcitedefaultseppunct}\relax
\EndOfBibitem
\bibitem{Aubert:2009wg}
\babar collaboration, B.~Aubert {\em et~al.},
  \ifthenelse{\boolean{articletitles}}{\emph{{Dalitz plot analysis of $\Bm \to
  \Dp\pim\pim$}}, }{}\href{http://dx.doi.org/10.1103/PhysRevD.79.112004}{Phys.\
  Rev.\  \textbf{D79} (2009) 112004},
  \href{http://arxiv.org/abs/0901.1291}{{\normalfont\ttfamily
  arXiv:0901.1291}}\relax
\mciteBstWouldAddEndPuncttrue
\mciteSetBstMidEndSepPunct{\mcitedefaultmidpunct}
{\mcitedefaultendpunct}{\mcitedefaultseppunct}\relax
\EndOfBibitem
\bibitem{Kuzmin:2006mw}
Belle collaboration, A.~Kuzmin {\em et~al.},
  \ifthenelse{\boolean{articletitles}}{\emph{{Study of $\Bzb \to D^0 \pi^+
  \pi^-$ decays}},
  }{}\href{http://dx.doi.org/10.1103/PhysRevD.76.012006}{Phys.\ Rev.\
  \textbf{D76} (2007) 012006},
  \href{http://arxiv.org/abs/hep-ex/0611054}{{\normalfont\ttfamily
  arXiv:hep-ex/0611054}}\relax
\mciteBstWouldAddEndPuncttrue
\mciteSetBstMidEndSepPunct{\mcitedefaultmidpunct}
{\mcitedefaultendpunct}{\mcitedefaultseppunct}\relax
\EndOfBibitem
\bibitem{LHCb-PAPER-2014-070}
LHCb collaboration, R.~Aaij {\em et~al.},
  \ifthenelse{\boolean{articletitles}}{\emph{{Dalitz plot analysis of $B^0\to
  \overline{D}^0\pi^+\pi^-$ decays}},
  }{}\href{http://dx.doi.org/10.1103/PhysRevD.92.032002}{Phys.\ Rev.\
  \textbf{D92} (2015) 032002},
  \href{http://arxiv.org/abs/1505.01710}{{\normalfont\ttfamily
  arXiv:1505.01710}}\relax
\mciteBstWouldAddEndPuncttrue
\mciteSetBstMidEndSepPunct{\mcitedefaultmidpunct}
{\mcitedefaultendpunct}{\mcitedefaultseppunct}\relax
\EndOfBibitem
\bibitem{LHCb-PAPER-2015-007}
LHCb collaboration, R.~Aaij {\em et~al.},
  \ifthenelse{\boolean{articletitles}}{\emph{{First observation and amplitude
  analysis of the $B^{-}\to D^{+}K^{-}\pi^{-}$ decay}},
  }{}\href{http://dx.doi.org/10.1103/PhysRevD.91.092002}{Phys.\ Rev.\
  \textbf{D91} (2015) 092002}, Erratum
  \href{http://dx.doi.org/10.1103/PhysRevD.93.119901}{ibid.\   \textbf{D93}
  (2016) 119901}, \href{http://arxiv.org/abs/1503.02995}{{\normalfont\ttfamily
  arXiv:1503.02995}}\relax
\mciteBstWouldAddEndPuncttrue
\mciteSetBstMidEndSepPunct{\mcitedefaultmidpunct}
{\mcitedefaultendpunct}{\mcitedefaultseppunct}\relax
\EndOfBibitem
\bibitem{LHCb-PAPER-2015-017}
LHCb collaboration, R.~Aaij {\em et~al.},
  \ifthenelse{\boolean{articletitles}}{\emph{{Amplitude analysis of $B^0\to
  \overline{D}^0K^+\pi^-$ decays}},
  }{}\href{http://dx.doi.org/10.1103/PhysRevD.92.012012}{Phys.\ Rev.\
  \textbf{D92} (2015) 012012},
  \href{http://arxiv.org/abs/1505.01505}{{\normalfont\ttfamily
  arXiv:1505.01505}}\relax
\mciteBstWouldAddEndPuncttrue
\mciteSetBstMidEndSepPunct{\mcitedefaultmidpunct}
{\mcitedefaultendpunct}{\mcitedefaultseppunct}\relax
\EndOfBibitem
\bibitem{Brodzicka:2007aa}
Belle collaboration, J.~Brodzicka {\em et~al.},
  \ifthenelse{\boolean{articletitles}}{\emph{{Observation of a new $D_{sJ}$
  meson in $\Bp \to \Dzb \Dz \Kp$ decays}},
  }{}\href{http://dx.doi.org/10.1103/PhysRevLett.100.092001}{Phys.\ Rev.\
  Lett.\  \textbf{100} (2008) 092001},
  \href{http://arxiv.org/abs/0707.3491}{{\normalfont\ttfamily
  arXiv:0707.3491}}\relax
\mciteBstWouldAddEndPuncttrue
\mciteSetBstMidEndSepPunct{\mcitedefaultmidpunct}
{\mcitedefaultendpunct}{\mcitedefaultseppunct}\relax
\EndOfBibitem
\bibitem{LHCb-PAPER-2014-035}
LHCb collaboration, R.~Aaij {\em et~al.},
  \ifthenelse{\boolean{articletitles}}{\emph{{Observation of overlapping
  spin-$1$ and spin-$3$ $\overline{D}^0 K^-$ resonances at mass
  $2.86$~GeV/$c^2$}},
  }{}\href{http://dx.doi.org/10.1103/PhysRevLett.113.162001}{Phys.\ Rev.\
  Lett.\  \textbf{113} (2014) 162001},
  \href{http://arxiv.org/abs/1407.7574}{{\normalfont\ttfamily
  arXiv:1407.7574}}\relax
\mciteBstWouldAddEndPuncttrue
\mciteSetBstMidEndSepPunct{\mcitedefaultmidpunct}
{\mcitedefaultendpunct}{\mcitedefaultseppunct}\relax
\EndOfBibitem
\bibitem{LHCb-PAPER-2014-036}
LHCb collaboration, R.~Aaij {\em et~al.},
  \ifthenelse{\boolean{articletitles}}{\emph{{Dalitz plot analysis of
  $B^0_s\to\overline{D}^0K^-\pi^+$ decays}},
  }{}\href{http://dx.doi.org/10.1103/PhysRevD.90.072003}{Phys.\ Rev.\
  \textbf{D90} (2014) 072003},
  \href{http://arxiv.org/abs/1407.7712}{{\normalfont\ttfamily
  arXiv:1407.7712}}\relax
\mciteBstWouldAddEndPuncttrue
\mciteSetBstMidEndSepPunct{\mcitedefaultmidpunct}
{\mcitedefaultendpunct}{\mcitedefaultseppunct}\relax
\EndOfBibitem
\bibitem{Lees:2014abp}
\babar collaboration, J.~P. Lees {\em et~al.},
  \ifthenelse{\boolean{articletitles}}{\emph{{Dalitz plot analyses of $B^0 \to
  D^- D^0 K^+$ and $B^+ \to \overline{D}^0 D^0 K^+$ decays}},
  }{}\href{http://dx.doi.org/10.1103/PhysRevD.91.052002}{Phys.\ Rev.\
  \textbf{D91} (2015) 052002},
  \href{http://arxiv.org/abs/1412.6751}{{\normalfont\ttfamily
  arXiv:1412.6751}}\relax
\mciteBstWouldAddEndPuncttrue
\mciteSetBstMidEndSepPunct{\mcitedefaultmidpunct}
{\mcitedefaultendpunct}{\mcitedefaultseppunct}\relax
\EndOfBibitem
\bibitem{LHCb-PAPER-2015-052}
LHCb collaboration, R.~Aaij {\em et~al.},
  \ifthenelse{\boolean{articletitles}}{\emph{{Study of $D_{sJ}^+$ mesons
  decaying to $D^{*+}K^0_S$ and $D^{*0}K^+$ final states}},
  }{}\href{http://dx.doi.org/10.1007/JHEP02(2016)133}{JHEP \textbf{02} (2015)
  133}, \href{http://arxiv.org/abs/1601.01495}{{\normalfont\ttfamily
  arXiv:1601.01495}}\relax
\mciteBstWouldAddEndPuncttrue
\mciteSetBstMidEndSepPunct{\mcitedefaultmidpunct}
{\mcitedefaultendpunct}{\mcitedefaultseppunct}\relax
\EndOfBibitem
\bibitem{PDG2014}
Particle Data Group, K.~A. Olive {\em et~al.},
  \ifthenelse{\boolean{articletitles}}{\emph{{\href{http://pdg.lbl.gov/}{Review
  of particle physics}}},
  }{}\href{http://dx.doi.org/10.1088/1674-1137/38/9/090001}{Chin.\ Phys.\
  \textbf{C38} (2014) 090001}\relax
\mciteBstWouldAddEndPuncttrue
\mciteSetBstMidEndSepPunct{\mcitedefaultmidpunct}
{\mcitedefaultendpunct}{\mcitedefaultseppunct}\relax
\EndOfBibitem
\bibitem{Chen:2015lpa}
B.~Chen, X.~Liu, and A.~Zhang,
  \ifthenelse{\boolean{articletitles}}{\emph{{Combined study of $2S$ and $1D$
  open-charm mesons with natural spin-parity}},
  }{}\href{http://dx.doi.org/10.1103/PhysRevD.92.034005}{Phys.\ Rev.\
  \textbf{D92} (2015) 034005},
  \href{http://arxiv.org/abs/1507.02339}{{\normalfont\ttfamily
  arXiv:1507.02339}}\relax
\mciteBstWouldAddEndPuncttrue
\mciteSetBstMidEndSepPunct{\mcitedefaultmidpunct}
{\mcitedefaultendpunct}{\mcitedefaultseppunct}\relax
\EndOfBibitem
\bibitem{Godfrey:2015dva}
S.~Godfrey and K.~Moats, \ifthenelse{\boolean{articletitles}}{\emph{{Properties
  of excited charm and charm-strange mesons}},
  }{}\href{http://dx.doi.org/10.1103/PhysRevD.93.034035}{Phys.\ Rev.\
  \textbf{D93} (2016) 034035},
  \href{http://arxiv.org/abs/1510.08305}{{\normalfont\ttfamily
  arXiv:1510.08305}}\relax
\mciteBstWouldAddEndPuncttrue
\mciteSetBstMidEndSepPunct{\mcitedefaultmidpunct}
{\mcitedefaultendpunct}{\mcitedefaultseppunct}\relax
\EndOfBibitem
\bibitem{lass}
LASS collaboration, D.~Aston {\em et~al.},
  \ifthenelse{\boolean{articletitles}}{\emph{{A study of $K^- \pi^+$ scattering
  in the reaction $K^- p \to K^- \pi^+ n$ at $11 \,Ge\kern -0.1em V\!/c$}},
  }{}\href{http://dx.doi.org/10.1016/0550-3213(88)90028-4}{Nucl.\ Phys.\
  \textbf{B296} (1988) 493}\relax
\mciteBstWouldAddEndPuncttrue
\mciteSetBstMidEndSepPunct{\mcitedefaultmidpunct}
{\mcitedefaultendpunct}{\mcitedefaultseppunct}\relax
\EndOfBibitem
\bibitem{Aitala:2005yh}
E791 collaboration, {E.\ M.\ ~Aitala \emph{et al.\ } and W.\ M.\ ~Dunwoodie},
  \ifthenelse{\boolean{articletitles}}{\emph{{Model independent measurement of
  S-wave $\Km\pip$ systems using $\Dp \to K\pi\pi$ decays from Fermilab E791}},
  }{}\href{http://dx.doi.org/10.1103/PhysRevD.73.032004}{Phys.\ Rev.\
  \textbf{D73} (2006) 032004}, Erratum
  \href{http://dx.doi.org/10.1103/PhysRevD.74.059901}{ibid.\   \textbf{D74}
  (2006) 059901},
  \href{http://arxiv.org/abs/hep-ex/0507099}{{\normalfont\ttfamily
  arXiv:hep-ex/0507099}}\relax
\mciteBstWouldAddEndPuncttrue
\mciteSetBstMidEndSepPunct{\mcitedefaultmidpunct}
{\mcitedefaultendpunct}{\mcitedefaultseppunct}\relax
\EndOfBibitem
\bibitem{Bonvicini:2008jw}
CLEO collaboration, G.~Bonvicini {\em et~al.},
  \ifthenelse{\boolean{articletitles}}{\emph{{Dalitz plot analysis of the $\Dp
  \to \Km\pip\pip$ decay}},
  }{}\href{http://dx.doi.org/10.1103/PhysRevD.78.052001}{Phys.\ Rev.\
  \textbf{D78} (2008) 052001},
  \href{http://arxiv.org/abs/0802.4214}{{\normalfont\ttfamily
  arXiv:0802.4214}}\relax
\mciteBstWouldAddEndPuncttrue
\mciteSetBstMidEndSepPunct{\mcitedefaultmidpunct}
{\mcitedefaultendpunct}{\mcitedefaultseppunct}\relax
\EndOfBibitem
\bibitem{Link:2009ng}
FOCUS collaboration, J.~M. Link {\em et~al.},
  \ifthenelse{\boolean{articletitles}}{\emph{{The $K^-\pi^+$ S-wave from the
  $D^+ \to K^-\pi^+\pi^+$ decay}},
  }{}\href{http://dx.doi.org/10.1016/j.physletb.2009.09.057}{Phys.\ Lett.\
  \textbf{B681} (2009) 14},
  \href{http://arxiv.org/abs/0905.4846}{{\normalfont\ttfamily
  arXiv:0905.4846}}\relax
\mciteBstWouldAddEndPuncttrue
\mciteSetBstMidEndSepPunct{\mcitedefaultmidpunct}
{\mcitedefaultendpunct}{\mcitedefaultseppunct}\relax
\EndOfBibitem
\bibitem{delAmoSanchez:2010fd}
\babar collaboration, P.~del Amo~Sanchez {\em et~al.},
  \ifthenelse{\boolean{articletitles}}{\emph{{Analysis of the $D^+ \to K^-
  \pi^+ e^+ \nu_e$ decay channel}},
  }{}\href{http://dx.doi.org/10.1103/PhysRevD.83.072001}{Phys.\ Rev.\
  \textbf{D83} (2011) 072001},
  \href{http://arxiv.org/abs/1012.1810}{{\normalfont\ttfamily
  arXiv:1012.1810}}\relax
\mciteBstWouldAddEndPuncttrue
\mciteSetBstMidEndSepPunct{\mcitedefaultmidpunct}
{\mcitedefaultendpunct}{\mcitedefaultseppunct}\relax
\EndOfBibitem
\bibitem{Lees:2015zzr}
\babar collaboration, J.~P. Lees {\em et~al.},
  \ifthenelse{\boolean{articletitles}}{\emph{{Measurement of the I=1/2 $K \pi$
  S-wave amplitude from Dalitz plot analyses of $\eta_c \to K \bar{K} \pi$ in
  two-photon interactions}},
  }{}\href{http://dx.doi.org/10.1103/PhysRevD.93.012005}{Phys.\ Rev.\
  \textbf{D93} (2016) 012005},
  \href{http://arxiv.org/abs/1511.02310}{{\normalfont\ttfamily
  arXiv:1511.02310}}\relax
\mciteBstWouldAddEndPuncttrue
\mciteSetBstMidEndSepPunct{\mcitedefaultmidpunct}
{\mcitedefaultendpunct}{\mcitedefaultseppunct}\relax
\EndOfBibitem
\bibitem{Aubert:2007dc}
\babar collaboration, B.~Aubert {\em et~al.},
  \ifthenelse{\boolean{articletitles}}{\emph{{Amplitude analysis of the decay
  $\Dz \to \Km \Kp \piz$}},
  }{}\href{http://dx.doi.org/10.1103/PhysRevD.76.011102}{Phys.\ Rev.\
  \textbf{D76} (2007) 011102},
  \href{http://arxiv.org/abs/0704.3593}{{\normalfont\ttfamily
  arXiv:0704.3593}}\relax
\mciteBstWouldAddEndPuncttrue
\mciteSetBstMidEndSepPunct{\mcitedefaultmidpunct}
{\mcitedefaultendpunct}{\mcitedefaultseppunct}\relax
\EndOfBibitem
\bibitem{Aubert:2008ao}
\babar collaboration, B.~Aubert {\em et~al.},
  \ifthenelse{\boolean{articletitles}}{\emph{{Dalitz plot analysis of $\Dsp \to
  \pi^{+} \pi^{-} \pi^{+}$}},
  }{}\href{http://dx.doi.org/10.1103/PhysRevD.79.032003}{Phys.\ Rev.\
  \textbf{D79} (2009) 032003},
  \href{http://arxiv.org/abs/0808.0971}{{\normalfont\ttfamily
  arXiv:0808.0971}}\relax
\mciteBstWouldAddEndPuncttrue
\mciteSetBstMidEndSepPunct{\mcitedefaultmidpunct}
{\mcitedefaultendpunct}{\mcitedefaultseppunct}\relax
\EndOfBibitem
\bibitem{LHCb-PAPER-2015-038}
LHCb collaboration, R.~Aaij {\em et~al.},
  \ifthenelse{\boolean{articletitles}}{\emph{{A model-independent confirmation
  of the $Z(4430)^-$ state}},
  }{}\href{http://dx.doi.org/10.1103/PhysRevD.92.112009}{Phys.\ Rev.\
  \textbf{D92} (2015) 112009},
  \href{http://arxiv.org/abs/1510.01951}{{\normalfont\ttfamily
  arXiv:1510.01951}}\relax
\mciteBstWouldAddEndPuncttrue
\mciteSetBstMidEndSepPunct{\mcitedefaultmidpunct}
{\mcitedefaultendpunct}{\mcitedefaultseppunct}\relax
\EndOfBibitem
\bibitem{LHCb-PAPER-2015-029}
LHCb collaboration, R.~Aaij {\em et~al.},
  \ifthenelse{\boolean{articletitles}}{\emph{{Observation of $J/\psi p$
  resonances consistent with pentaquark states in $\Lambda_b^0\to J/\psi pK^-$
  decays}}, }{}\href{http://dx.doi.org/10.1103/PhysRevLett.115.072001}{Phys.\
  Rev.\ Lett.\  \textbf{115} (2015) 072001},
  \href{http://arxiv.org/abs/1507.03414}{{\normalfont\ttfamily
  arXiv:1507.03414}}\relax
\mciteBstWouldAddEndPuncttrue
\mciteSetBstMidEndSepPunct{\mcitedefaultmidpunct}
{\mcitedefaultendpunct}{\mcitedefaultseppunct}\relax
\EndOfBibitem
\bibitem{Kolomeitsev:2003ac}
E.~E. Kolomeitsev and M.~F.~M. Lutz,
  \ifthenelse{\boolean{articletitles}}{\emph{{On heavy light meson resonances
  and chiral symmetry}},
  }{}\href{http://dx.doi.org/10.1016/j.physletb.2003.10.118}{Phys.\ Lett.\
  \textbf{B582} (2004) 39},
  \href{http://arxiv.org/abs/hep-ph/0307133}{{\normalfont\ttfamily
  arXiv:hep-ph/0307133}}\relax
\mciteBstWouldAddEndPuncttrue
\mciteSetBstMidEndSepPunct{\mcitedefaultmidpunct}
{\mcitedefaultendpunct}{\mcitedefaultseppunct}\relax
\EndOfBibitem
\bibitem{Vijande:2006hj}
J.~Vijande, F.~Fernandez, and A.~Valcarce,
  \ifthenelse{\boolean{articletitles}}{\emph{{Open-charm meson spectroscopy}},
  }{}\href{http://dx.doi.org/10.1103/PhysRevD.73.034002}{Phys.\ Rev.\
  \textbf{D73} (2006) 034002}, Erratum
  \href{http://dx.doi.org/10.1103/PhysRevD.74.059903}{ibid.\   \textbf{D74}
  (2006) 059903},
  \href{http://arxiv.org/abs/hep-ph/0601143}{{\normalfont\ttfamily
  arXiv:hep-ph/0601143}}\relax
\mciteBstWouldAddEndPuncttrue
\mciteSetBstMidEndSepPunct{\mcitedefaultmidpunct}
{\mcitedefaultendpunct}{\mcitedefaultseppunct}\relax
\EndOfBibitem
\bibitem{Guo:2006fu}
F.-K. Guo {\em et~al.}, \ifthenelse{\boolean{articletitles}}{\emph{{Dynamically
  generated $0^+$ heavy mesons in a heavy chiral unitary approach}},
  }{}\href{http://dx.doi.org/10.1016/j.physletb.2006.08.064}{Phys.\ Lett.\
  \textbf{B641} (2006) 278},
  \href{http://arxiv.org/abs/hep-ph/0603072}{{\normalfont\ttfamily
  arXiv:hep-ph/0603072}}\relax
\mciteBstWouldAddEndPuncttrue
\mciteSetBstMidEndSepPunct{\mcitedefaultmidpunct}
{\mcitedefaultendpunct}{\mcitedefaultseppunct}\relax
\EndOfBibitem
\bibitem{Gamermann:2006nm}
D.~Gamermann, E.~Oset, D.~Strottman, and M.~J. Vicente~Vacas,
  \ifthenelse{\boolean{articletitles}}{\emph{{Dynamically generated open and
  hidden charm meson systems}},
  }{}\href{http://dx.doi.org/10.1103/PhysRevD.76.074016}{Phys.\ Rev.\
  \textbf{D76} (2007) 074016},
  \href{http://arxiv.org/abs/hep-ph/0612179}{{\normalfont\ttfamily
  arXiv:hep-ph/0612179}}\relax
\mciteBstWouldAddEndPuncttrue
\mciteSetBstMidEndSepPunct{\mcitedefaultmidpunct}
{\mcitedefaultendpunct}{\mcitedefaultseppunct}\relax
\EndOfBibitem
\bibitem{Alves:2008zz}
LHCb collaboration, A.~A. Alves~Jr.\ {\em et~al.},
  \ifthenelse{\boolean{articletitles}}{\emph{{The \lhcb detector at the LHC}},
  }{}\href{http://dx.doi.org/10.1088/1748-0221/3/08/S08005}{JINST \textbf{3}
  (2008) S08005}\relax
\mciteBstWouldAddEndPuncttrue
\mciteSetBstMidEndSepPunct{\mcitedefaultmidpunct}
{\mcitedefaultendpunct}{\mcitedefaultseppunct}\relax
\EndOfBibitem
\bibitem{LHCb-DP-2014-002}
LHCb collaboration, R.~Aaij {\em et~al.},
  \ifthenelse{\boolean{articletitles}}{\emph{{LHCb detector performance}},
  }{}\href{http://dx.doi.org/10.1142/S0217751X15300227}{Int.\ J.\ Mod.\ Phys.\
  \textbf{A30} (2015) 1530022},
  \href{http://arxiv.org/abs/1412.6352}{{\normalfont\ttfamily
  arXiv:1412.6352}}\relax
\mciteBstWouldAddEndPuncttrue
\mciteSetBstMidEndSepPunct{\mcitedefaultmidpunct}
{\mcitedefaultendpunct}{\mcitedefaultseppunct}\relax
\EndOfBibitem
\bibitem{BBDT}
V.~V. Gligorov and M.~Williams,
  \ifthenelse{\boolean{articletitles}}{\emph{{Efficient, reliable and fast
  high-level triggering using a bonsai boosted decision tree}},
  }{}\href{http://dx.doi.org/10.1088/1748-0221/8/02/P02013}{JINST \textbf{8}
  (2013) P02013}, \href{http://arxiv.org/abs/1210.6861}{{\normalfont\ttfamily
  arXiv:1210.6861}}\relax
\mciteBstWouldAddEndPuncttrue
\mciteSetBstMidEndSepPunct{\mcitedefaultmidpunct}
{\mcitedefaultendpunct}{\mcitedefaultseppunct}\relax
\EndOfBibitem
\bibitem{Sjostrand:2006za}
T.~Sj\"{o}strand, S.~Mrenna, and P.~Skands,
  \ifthenelse{\boolean{articletitles}}{\emph{{PYTHIA 6.4 physics and manual}},
  }{}\href{http://dx.doi.org/10.1088/1126-6708/2006/05/026}{JHEP \textbf{05}
  (2006) 026}, \href{http://arxiv.org/abs/hep-ph/0603175}{{\normalfont\ttfamily
  arXiv:hep-ph/0603175}}\relax
\mciteBstWouldAddEndPuncttrue
\mciteSetBstMidEndSepPunct{\mcitedefaultmidpunct}
{\mcitedefaultendpunct}{\mcitedefaultseppunct}\relax
\EndOfBibitem
\bibitem{Sjostrand:2007gs}
T.~Sj\"{o}strand, S.~Mrenna, and P.~Skands,
  \ifthenelse{\boolean{articletitles}}{\emph{{A brief introduction to PYTHIA
  8.1}}, }{}\href{http://dx.doi.org/10.1016/j.cpc.2008.01.036}{Comput.\ Phys.\
  Commun.\  \textbf{178} (2008) 852},
  \href{http://arxiv.org/abs/0710.3820}{{\normalfont\ttfamily
  arXiv:0710.3820}}\relax
\mciteBstWouldAddEndPuncttrue
\mciteSetBstMidEndSepPunct{\mcitedefaultmidpunct}
{\mcitedefaultendpunct}{\mcitedefaultseppunct}\relax
\EndOfBibitem
\bibitem{LHCb-PROC-2010-056}
I.~Belyaev {\em et~al.}, \ifthenelse{\boolean{articletitles}}{\emph{{Handling
  of the generation of primary events in Gauss, the LHCb simulation
  framework}}, }{}\href{http://dx.doi.org/10.1088/1742-6596/331/3/032047}{{J.\
  Phys.\ Conf.\ Ser.\ } \textbf{331} (2011) 032047}\relax
\mciteBstWouldAddEndPuncttrue
\mciteSetBstMidEndSepPunct{\mcitedefaultmidpunct}
{\mcitedefaultendpunct}{\mcitedefaultseppunct}\relax
\EndOfBibitem
\bibitem{Lange:2001uf}
D.~J. Lange, \ifthenelse{\boolean{articletitles}}{\emph{{The EvtGen particle
  decay simulation package}},
  }{}\href{http://dx.doi.org/10.1016/S0168-9002(01)00089-4}{Nucl.\ Instrum.\
  Meth.\  \textbf{A462} (2001) 152}\relax
\mciteBstWouldAddEndPuncttrue
\mciteSetBstMidEndSepPunct{\mcitedefaultmidpunct}
{\mcitedefaultendpunct}{\mcitedefaultseppunct}\relax
\EndOfBibitem
\bibitem{Golonka:2005pn}
P.~Golonka and Z.~Was, \ifthenelse{\boolean{articletitles}}{\emph{{PHOTOS Monte
  Carlo: A precision tool for QED corrections in $Z$ and $W$ decays}},
  }{}\href{http://dx.doi.org/10.1140/epjc/s2005-02396-4}{Eur.\ Phys.\ J.\
  \textbf{C45} (2006) 97},
  \href{http://arxiv.org/abs/hep-ph/0506026}{{\normalfont\ttfamily
  arXiv:hep-ph/0506026}}\relax
\mciteBstWouldAddEndPuncttrue
\mciteSetBstMidEndSepPunct{\mcitedefaultmidpunct}
{\mcitedefaultendpunct}{\mcitedefaultseppunct}\relax
\EndOfBibitem
\bibitem{Allison:2006ve}
Geant4 collaboration, J.~Allison {\em et~al.},
  \ifthenelse{\boolean{articletitles}}{\emph{{Geant4 developments and
  applications}}, }{}\href{http://dx.doi.org/10.1109/TNS.2006.869826}{IEEE
  Trans.\ Nucl.\ Sci.\  \textbf{53} (2006) 270}\relax
\mciteBstWouldAddEndPuncttrue
\mciteSetBstMidEndSepPunct{\mcitedefaultmidpunct}
{\mcitedefaultendpunct}{\mcitedefaultseppunct}\relax
\EndOfBibitem
\bibitem{Agostinelli:2002hh}
Geant4 collaboration, S.~Agostinelli {\em et~al.},
  \ifthenelse{\boolean{articletitles}}{\emph{{Geant4: a simulation toolkit}},
  }{}\href{http://dx.doi.org/10.1016/S0168-9002(03)01368-8}{Nucl.\ Instrum.\
  Meth.\  \textbf{A506} (2003) 250}\relax
\mciteBstWouldAddEndPuncttrue
\mciteSetBstMidEndSepPunct{\mcitedefaultmidpunct}
{\mcitedefaultendpunct}{\mcitedefaultseppunct}\relax
\EndOfBibitem
\bibitem{LHCb-PROC-2011-006}
M.~Clemencic {\em et~al.}, \ifthenelse{\boolean{articletitles}}{\emph{{The
  \lhcb simulation application, Gauss: Design, evolution and experience}},
  }{}\href{http://dx.doi.org/10.1088/1742-6596/331/3/032023}{{J.\ Phys.\ Conf.\
  Ser.\ } \textbf{331} (2011) 032023}\relax
\mciteBstWouldAddEndPuncttrue
\mciteSetBstMidEndSepPunct{\mcitedefaultmidpunct}
{\mcitedefaultendpunct}{\mcitedefaultseppunct}\relax
\EndOfBibitem
\bibitem{Feindt2006190}
M.~Feindt and U.~Kerzel, \ifthenelse{\boolean{articletitles}}{\emph{{The
  NeuroBayes neural network package}},
  }{}\href{http://dx.doi.org/10.1016/j.nima.2005.11.166}{Nucl.\ Instrum.\
  Meth.\ A \textbf{559} (2006) 190}\relax
\mciteBstWouldAddEndPuncttrue
\mciteSetBstMidEndSepPunct{\mcitedefaultmidpunct}
{\mcitedefaultendpunct}{\mcitedefaultseppunct}\relax
\EndOfBibitem
\bibitem{Pivk:2004ty}
M.~Pivk and F.~R. Le~Diberder,
  \ifthenelse{\boolean{articletitles}}{\emph{{sPlot: A statistical tool to
  unfold data distributions}},
  }{}\href{http://dx.doi.org/10.1016/j.nima.2005.08.106}{Nucl.\ Instrum.\
  Meth.\  \textbf{A555} (2005) 356},
  \href{http://arxiv.org/abs/physics/0402083}{{\normalfont\ttfamily
  arXiv:physics/0402083}}\relax
\mciteBstWouldAddEndPuncttrue
\mciteSetBstMidEndSepPunct{\mcitedefaultmidpunct}
{\mcitedefaultendpunct}{\mcitedefaultseppunct}\relax
\EndOfBibitem
\bibitem{LHCb-DP-2012-003}
M.~Adinolfi {\em et~al.},
  \ifthenelse{\boolean{articletitles}}{\emph{{Performance of the \lhcb RICH
  detector at the LHC}},
  }{}\href{http://dx.doi.org/10.1140/epjc/s10052-013-2431-9}{Eur.\ Phys.\ J.\
  \textbf{C73} (2013) 2431},
  \href{http://arxiv.org/abs/1211.6759}{{\normalfont\ttfamily
  arXiv:1211.6759}}\relax
\mciteBstWouldAddEndPuncttrue
\mciteSetBstMidEndSepPunct{\mcitedefaultmidpunct}
{\mcitedefaultendpunct}{\mcitedefaultseppunct}\relax
\EndOfBibitem
\bibitem{LHCb-PAPER-2012-048}
LHCb collaboration, R.~Aaij {\em et~al.},
  \ifthenelse{\boolean{articletitles}}{\emph{{Measurements of the
  $\Lambda_b^0$, $\Xi_b^-$, and $\Omega_b^-$ baryon masses}},
  }{}\href{http://dx.doi.org/10.1103/PhysRevLett.110.182001}{Phys.\ Rev.\
  Lett.\  \textbf{110} (2013) 182001},
  \href{http://arxiv.org/abs/1302.1072}{{\normalfont\ttfamily
  arXiv:1302.1072}}\relax
\mciteBstWouldAddEndPuncttrue
\mciteSetBstMidEndSepPunct{\mcitedefaultmidpunct}
{\mcitedefaultendpunct}{\mcitedefaultseppunct}\relax
\EndOfBibitem
\bibitem{LHCb-PAPER-2013-011}
LHCb collaboration, R.~Aaij {\em et~al.},
  \ifthenelse{\boolean{articletitles}}{\emph{{Precision measurement of $D$
  meson mass differences}},
  }{}\href{http://dx.doi.org/10.1007/JHEP06(2013)065}{JHEP \textbf{06} (2013)
  065}, \href{http://arxiv.org/abs/1304.6865}{{\normalfont\ttfamily
  arXiv:1304.6865}}\relax
\mciteBstWouldAddEndPuncttrue
\mciteSetBstMidEndSepPunct{\mcitedefaultmidpunct}
{\mcitedefaultendpunct}{\mcitedefaultseppunct}\relax
\EndOfBibitem
\bibitem{Hulsbergen:2005pu}
W.~D. Hulsbergen, \ifthenelse{\boolean{articletitles}}{\emph{{Decay chain
  fitting with a Kalman filter}},
  }{}\href{http://dx.doi.org/10.1016/j.nima.2005.06.078}{Nucl.\ Instrum.\
  Meth.\  \textbf{A552} (2005) 566},
  \href{http://arxiv.org/abs/physics/0503191}{{\normalfont\ttfamily
  arXiv:physics/0503191}}\relax
\mciteBstWouldAddEndPuncttrue
\mciteSetBstMidEndSepPunct{\mcitedefaultmidpunct}
{\mcitedefaultendpunct}{\mcitedefaultseppunct}\relax
\EndOfBibitem
\bibitem{Skwarnicki:1986xj}
T.~Skwarnicki, {\em {A study of the radiative cascade transitions between the
  Upsilon-prime and Upsilon resonances}}, PhD thesis, Institute of Nuclear
  Physics, Krakow, 1986,
  {\href{http://inspirehep.net/record/230779/}{DESY-F31-86-02}}\relax
\mciteBstWouldAddEndPuncttrue
\mciteSetBstMidEndSepPunct{\mcitedefaultmidpunct}
{\mcitedefaultendpunct}{\mcitedefaultseppunct}\relax
\EndOfBibitem
\bibitem{LHCb-PAPER-2015-012}
LHCb collaboration, R.~Aaij {\em et~al.},
  \ifthenelse{\boolean{articletitles}}{\emph{{Search for the decay
  $B^0_s\to\overline{D}^0f_0(980)$}},
  }{}\href{http://dx.doi.org/10.1007/JHEP08(2015)005}{JHEP \textbf{08} (2015)
  005}, \href{http://arxiv.org/abs/1505.01654}{{\normalfont\ttfamily
  arXiv:1505.01654}}\relax
\mciteBstWouldAddEndPuncttrue
\mciteSetBstMidEndSepPunct{\mcitedefaultmidpunct}
{\mcitedefaultendpunct}{\mcitedefaultseppunct}\relax
\EndOfBibitem
\bibitem{Dalitz:1953cp}
R.~H. Dalitz, \ifthenelse{\boolean{articletitles}}{\emph{{On the analysis of
  tau-meson data and the nature of the tau-meson}},
  }{}\href{http://dx.doi.org/10.1080/14786441008520365}{Phil.\ Mag.\
  \textbf{44} (1953) 1068}\relax
\mciteBstWouldAddEndPuncttrue
\mciteSetBstMidEndSepPunct{\mcitedefaultmidpunct}
{\mcitedefaultendpunct}{\mcitedefaultseppunct}\relax
\EndOfBibitem
\bibitem{Fleming:1964zz}
G.~N. Fleming, \ifthenelse{\boolean{articletitles}}{\emph{{Recoupling effects
  in the isobar model. 1. General formalism for three-pion scattering}},
  }{}\href{http://dx.doi.org/10.1103/PhysRev.135.B551}{Phys.\ Rev.\
  \textbf{135} (1964) B551}\relax
\mciteBstWouldAddEndPuncttrue
\mciteSetBstMidEndSepPunct{\mcitedefaultmidpunct}
{\mcitedefaultendpunct}{\mcitedefaultseppunct}\relax
\EndOfBibitem
\bibitem{Morgan:1968zza}
D.~Morgan, \ifthenelse{\boolean{articletitles}}{\emph{{Phenomenological
  analysis of $I=\frac{1}{2}$ single-pion production processes in the energy
  range 500 to 700 MeV}},
  }{}\href{http://dx.doi.org/10.1103/PhysRev.166.1731}{Phys.\ Rev.\
  \textbf{166} (1968) 1731}\relax
\mciteBstWouldAddEndPuncttrue
\mciteSetBstMidEndSepPunct{\mcitedefaultmidpunct}
{\mcitedefaultendpunct}{\mcitedefaultseppunct}\relax
\EndOfBibitem
\bibitem{Herndon:1973yn}
D.~Herndon, P.~Soding, and R.~J. Cashmore,
  \ifthenelse{\boolean{articletitles}}{\emph{{A generalised isobar model
  formalism}}, }{}\href{http://dx.doi.org/10.1103/PhysRevD.11.3165}{Phys.\
  Rev.\  \textbf{D11} (1975) 3165}\relax
\mciteBstWouldAddEndPuncttrue
\mciteSetBstMidEndSepPunct{\mcitedefaultmidpunct}
{\mcitedefaultendpunct}{\mcitedefaultseppunct}\relax
\EndOfBibitem
\bibitem{blatt-weisskopf}
J.~Blatt and V.~E. Weisskopf, {\em Theoretical nuclear physics}, J. Wiley (New
  York), 1952.
\newblock {page 362}\relax
\mciteBstWouldAddEndPuncttrue
\mciteSetBstMidEndSepPunct{\mcitedefaultmidpunct}
{\mcitedefaultendpunct}{\mcitedefaultseppunct}\relax
\EndOfBibitem
\bibitem{Aubert:2005ce}
\babar collaboration, B.~Aubert {\em et~al.},
  \ifthenelse{\boolean{articletitles}}{\emph{{Dalitz-plot analysis of the
  decays $B^\pm \to K^\pm \pi^\mp \pi^\pm$}},
  }{}\href{http://dx.doi.org/10.1103/PhysRevD.72.072003}{Phys.\ Rev.\
  \textbf{D72} (2005) 072003}, Erratum
  \href{http://dx.doi.org/10.1103/PhysRevD.74.099903}{ibid.\   \textbf{D74}
  (2006) 099903},
  \href{http://arxiv.org/abs/hep-ex/0507004}{{\normalfont\ttfamily
  arXiv:hep-ex/0507004}}\relax
\mciteBstWouldAddEndPuncttrue
\mciteSetBstMidEndSepPunct{\mcitedefaultmidpunct}
{\mcitedefaultendpunct}{\mcitedefaultseppunct}\relax
\EndOfBibitem
\bibitem{Zemach:1963bc}
C.~Zemach, \ifthenelse{\boolean{articletitles}}{\emph{{Three pion decays of
  unstable particles}},
  }{}\href{http://dx.doi.org/10.1103/PhysRev.133.B1201}{Phys.\ Rev.\
  \textbf{133} (1964) B1201}\relax
\mciteBstWouldAddEndPuncttrue
\mciteSetBstMidEndSepPunct{\mcitedefaultmidpunct}
{\mcitedefaultendpunct}{\mcitedefaultseppunct}\relax
\EndOfBibitem
\bibitem{Zemach:1968zz}
C.~Zemach, \ifthenelse{\boolean{articletitles}}{\emph{{Use of angular-momentum
  tensors}}, }{}\href{http://dx.doi.org/10.1103/PhysRev.140.B97}{Phys.\ Rev.\
  \textbf{140} (1965) B97}\relax
\mciteBstWouldAddEndPuncttrue
\mciteSetBstMidEndSepPunct{\mcitedefaultmidpunct}
{\mcitedefaultendpunct}{\mcitedefaultseppunct}\relax
\EndOfBibitem
\bibitem{Laura++}
{{\tt Laura++} Dalitz plot fitting package, \url{http://laura.hepfo
  rge.net}}\relax
\mciteBstWouldAddEndPuncttrue
\mciteSetBstMidEndSepPunct{\mcitedefaultmidpunct}
{\mcitedefaultendpunct}{\mcitedefaultseppunct}\relax
\EndOfBibitem
\bibitem{Achasov:2003xn}
N.~N. Achasov and G.~N. Shestakov,
  \ifthenelse{\boolean{articletitles}}{\emph{{$\pi\pi$ scattering S wave from
  the data on the reaction $\pim \proton \to \piz\piz\neutron$}},
  }{}\href{http://dx.doi.org/10.1103/PhysRevD.67.114018}{Phys.\ Rev.\
  \textbf{D67} (2003) 114018},
  \href{http://arxiv.org/abs/hep-ph/0302220}{{\normalfont\ttfamily
  arXiv:hep-ph/0302220}}\relax
\mciteBstWouldAddEndPuncttrue
\mciteSetBstMidEndSepPunct{\mcitedefaultmidpunct}
{\mcitedefaultendpunct}{\mcitedefaultseppunct}\relax
\EndOfBibitem
\bibitem{Pelaez:2004vs}
J.~R. Pelaez and F.~J. Yndurain,
  \ifthenelse{\boolean{articletitles}}{\emph{{The pion-pion scattering
  amplitude}}, }{}\href{http://dx.doi.org/10.1103/PhysRevD.71.074016}{Phys.\
  Rev.\  \textbf{D71} (2005) 074016},
  \href{http://arxiv.org/abs/hep-ph/0411334}{{\normalfont\ttfamily
  arXiv:hep-ph/0411334}}\relax
\mciteBstWouldAddEndPuncttrue
\mciteSetBstMidEndSepPunct{\mcitedefaultmidpunct}
{\mcitedefaultendpunct}{\mcitedefaultseppunct}\relax
\EndOfBibitem
\bibitem{Kaminski:2006qe}
R.~Kaminski, J.~R. Pelaez, and F.~J. Yndurain,
  \ifthenelse{\boolean{articletitles}}{\emph{{The pion-pion scattering
  amplitude. III. Improving the analysis with forward dispersion relations and
  Roy equations}},
  }{}\href{http://dx.doi.org/10.1103/PhysRevD.77.054015}{Phys.\ Rev.\
  \textbf{D77} (2008) 054015},
  \href{http://arxiv.org/abs/0710.1150}{{\normalfont\ttfamily
  arXiv:0710.1150}}\relax
\mciteBstWouldAddEndPuncttrue
\mciteSetBstMidEndSepPunct{\mcitedefaultmidpunct}
{\mcitedefaultendpunct}{\mcitedefaultseppunct}\relax
\EndOfBibitem
\bibitem{GarciaMartin:2011cn}
R.~Garcia-Martin {\em et~al.}, \ifthenelse{\boolean{articletitles}}{\emph{{The
  pion-pion scattering amplitude. IV: Improved analysis with once subtracted
  Roy-like equations up to $1100 \mev$}},
  }{}\href{http://dx.doi.org/10.1103/PhysRevD.83.074004}{Phys.\ Rev.\
  \textbf{D83} (2011) 074004},
  \href{http://arxiv.org/abs/1102.2183}{{\normalfont\ttfamily
  arXiv:1102.2183}}\relax
\mciteBstWouldAddEndPuncttrue
\mciteSetBstMidEndSepPunct{\mcitedefaultmidpunct}
{\mcitedefaultendpunct}{\mcitedefaultseppunct}\relax
\EndOfBibitem
\bibitem{Williams:2010vh}
M.~Williams, \ifthenelse{\boolean{articletitles}}{\emph{{How good are your
  fits? Unbinned multivariate goodness-of-fit tests in high energy physics}},
  }{}\href{http://dx.doi.org/10.1088/1748-0221/5/09/P09004}{JINST \textbf{5}
  (2010) P09004}, \href{http://arxiv.org/abs/1006.3019}{{\normalfont\ttfamily
  arXiv:1006.3019}}\relax
\mciteBstWouldAddEndPuncttrue
\mciteSetBstMidEndSepPunct{\mcitedefaultmidpunct}
{\mcitedefaultendpunct}{\mcitedefaultseppunct}\relax
\EndOfBibitem
\end{mcitethebibliography}

\ifx\mcitethebibliography\mciteundefinedmacro
\PackageError{LHCb.bst}{mciteplus.sty has not been loaded}
{This bibstyle requires the use of the mciteplus package.}\fi
\providecommand{\href}[2]{#2}

\clearpage
\centerline{\large\bf LHCb collaboration}
\begin{flushleft}
\small
R.~Aaij$^{40}$,
B.~Adeva$^{39}$,
M.~Adinolfi$^{48}$,
Z.~Ajaltouni$^{5}$,
S.~Akar$^{6}$,
J.~Albrecht$^{10}$,
F.~Alessio$^{40}$,
M.~Alexander$^{53}$,
S.~Ali$^{43}$,
G.~Alkhazov$^{31}$
% ...

{\footnotesize \it
% ...
$ ^{5}$Clermont Universit{\'e}, Universit{\'e} Blaise Pascal, CNRS/IN2P3, LPC, Clermont-Ferrand, France\\
$ ^{6}$CPPM, Aix-Marseille Universit{\'e}, CNRS/IN2P3, Marseille, France\\
% ...
$ ^{10}$Fakult{\"a}t Physik, Technische Universit{\"a}t Dortmund, Dortmund, Germany\\
% ...
$ ^{31}$Petersburg Nuclear Physics Institute (PNPI), Gatchina, Russia\\
% ...
$ ^{39}$Universidad de Santiago de Compostela, Santiago de Compostela, Spain\\
$ ^{40}$European Organization for Nuclear Research (CERN), Geneva, Switzerland\\
% ...
$ ^{43}$Nikhef National Institute for Subatomic Physics, Amsterdam, The Netherlands\\
% ...
$ ^{48}$H.H. Wills Physics Laboratory, University of Bristol, Bristol, United Kingdom\\
% ...
$ ^{53}$School of Physics and Astronomy, University of Glasgow, Glasgow, United Kingdom\\
% ...
}
\end{flushleft}

\end{document}